\documentclass[12pt]{article}
\usepackage{amssymb}
\usepackage{graphicx}
\usepackage{amsmath}

\def\ee{\end{eqnarray}}

\newcommand{\nn}{\nonumber}

\def\D{\mathcal{D}}

\def\=:{=\hspace{-.7em}\raisebox{1.1ex}{.}\hspace{.1em}\raisebox{-0.2ex}{.} }
\newcommand{\NF}{N_{\rm F}}
\newcommand{\NC}{N_{\rm C}}

\newcommand {\beq}{\begin{eqnarray}}
\newcommand {\eeq}{\end{eqnarray}}
\newcommand {\non}{\nonumber\\}

\newcommand {\1}[1]{\frac{1}{#1}}

\newcommand {\thb}{\bar{\theta}}

\newcommand {\sig}{\sigma}

\newcommand {\del}{\partial}
\newcommand {\dagg}{^{\dagger}}

\newcommand{\vs}[1]{\vspace{#1 mm}}
\newcommand{\hs}[1]{\hspace{#1 mm}}

\newcommand {\Nf}{N_{\rm F}} 
\newcommand {\Nc}{N_{\rm C}} 

\makeatletter
\@addtoreset{equation}{section}

\renewcommand{\thefootnote}{\fnsymbol{footnote}}
\makeatother

\begin{document}
\thispagestyle{empty}
\begin{flushright}
TIT/HEP--531 \\
RIKEN-TH--32 \\
{\tt hep-th/0412024} \\
December, 2004 \\
\end{flushright}
\vspace{3mm}

\begin{center}
{\Large \bf 
D-brane Construction for Non-Abelian Walls} 
\\[12mm]
\vspace{5mm}

\normalsize
 {\large \bf 
Minoru~Eto}$^1$
\footnote{\it  e-mail address: 
meto@th.phys.titech.ac.jp
}, 
  {\large \bf 
Youichi~Isozumi}$^1$
\footnote{\it  e-mail address: 
isozumi@th.phys.titech.ac.jp
}, 
  {\large \bf 
Muneto~Nitta}$^1$
\footnote{\it  e-mail address: 
nitta@th.phys.titech.ac.jp
}, 
  {\large \bf 
 Keisuke~Ohashi}$^1$\footnote{\it  e-mail address: 
keisuke@th.phys.titech.ac.jp
}, 
{\large \bf 
 Kazutoshi~Ohta}$^2$\footnote{\it  e-mail address: 
k-ohta@riken.jp
}, 
~and~~  {\large \bf 
Norisuke~Sakai}$^1$
\footnote{\it  e-mail address: 
nsakai@th.phys.titech.ac.jp
} 

\vskip 1.5em

$^1$ {\it Department of Physics, Tokyo Institute of 
Technology \\
Tokyo 152-8551, JAPAN}
 
 and 
 
$^2$ {\it Theoretical Physics Laboratory\\
The Institute of Physical and Chemical Research (RIKEN)\\
2-1 Hirosawa, Wako, Saitama 351-0198, JAPAN}\\
\vspace{15mm}
{\bf Abstract}\\[5mm]
{\parbox{13cm}{\hspace{5mm}

Supersymmetric $U(\Nc)$ gauge theory 
with $\Nf$ massive hypermultiplets 
in the fundamental representation 
is given by the brane configuration 
made of $\Nc$ fractional D$p$-branes stuck at 
the ${\bf Z}_2$ orbifold singularity 
on $\Nf$ separated D$(p+4)$-branes. 
We show that non-Abelian walls in this theory 
are realized as kinky fractional D$p$-branes 
interpolating between D$(p+4)$-branes. 
Wall solutions and their duality between 
$\Nc$ and $\Nf - \Nc$ imply extensions of 
the s-rule and the Hanany-Witten effect
in brane dynamics.
We also find that the reconnection of fractional D-branes 
occurs in this system.
Diverse phenomena in non-Abelian walls found 
in field theory can be understood 
very easily by this brane configuration.

}}
\end{center}
\vfill
\newpage
\setcounter{page}{1}
\setcounter{footnote}{0}
\renewcommand{\thefootnote}{\arabic{footnote}}

\section{Introduction}\label{INTRO}

Yang-Mills-Higgs systems admit 
stable topological solitons; 
instantons, monopoles, vortices and domain walls. 
When they saturate energy bounds 
in supersymmetric (SUSY) gauge theories, 
they are called 
Bogomol'nyi-Prasad-Sommerfield (BPS) states 
preserving some fraction of SUSY.  
There are no forces between BPS solitons in general 
and all of these solutions constitute moduli spaces. 
The ADHM construction for instantons~\cite{ADHM} and 
the Nahm construction for BPS monopoles~\cite{Nahm} 
were given long time ago 
and have been applied to a lot of subjects in 
physics and mathematics. 
In particular monopoles and instantons 
played crucial roles in non-perturbative dynamics in 
${\cal N}=2$ and ${\cal N}=1$ 
SUSY gauge theories in four dimensions~\cite{SW}. 
Contrary to these developments, 
it was, however, the last year that 
the moduli space of multiple vortices was 
constructed by Hanany and Tong 
in non-Abelian gauge theories~\cite{HT,HT2,ENS}
following the discovery of 
the moduli space for a single vortex~\cite{Auzzi:2003fs}.
There have been few discussions for 
explicit constructions of moduli spaces of walls; 
notable exceptions are the cases of 
${\cal N}=1$ generalized Wess-Zumino models~\cite{Shifman:1997hg}, 
${\cal N}=1$ SUSY gauge theories~\cite{Acharya:2001dz,Ritz:2002fm} 
and ${\cal N}=2$ SUSY gauge theories with {\it Abelian} 
gauge group~\cite{GTT2,To}. 
Construction for wall solutions and their moduli space 
in ${\cal N}=2$ SUSY non-Abelian gauge theory 
with eight supercharges was desired.

Recently a systematic method to construct all exact wall 
solutions in $U(\Nc)$ gauge theory with 
$\Nf$ hypermultiplets in the fundamental representation, 
which we call {\it non-Abelian walls}, 
has been given in \cite{INOS1,INOS2} 
(see \cite{INOS4} for a review). 
Their moduli space has been completely determined to be 
the complex Grassmann manifold 
$G_{\Nf,\Nc} \simeq 
SU(\Nf)/[SU(\Nc) \times SU(\Nf-\Nc) \times U(1)]$ 
endowed with a deformed metric. 
One of interesting aspects is that this moduli space is 
a total space containing all topological sectors 
with different boundary conditions and with 
different number of walls.
Moreover the construction method 
has been generalized in \cite{INOS3} to exact 
solutions for a set of 1/4 BPS equations 
which are made of walls, vortices and monopoles 
in the Higgs phase \cite{wv,vm,HT2,Kneipp,Auzzi:2004yg}, 
and to those for another set of 1/4 BPS equations~\cite{HT2} 
for vortices and instantons in the Higgs phase~\cite{EINOS1}.

D-branes in string theory realize SUSY gauge theories 
on their world-volume at low energy. 
Using this property non-perturbative dynamics for SUSY gauge theories 
was clarified by brane dynamics in several brane configurations 
and vice versa~\cite{HW}--\cite{Giveon:1998sr}.
Also D-brane configurations are very useful tools to 
describe various aspects of 
BPS solitons in SUSY gauge theories. 
Instantons are realized by a D$p$-D$(p+4)$ system~\cite{Wi}  
where $k$ D$p$-branes are regarded as $k$-instantons in 
the effective $U(N)$ gauge theory on $N$ D$(p+4)$-branes.\footnote{
We call solitons with co-dimension four as instantons.
} 
The ADHM conditions are obtained as 
the SUSY vacuum conditions in the effective $U(k)$ gauge theory 
on the D$p$-branes.
Taking a T-dual of this configuration we obtain one made of 
$k$ D$(p+1)$-branes ending on $N$ D$(p+3)$-branes.
This is a brane configuration for BPS 
monopoles~\cite{brane-monopole} giving 
the Nahm construction in the same way. 
The brane configuration for vortices was 
realized by Hanany and Tong~\cite{HT,HT2} on 
D$p$-brane world-volume in a D$p$-D($p+2$)-D$(p+4)$-NS$5$ system. 
There are several advantages to consider brane configurations 
for analysis of BPS solitons. 
One advantage is that we can interpret the ADHM or other conditions 
for the moduli space of solitons  
by the F- and D-flatness conditions of SUSY vacua as stated above. 
For vortices the brane configuration is the only available method 
to construct their moduli space 
although there exists some ambiguity 
in the moduli metric~\cite{HT,ENS}.  
The other advantage is for instance that 
these brane configurations give solitons in non-commutative space
if we turn on an NS-NS $B$-field as 
a background~\cite{NS,Hashimoto:1999zw,Gross:2000ss}.

\begin{figure}[thb]
\begin{center}
\includegraphics[width=7cm,clip]{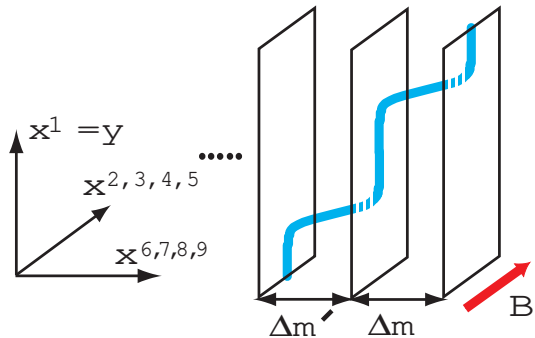}
\end{center}
\caption{\small
A kinky D-brane configuration for Abelian walls. 
The theory is realized on a single D$p$-brane in
the background of $\Nf$ D($p+4$)-branes with 
separated positions and the $B$-field. 
Multiple walls in SUSY $U(1)$ gauge theory are realized 
as a single kinky D-brane interpolating between 
the separated D($p+4$)-branes~\cite{LT,GTT2}.
The extra coordinate is denoted by $y = x^1$.
}
\label{fig1}
\end{figure}

In this paper we give a D-brane configuration 
for non-Abelian walls found in \cite{INOS1,INOS2}, 
generalizing the kinky D-brane in 
the type II string theories 
given by Lambert and Tong~\cite{LT,GTT2} 
for Abelian walls in SUSY $U(1)$ gauge theory 
(with eight supercharges) containing 
$\Nf$ massive hypermultiplets 
in the fundamental representation 
and a Fayet-Iliopoulos (FI) term (see Fig.~\ref{fig1}). 
Here we recall their discussion \cite{LT} briefly. 
In the case of massless hypermultiplets, 
the moduli space of vacua in the original theory 
is realized as a D$p$-brane world-volume theory in 
a D$p$-D($p+4$) system in type II string theories 
with a constant $B$-field 
perpendicular to the single D$p$-brane 
along the coincident $N_{\rm F}$ D$(p+4)$-branes. 
The effective action for 
the D$(p+4)$-branes with the $B$-field is 
a non-commutative $U(N_{\rm F})$ gauge theory whereas 
the D$p$-brane inside them can be 
regarded as a non-commutative instanton~\cite{NS}. 
The effective theory on the D$p$-brane is 
the SUSY $U(1)$ gauge theory with eight supercharges
containing $\Nf$ massless hypermultiplets and 
the FI-term proportional to the $B$-field. 
The moduli space of vacua in the D$p$-brane theory coincides with 
the ADHM moduli space for the single non-commutative instanton. 
Turning on masses for hypermultiplets 
correspond to placing D($p+4$)-branes at separated positions 
where mass differences determine 
distances of adjacent D($p+4$)-branes.  
(Here we consider real masses most suitable to consider 1/2 BPS walls 
although mass parameters can become 
complex for $p=3$, triplet for $p=2$ and quartet for $p=1$.) 
The effective theory on the single D$p$-brane has
a potential due to the FI-term, 
from which the theory contains $N_{\rm F}$ discrete 
degenerate vacua corresponding to 
$N_{\rm F}$ positions of the D$(p+4)$-branes~\cite{LT}. 
Multiple domain wall configuration is realized as 
a kinky D$p$-brane interpolating between  
one D($p+4$)-brane to another D($p+4$)-brane 
successively as illustrated in Fig.~\ref{fig1},  
where transverse modes of the D$p$-brane 
perpendicular to D($p+4$)-brane world-volume 
are parametrized by the scalar field components 
in the vector multiplets 
and those along the D($p+4$)-branes are decoupled 
because of single D$p$-brane.
The 1/2 BPS condition allows the D$p$-brane to curve 
into only one direction as illustrated in Fig.~\ref{fig1}, 
and this is why it is enough to consider 
real masses for 1/2 BPS states 
even in higher dimensions.

Naive extension to multiple $\Nc$ D$p$-branes 
gives SUSY $U(N_{\rm C})$ gauge theory 
with eight supercharges containing 
an adjoint hypermultiplet in addition to
$N_{\rm F}$ fundamental hypermultiplets 
on their world volume. 
The adjoint hypermultiplet describes 
positions of the D$p$-branes along 
the D$(p+4)$-brane world-volume 
which is not decoupled unless $\Nc=1$ discussed above;
on the other hand, walls in \cite{INOS1,INOS2} 
are constructed in SUSY gauge theory 
without adjoint hypermultiplets. 
To remove that unwanted adjoint hypermultiplet 
we divide the orthogonal space ${\bf C}^2$ of 
the D$p$-branes along  
the D($p+4$)-brane world-volume by ${\bf Z}_2$. 
The fixed point of the ${\bf Z}_2$-action becomes 
an orbifold singularity of ${\bf C}^2/{\bf Z}_2$.
Fractional D$p$-branes mean D$p$-branes
with the fractional tension of 
the ordinary D$p$-branes,
which are stuck at the singularity 
and a part of whose world-volume is wrapped
on the shrunken cycles
of the orbifold.
Since they do not have degree of freedom 
to move in the D$(p+4)$-brane world-volume,  
their world-volume theory does not contain 
adjoint hypermultiplets
corresponding to these directions. 
We thus realize the SUSY $U(\Nc)$ gauge theory 
with $\Nf$ fundamental hypermultiplets 
without adjoint hypermultiplets on fractional D$p$-branes 
stuck at the orbifold singularity. 
Ground states of the fractional D$p$-branes can be 
understood as Yang-Mills instantons 
in the orbifold ${\bf C}^2/{\bf Z}_2$ whose singularity 
is blown up by $S^2$ to the $A_1$ gravitational instanton, 
namely, the Eguchi-Hanson space~\cite{Eguchi:1978xp}. 
Yang-Mills instantons in gravitational instantons 
were constructed in \cite{KN} and 
were later interpreted by D-brane configurations~\cite{DM}. 
Non-Abelian walls are realized  
as multiple kinky fractional D$p$-branes in such a system.

This paper is organized as follows. 
We give a brief review of the field theoretical results 
for non-Abelian walls in Sec.~\ref{NAW}.  
We discuss the vacuum structure of our theory
by a brane configuration of a D$p$-D($p+4$) system 
with ${\bf C}^2/{\bf Z}_2$ in Sec.~\ref{MSOV}. 
Taking a T-dual it becomes a brane configuration 
of Hanany-Witten~\cite{HW} made of 
$\Nc$ D($p+1$)-branes stretched between 
two parallel NS$5$-branes 
with orthogonal $\Nf$ D($p+3$)-branes. 
There the s-rule (see \cite{Giveon:1998sr} for review) is essential 
to count the vacua, and 
the Hanany-Witten effect~\cite{HW,NOYY2} 
explains the duality between two gauge theories 
with $\Nc \leftrightarrow \Nf - \Nc$ 
for vacuum states. 
Sec.~\ref{NWAMKD} contains the main results in this paper.
There we realize non-Abelian walls 
as multiple kinky fractional D$p$-branes in 
the original brane configuration before taking T-duality. 
The brane picture is very useful to read 
the vacua which appear when two adjacent walls are 
separated enough. 
We show that the tension of the fractional D$p$-branes 
correctly reproduce the tension of walls. 
We also calculate the tension of walls as 
the tension of the D($p+1$)-branes 
in the T-dualized configuration.
We find that 
the s-rule and the Hanany-Witten effect 
are extended to kinky D-brane configurations 
to understand the dynamics of non-Abelian 
walls found in \cite{INOS1,INOS2}. 
The reconnection (recombination) of D$p$-branes~\cite{recombination} 
is found to occur in our kinky D$p$-brane configurations. 
Dimensions of the wall moduli space 
and its topological sectors are 
calculated in the brane picture. 
Sec.~\ref{CD} is devoted to conclusion 
and discussions such as a brane configuration 
for 1/4 BPS states of wall, vortices and monopoles,  
a brane configuration for non-BPS walls, 
extension to degenerate masses 
or complex or more general masses for hypermultiplets and
a brane configuration for a wall junction.

\section{Non-Abelian Walls} \label{NAW}
In this section we briefly present 
field theoretical description 
of non-Abelian walls~\cite{INOS1,INOS2,INOS4}. 
We consider five-dimensional ($D=4+1$) theory 
because it is the maximal dimension 
with massive hypermultiplets admitting domain walls. 
We denote space-time indices by 
$M,N, \cdots=0,1,2,3,4$ with the metric 
$\eta_{MN}={\rm diag}(+1,-1,-1,-1,-1)$. 
Physical bosonic fields in the vector multiplet are 
gauge field $W_M = W_M^I T_I$, 
and a real scalar field $\Sigma = \Sigma^I T_I$ 
in the adjoint representation of the $U(N_{\rm C})$ gauge group,
with $T_I$ ($I=0,1,\cdots,N^2-1$) generators 
for the Lie algebra of $U(N_{\rm C})$.
Physical bosonic fields in the hypermultiplets are 
an $SU(2)_R$ doublet of complex scalars 
represented by two 
$\Nc \times \Nf$ matrices $(H^i)^{rA} \equiv (H^{irA})$ 
with $SU(2)_R$ $i=1,2$,  
color $r=1, \cdots, N_{\rm C}$ and flavor 
$A=1, \cdots, N_{\rm F}$ indices. 

The bosonic Lagrangian is given by 
\begin{equation}
{\cal L} 
=
-\frac{1}{2g^2}{\rm Tr}\left( F_{MN}F^{MN}\right)
+\frac{1}{g^2}{\rm Tr}
\left({\cal D}_M \Sigma {\cal D}^M \Sigma \right)
+{\rm Tr}
\left[ {\cal D}^M H^i ({\cal D}_M H^i)^\dagger \right] -V,
  \label{Lagrangian}
\end{equation}
where $g$ is the gauge coupling constant 
taken common for $U(1)$ and $SU(N)$ gauge groups. 
The covariant derivatives are defined by 
${\cal D}_M \Sigma = \partial_M \Sigma + i[ W_M , \Sigma ]$, 
${\cal D}_M H^{i}=(\partial_M + iW_M)H^{i} $, 
and the field strength by
$F_{MN}=\frac{1}{i}[{\cal D}_M , {\cal D}_N]
=\partial_M W_N -\partial_N W_M + i[W_M, W_N]$. 
The potential $V$ is given by 
\begin{eqnarray}
V
&\!\!\!=&\!\!\! 
\frac{g^2}{4}
{\rm Tr}
\left[
\left(
H^{1}  H^{1\dagger}  - H^{2} H^{2\dagger} 
- c\mathbf{1}_{N_{\rm C}}
\right)^2 +
4H^2H^{1\dagger} H^1H^{2\dagger} 
\right] 
\nonumber\\
&
&
+{\rm Tr}\left[
 (\Sigma H^i - H^i M) 
 (\Sigma H^i - H^i M)^\dagger 
 \right],
\end{eqnarray}
where we have taken a triplet of the Fayet-Iliopoulos (FI) 
parameters to the third direction as $(0,0,c)$, 
using the $SU(2)_R$ rotation 
without loss of generality.  
Here the mass matrix is defined by 
$(M)^A{}_B\equiv m_A\delta ^A{}_B$. 
We consider non-degenerate masses $m_A \neq m_B$ for $A \neq B$ and 
order them as $m_{A+1} < m_{A}$.  
The flavor group $SU(\Nf)$ is explicitly broken 
to $U(1)^{\Nf-1}$ by these non-degenerate masses. 
For (partially) degenerate masses this flavor group is 
(partially) recovered.

In the massless case ($m_A=0$ for all $A$) 
the moduli space of vacua 
becomes the Higgs branch given by  
a hyper-K\"ahler quotient~\cite{HKLR} 
resulting  
the cotangent bundle over the complex Grassmann manifold~\cite{LR,ANS}
\beq
 {\cal M}^{M=0}_{\rm vacua} 
  \simeq T^* G_{\NF,\NC} 
  = T^* \left[{SU(\Nf) \over SU(\Nc) 
   \times SU(\tilde \Nc) \times U(1)} \right]\;, 
    \label{massless-Higgs}
\eeq
with $\tilde \Nc \equiv \Nf - \Nc$, 
on which flavor symmetry $SU(\Nf)$ acts 
as a tri-holomorphic isometry. 
Turning on masses for hypermultiplets, 
most points on 
${\cal M}^{M=0}_{\rm vacua}$ are lifted leaving 
the discrete SUSY vacua given by 
\begin{eqnarray}
 && H^{1rA}=\sqrt{c}\,\delta ^{A_r}{}_A,\quad H^{2rA}=0, \non
 &&\Sigma ={\rm diag.}(m_{A_1},\,m_{A_2},\,\cdots,\,
           m_{A_{N_{\rm C}}}).  \label{massive-vacua}
\end{eqnarray}
Note that there are no non-baryonic Higgs 
branches~\cite{Carlino:2000ff} due to 
the FI-parameters. 
These SUSY vacua (the baryonic Higgs branch) can be labeled by 
a set of non-zero elements $A_r$ as
\beq
 \langle A_1,\cdots, A_{\NC}\rangle , \label{vacua-labels}
\eeq
and therefore the number of SUSY vacua is obtained as 
\beq
 {}_{\Nf} C_{\Nc} = {\Nf ! \over \Nc ! \tilde \Nc!} . 
  \label{number-vacua}
\eeq
For partially degenerate masses continuously degenerate 
vacua appear.

The BPS equations are obtained by either 
the 1/2 BPS condition on the fermion fields or 
the Bogomol'nyi completion on energy density as  
\begin{eqnarray}
 && \D_y \Sigma =  
 {g^2\over 2}\left(c{\bf 1}_{N_{\rm C}}-H^1H^1{}^\dagger 
   +H^2H^2{}^\dagger \right), \hs{5} 
 0 = - g^2 H^2H^1{}^\dagger , \non
 && \D_y H^1 = -\Sigma H^1 + H^1 M, \hs{5}
    \D_y H^2 = \Sigma H^2 -H^2 M ,\label{BPSeqs}
\end{eqnarray}
where we have denoted the extra dimension perpendicular to 
walls by $y$. 
The tension of multiple BPS walls, 
interpolating between a vacuum labeled by 
$\langle A_1,\cdots, A_{\NC}\rangle$ at $y \to + \infty$ 
and a vacuum 
$\langle B_1,\cdots, B_{\NC}\rangle$ at $y \to - \infty$, 
is obtained as
\begin{eqnarray}
T_{\rm w} =
 c \left[{\rm Tr}\Sigma \right]^{y = +\infty}_{y = -\infty}
\!\!=\!c \left(\sum_{r=1}^{N_{\rm C}}m_{A_r}
-\sum_{r=1}^{N_{\rm C}}m_{B_r}\right) . 
\label{eq:tension}
\end{eqnarray}

To solve the BPS equations (\ref{BPSeqs}), 
it is convenient to define a matrix function $S (y)$ 
taking values in $GL(\Nc,{\bf C})$ by 
\begin{eqnarray} 
  \Sigma + iW_y \equiv S^{-1}\partial_y S \,.
  \label{def-S}
\end{eqnarray}
Then the hypermultiplets can be solved  
from the third and forth equations in (\ref{BPSeqs}) 
as 
\begin{eqnarray}
  H^1 (y) = S^{-1}(y)    H_0^1 e^{My},\quad 
  H^2 (y) = S^\dagger(y) H_0^2 e^{-My} 
 \label{sol-H}
\end{eqnarray}
with $H^i_0$ ($i=1,2$) $\Nc \times \Nf$ constant complex matrices.  
In our choice of the FI-parameter, 
the vacua are obtained with $H^2=0$ 
as Eq.~(\ref{massive-vacua}).
It was shown in Appendix C in \cite{INOS2} that 
$H_0^2$ must vanish for 
any wall configuration interpolating between these vacua:  
\beq
 H_0 \equiv H_0^1 \neq 0, \hs{5} 
 H_0^2 = 0  \label{H2=0}
\eeq
with $H_0$ being a rank $\Nc$ matrix, which we call 
the {\it moduli matrix}.

Defining an $\Nc \times \Nc$ gauge invariant matrix
\begin{eqnarray}
 \Omega  \equiv SS^\dagger , 
\label{def-Omega}
\end{eqnarray}
the first equation in (\ref{BPSeqs}) can be rewritten as
\begin{eqnarray}
 \partial_y^2 \Omega -\partial_y \Omega 
 \Omega^{-1} \partial_y \Omega = g^2 
 \left(c\, \Omega - H_0 \,e^{2My} H_0{}^\dagger 
    \right).\label{diff-eq-S}
\end{eqnarray}
Once a solution $\Omega (y)$ of this equation is obtained, 
$S$ can be calculated by Eq.~(\ref{def-Omega}) 
with fixing a gauge. 
Then all physical quantities 
$H^i(y)$, $\Sigma(y)$ and $W_y(y)$ 
can be obtained from $S$ by Eqs.~(\ref{def-S}) and (\ref{sol-H}).

A set $(S',H_0^1{}')$ and another set 
$(S,H_0)$ related by the following relation 
give the same physical quantities: 
\begin{eqnarray}
 S' = VS,\quad 
 H_0{}'=V H_0,\quad 
  \label{art-sym}
\end{eqnarray} 
with $V\in GL (N_{\rm C},{\bf C})$.
We call this transformation by $GL (N_{\rm C},{\bf C})$ 
the {\it world-volume symmetry}.

All moduli parameters in wall solutions 
are contained in the moduli matrix $H_0$. 
However the world-volume symmetry (\ref{art-sym}) enforces 
a equivalence relation on $H_0$, 
and not all of parameters in $H_0$ are independent. 
Therefore the moduli space ${\cal M}_{\rm wall}$ 
for wall configurations 
are obtained as the complex Grassmann manifold
\begin{eqnarray}
 {\cal M}_{\rm wall}
 \simeq \{H_0 | H_0 \sim V H_0, V \in GL(N_{\rm C},{\bf C})\} 
 \simeq G_{N_{\rm F},N_{\rm C}}
 \simeq {SU(N_{\rm F}) \over 
 SU(N_{\rm C}) \times SU(\tilde N_{\rm C}) \times U(1)}\,.
  \label{Gr}
\end{eqnarray} 
In particular its real dimension is 
\beq
 \dim {\cal M}_{\rm wall} = 2 \Nc \tilde \Nc 
  = 2 \Nc (\Nf - \Nc). 
  \label{dim-wall-mod0}
\eeq
This moduli space contains all topological sectors 
with different possible boundary conditions:  
\beq
 {\cal M}_{\rm wall} 
 = \sum_{\rm BPS} {\cal M}^{\langle A_1,\cdots, A_{\NC} \rangle 
                 \leftarrow \langle B_1,\cdots, B_{\NC} \rangle}
\eeq
where 
${\cal M}^{\langle A_1,\cdots, A_{\NC} \rangle 
\leftarrow \langle B_1,\cdots, B_{\NC} \rangle}$
is defined as the topological sector with boundary conditions, 
$\langle A_1,\cdots, A_{\NC}\rangle$ at $y \to + \infty$ 
and  
$\langle B_1,\cdots, B_{\NC}\rangle$ at $y \to - \infty$. 
Therefore we call ${\cal M}_{\rm wall}$ the {\it total moduli space 
for walls}.

To obtain a (multiple) wall 
configuration in the topological sector 
${\cal M}^{\langle A_1,\cdots, A_{\NC} \rangle
\leftarrow \langle B_1,\cdots, B_{\NC} \rangle}$ 
we need to fix the world-volume symmetry (\ref{art-sym}) 
in a proper way. 
Any moduli matrix for such a configuration in that topological sector 
can be uniquely fixed as the form of
\begin{eqnarray}
&&\hspace{5.2em}A_1\hspace{3.1em}A_r
 \hspace{2.7em}\stackrel{y\rightarrow \infty }{\longleftarrow }
 \hspace{2.6em}B_1\hspace{4.2em}B_r \nn\\
 H_0&=&\sqrt{c}\left(
\begin{array}{cccccccccccc}
\cdots 0&1&*\cdots &*&\cdots  & &\cdots *&e^{v_1} &0\cdots     \\
 & &\vdots  & &        & &\vdots  &      &                 &&& \\
 & &\cdots 0&1&*\cdots & &\cdots  &    &\cdots*&e^{v_r}&0\cdots\\
 & &\vdots  & &        & &\vdots  &      &                 &&& \\
 & &  & &\cdots 0&1&*\cdots &\cdots *&e^{v_{N_{\rm C}}}&0\cdots\\
\end{array}\right)  {{\small <r}} \, , 
  \label{standard-form}\\
&&\hspace{13.8em}A_{N_{\rm C}}\hspace{5.2em} B_{N_{\rm C}} \nn
\end{eqnarray}
where in the $r$-th row 
the left-most non-zero $(r, A_r)$-elements are fixed to be one,  
the right-most non-zero $(r,B_r)$-elements  
are denoted by 
$e^{v_r} (\in {\bf C}^* \equiv {\bf C} - \{0\} \simeq {\bf R} \times S^1)$, 
and elements between 
them denoted by $* (\in {\bf C})$ are 
complex parameters which can be zero. 
We call this matrix the {\it standard form} of $H_0$. 
Here $A_r$ are ordered as $A_r<A_{r+1}$ 
using the world-volume symmetry (\ref{art-sym}), 
but $B_r$ cannot be ordered in general.
When $B_r$ are not ordered some elements between 1 and $e^{v_r}$ 
must be eliminated to fix the world-volume symmetry 
(\ref{art-sym}) completely, 
where the positions of zeros can be uniquely determined 
(see Appendix B in \cite{INOS2}).
On the other hand, when $B_r$ happen to be ordered all elements 
can be non-zero and we call that moduli matrix 
the {\it generic moduli matrix} in that topological sector. 
This implies that the generic region of 
the topological sector 
${\cal M}^{\langle A_1,\cdots, A_{\NC} \rangle 
\leftarrow \langle B_1,\cdots, B_{\NC} \rangle}$ 
is covered by the moduli parameters in 
the generic moduli matrix, 
whereas those of 
the other moduli matrices cover 
regions with smaller dimensions  
which are boundaries of the generic region.
Each topological sector is completely covered by 
patches 
$U^{\langle A_1, \cdots, A_{\Nc} 
\leftarrow B_{1}, \cdots, B_{\Nc} \rangle}$
of moduli matrices in the standard form 
(\ref{standard-form}) with 
all allowed ordering of $B_r$ with $A_r < B_r$.\footnote{
Here we put an arrow inside a bracket because 
we cannot arrange $B_r$ to be ordered 
in one standard form (\ref{standard-form})
and the ordering of $B_r$ is important.
} 
The topological sector can be decomposed as
\beq
 {\cal M}^{\langle A_1, \cdots, A_{\Nc} \rangle 
 \leftarrow  \langle B_1, \cdots, B_{\Nc} \rangle} 
 = \sum_{\sig [A_r \leq B_{\sig (r)}]} 
  U^{\langle A_1, \cdots, A_{\Nc} 
     \leftarrow B_{\sig(1)}, \cdots, B_{\sig(\Nc)} \rangle}
 \label{decomposed}
\eeq 
where 
$\sig$ denotes an element of the permutation group 
acting on the ordered $B_r$, 
and the sum is taken over all allowed permutations 
under the condition $A_r \leq B_{\sig (r)}$ for 
each $r$-th row. 
Here each pair of $U$'s does not have overlap 
because the standard form is unique once 
the ordering for $B_r$ is given.

It is in general so difficult to solve Eq.~(\ref{diff-eq-S})
explicitly. 
For finite gauge coupling with particular values 
some exact solutions are obtained 
for $U(1)$ gauge theories~\cite{LT,KS,IOS1,IOS2}.
We can obtain exact solutions with full generic 
moduli for non-Abelian gauge theories 
by taking the strong gauge coupling limit 
$g \to \infty$. 
In this limit Eq.~(\ref{diff-eq-S}) can be algebraically 
solved to give
\beq
 \Omega_{g \to \infty} 
 \equiv \Omega_0 = c^{-1} H_0 e^{2 M y} H_0{}^\dagger \;.
  \label{infinite}
\eeq
The original Lagrangian (\ref{Lagrangian}) reduces to 
a hyper-K\"ahler nonlinear sigma model on the target space 
$T^* G_{\NF,\NC} (\simeq {\cal M}^{M=0}_{\rm vacua})$ 
in Eq.~(\ref{massless-Higgs}) 
with the potential term~\cite{ANS}. 
This method to obtain hyper-K\"ahler manifold is known as 
a hyper-K\"ahler quotient~\cite{LR,HKLR}. 
The potential term can be written as square of a 
tri-holomorphic Killing vector~\cite{AF} 
generated by $\sum_A m_A t_A$ with 
$t_A$ a $U(1) [\subset U(1)^{\Nf}]$ generator acting on $H^{irA}$ 
(the overall $U(1)$ is gauged away). 
This type of model 
is called a massive hyper-K\"ahler nonlinear sigma model 
and is known to admit various (composite) 
BPS solitons like (multiple) walls~\cite{AT,GTT2,To,ANNS}, 
strings ending on a wall~\cite{wv} 
and stretched between multiple walls~\cite{INOS3}, 
intersecting walls~\cite{GTT1} and 
intersecting strings (lumps)~\cite{Naganuma:2001pu}.

\section{Brane Configuration for Moduli Space of Vacua}\label{MSOV}

\subsection{Massless Hypermultiplets}

\begin{figure}[thb]
\begin{center}
\includegraphics[width=7cm,clip]{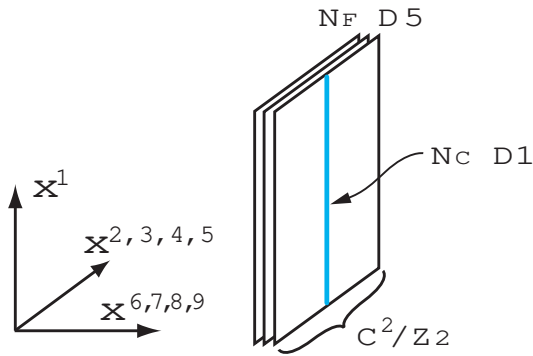}
\end{center}
\caption{\small
A D-brane configuration for 
massless hypermultiplets in the fundamental representation.
}
\label{fig3}
\end{figure}

First we discuss the case that 
all hypermultiplets are massless. 
Consider a D$p$-D$(p+4)$ system with $\NC$ D$p$-branes 
and $\NF$ coincident D$(p+4)$-branes in type IIA/IIB string theory. 
Moreover we divide the orthogonal space ${\bf C}^2$ 
perpendicular to the D$p$-branes 
inside the D$(p+4)$-brane world-volume by ${\bf Z}_2$ 
(to remove unwanted adjoint hypermultiplets)
and turn on a constant self-dual NS-NS $B$-field 
on ${\bf C}^2/{\bf Z}_2$. 
The ALE space of the $A_1$-type, the Eguchi-Hanson space, 
is obtained by blowing up the orbifold singularity by 
inserting $S^2$. 
The positions of fractional D$p$-branes 
inside the D$(p+4)$-branes  
are localized at the fixed point of the orbifold singularity 
of ${\bf C}^2/{\bf Z}_2$ because they are 
D$(p+2)$-branes wrapping around $S^2$ which blows up 
the singularity. 
In the case of $p=1$ in the type IIB string theory, 
we obtain the following configuration 
(see Fig.~\ref{fig3})
\beq
 \mbox{$\NC$ D$1$:} && 01     \non
 \mbox{$\NF$ D$5$:} && 012345 \non
 \mbox{${\bf C}^2/{\bf Z}_2$ ALE:}        && \hs{4.5} 2345
\eeq

The $\Nc$ D$p$-branes are interpreted as 
instantons (with the instanton number $\Nc$) on the ALE space 
in the effective $U(\Nf)$ gauge theory on 
the D$(p+4)$-branes. 
The moment maps of $U(\Nc)$ gauge group 
for the ADHM moduli space of instantons 
in the ALE space are found 
by Kronheimer and Nakajima \cite{KN} to be 
\beq
 \mu_{\bf C} = I J  , \hs{5} 
 \mu_{\bf R} = I I\dagg - J\dagg J 
\eeq
with $I$ and $J$ 
being $\NC \times \NF$ and $\NF \times \NC$ matrices 
with complex components, respectively. 
A set of $I$ and $J$ can be interpreted as 
an $\NC \times \NF$ matrix of hypermultiplets arisen from 
a string stretched between the D$p$-branes and the D($p+4$)-branes.
Then the ADHM moduli space is obtained as 
a hyper-K\"ahler quotient~\cite{HKLR}, to yield 
\beq 
 {\cal M}^{M=0}_{\rm vacua} \simeq 
  \{ (I,J) | \mu_{\bf C} =0, \mu_{\bf R} = c \} / U(\NC) \simeq 
  T^* G_{\NF,\NC}  \label{IJGrassmann}
\eeq
where $c$ is interpreted as the FI-parameter of  
the effective gauge theory on the fractional D$p$-branes.
Eq.~(\ref{IJGrassmann}) is understood as 
the moduli space of vacua in the effective theory on 
the fractional D$p$-branes which coincides with 
the field theoretical result (\ref{massless-Higgs})~\cite{ANS}. 
Let $g_s$ and $l_s = \sqrt {\alpha'}$ 
the string coupling constant and the string length, 
respectively. 
We now express the parameters in 
the effective gauge theory 
on the fractional D$p$-branes 
in terms of them.
The FI-parameter $c$  
blows up the ${\bf Z}_2$ orbifold singularity 
by inserting $S^2$ with the area 
\beq
 A = {\rm Area}\, (S^2) \sim c \,\, l_s^{p+1}  
  \label{radius}
\eeq
where we will determine the $g_s$ dependence 
in Eq.~(\ref{S2area}), below.
The gauge coupling constant $g$ is given by 
\beq
 \1{g^2} = b \, \tau_{p+2} \, l_s^2 
   = {b \over \, g_s l_s^{p-3} }
\eeq
with $\tau_{p+2} = 1/ g_s l_s^{p+3}$ the D($p+2$)-brane tension and 
$b$ the $B$-field flux integrated over the $S^2$, 
$b \sim A B_{ij}$. 
The decoupling limit of gravity and higher derivative corrections 
(the gauge theory limit) is taken as 
$l_s \to 0$ with keeping $g^2$ and $c$ fixed 
(the limit of $g_s \to 0$ or $\infty$ depends on $p$).

\begin{figure}[thb]
\begin{center}
\includegraphics[width=8cm,clip]{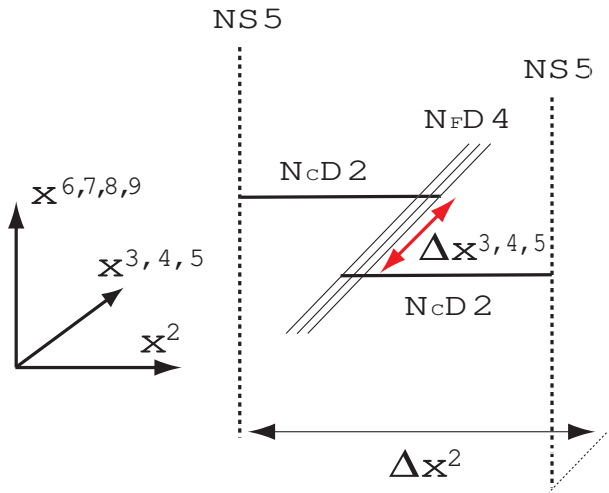}
\end{center}
\caption{\small
The T-dualized brane configuration for massless hypermultiplets.
The triplet of the FI-parameters correspond to separations 
$\Delta x^{3,4,5}$ of the D$2$-branes along the D$4$-branes. 
Separation $\Delta x^{2}$ of 
the two NS$5$-branes along the $x^2$-coordinate 
is proportional to the gauge coupling constant squared 
on the effective theory on the D$2$-branes.
}
\label{fig4}
\end{figure}

Second, we consider the T-dual picture of this configuration. 
Compactifying the $x^2$-direction ${\bf C}^2 / {\bf Z}_2$ becomes 
the Taub-NUT space and then 
taking T-duality along that direction it becomes two NS$5$-branes. 
The D$p$- and D$(p+4)$-branes are mapped to 
D$(p+1)$- and D$(p+3)$-branes, respectively.
This brane configuration with $p=3$ is the same as 
the theory admitting vortices~\cite{HT2}.
In the case of $p=1$ the brane configuration is given as
\beq
 \mbox{$\NC$ D$2$:} && 012     \non
 \mbox{$\NF$ D$4$:} && 01\hs{2} 345 \non
 \mbox{2 NS$5$:}      && 01\hs{8} 6789  \;.
\eeq
This configuration is illustrated in Fig.~\ref{fig4}.
Here the positions of the two NS$5$-branes have 
the following physical meanings.
The distance $\Delta x^2$ 
between the NS$5$-branes along the $x^2$-coordinate 
is proportional 
to inverse of the square of 
the gauge coupling constant $g$ in the effective gauge theory 
on the D($p+1$)-branes:
\beq
 {1\over g^2} 
 = |\Delta x^2|  \tau_{p+1} \,l_s^4  
 = { |\Delta x^2| \over g_s l_s^{p-2}}   
   \label{gauge-coupling}
\eeq
with $g_s$ and $l_s = \sqrt {\alpha'}$ 
the string coupling constant and the string length in 
the T-dualized type II sting theory, 
respectively.\footnote{
The gauge theory limit in this T-dualized 
type II string theory 
is different from the one 
before taking T-duality.
} 
The triplet of the FI-parameters correspond 
to the distance between the two NS$5$-branes 
in the coordinates $x^3$, $x^4$ and $x^5$:
\beq
 c^a = \1{g_s l_s^p } 
   (\Delta x^3, \Delta x^4, \Delta x^5) . \label{FI-brane}
\eeq
In the limit of vanishing FI-parameters, 
there can appear non-baryonic Higgs branches realized by 
D($p+1$)-branes stretched 
between two NS$5$-branes directly~\cite{HW,HOO,NOYY}.
Due to non-vanishing FI-parameters,  
these D($p+1$)-branes cannot be supersymmetric
without total gauge symmetry breaking and 
therefore non-baryonic branches are prohibited.

\begin{figure}[thb]
\begin{center}
\includegraphics[width=14cm,clip]{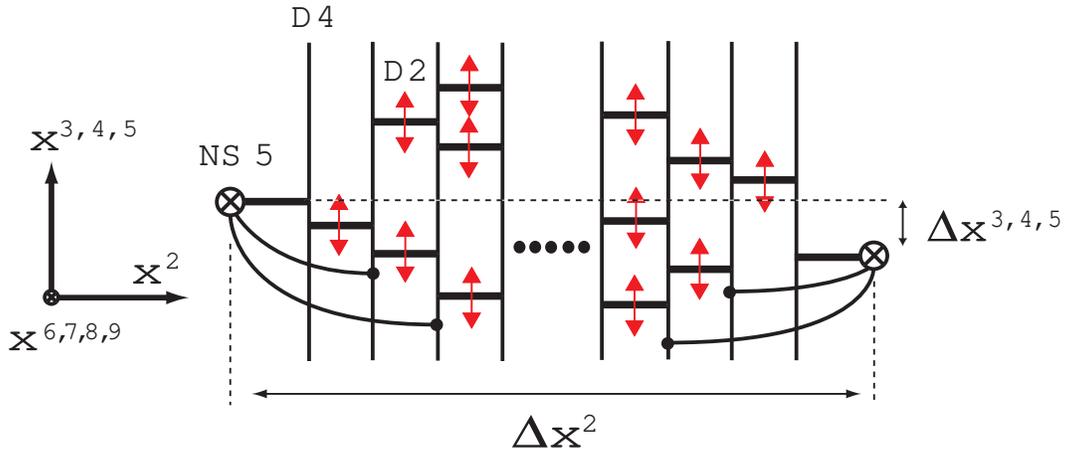}
\end{center}
\caption{\small
The moduli space for vacua. 
Its dimension can be counted as the D$2$-brane degree of freedom 
in the $x^{3,4,5}$-positions and the position on 
$S^1$ in the M-theory. 
The s-rule plays a crucial role for this counting.
}
\label{fig5}
\end{figure}

Strings connecting $\NC$ D$(p+1)$-branes and 
$\NF$ D$(p+3)$-branes give 
the $\NC \times \NF$ matrix of hypermultiplets. 
The effective gauge theory on the world-volume of D$p$-branes 
is insensitive to the positions of 
the $\NF$ D$(p+3)$-branes in the $x^2$-coordinate, 
and so they are arbitrary. 
The moduli space of the vacua 
${\cal M}^{M=0}_{\rm vacua}$ of the theory 
is identical to the one 
$T^* G_{\Nf,\Nc}$ in Eq.~(\ref{IJGrassmann}) 
before taking the T-duality. 
We can count the dimension of 
${\cal M}^{M=0}_{\rm vacua}$ easily, 
if the configuration is viewed from 
the $x^{6,7,8,9}$-directions 
as in Fig.~\ref{fig5}.
The D($p+1$)-branes stretched between D($p+3$)-branes can move freely 
in the $x^3,x^4,x^5$-coordinates.
Here we have a very important rule called 
the {\it s-rule} (see for review and references in \cite{Giveon:1998sr})
 which requires that 
{\it at most one D$(p+1)$ can be stretched between a D$(p+3)$ brane 
and a NS$5$-brane}.  
Thanks to this rule some D($p+1$)-brane cannot move 
in the $x^3,x^4,x^5$-coordinates because they are 
connected to one of NS$5$-branes directly, 
and therefore the dimension of the moduli space of vacua 
can be calculated by counting 
the freedom of the D($p+1$)-branes as~\cite{HOO}
\beq
 \dim {\cal M}^{M=0}_{\rm vacua} 
&=& 4 \left[ 2 \sum_{k=1}^{N_{\rm C}} k 
       +  N_{\rm C} (N_{\rm F} - 1 - 2 N_{\rm C}) \right] \non 
  &=& 4 \NC (\NF - \NC) = \dim T^* G_{\NF,\NC}  
      \label{dimension}
\eeq
which coincides again with 
the field theoretical result (\ref{massless-Higgs}).
The factor four comes from the freedom of the D($p+1$)-brane 
endpoint positions  
on the world-volume $x^3$, $x^4$ and $x^5$ 
of the D($p+3$)-branes and position on $S^1$ of the eleventh 
direction when 
this system is promoted from the type IIA theory 
to the M-theory.

If we did not divide ${\bf C}^2$ by ${\bf Z}_2$ 
there would be no NS$5$-branes and therefore 
the s-rule does not play any roles in its T-dual picture. 
Then the moduli space dimension becomes $4 \NC \NF$, 
and the difference $4 \NC^2$ 
with the dimension (\ref{dimension}) 
is supplied from adjoint hypermultiplets 
which are projected out by ${\bf Z}_2$ in our case.
Therefore the s-rule was essential to get correct results 
for our ${\bf Z}_2$ case.

\subsection{Massive Hypermultiplets}

\begin{figure}[thb]
\begin{center}
\includegraphics[width=7cm,clip]{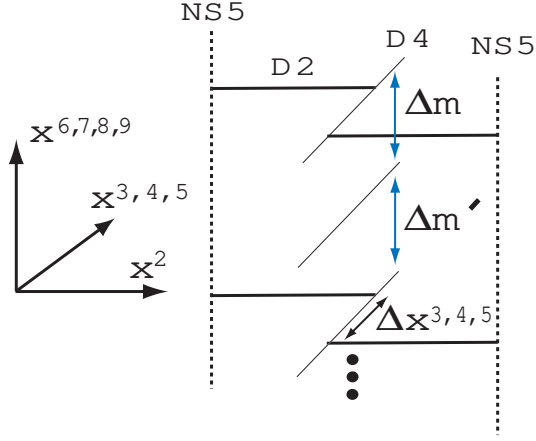}
\end{center}
\caption{\small
The brane configuration for 
massive hypermultiplets.
Hypermultiplets coming from 
strings stretched between D$2$- and D$4$-branes 
become massive by placing D$4$-branes with distances 
in the $x^{6,7,8,9}$-coordinates. 
$\Delta m$ and $\Delta m'$ denote mass differences 
for the hypermultiplets. 
The relation with coordinates is 
$\Delta \vec{m} 
= l_s^{-2} (\Delta x^6,\Delta x^7,\Delta x^8,\Delta x^9)$.
}
\label{fig6}
\end{figure}

For the presence of domain walls in the theory 
we need to give mass differences to the hypermultiplets. 
We consider the case of non-degenerate masses  
and briefly discuss the partially degenerate masses 
in Sec.~\ref{CD}. 
In the T-dual picture, masses for hypermultiplets are 
obtained by sliding D$(p+3)$-branes along the NS$5$-brane 
world-volume as shown in Fig.~\ref{fig6}. 
Note that the freedom of sliding and therefore 
the number of mass parameters for one hypermultiplet
depend on the dimension $p$.
There exist four degrees of freedom in the case of $p=1$,
\beq
 \Delta \vec{m} 
 = \1{l_s^2} (\Delta x^6,\Delta x^7,\Delta x^8,\Delta x^9) .
 \label{mass-brane}
\eeq
However we simply exploit one freedom for instance 
the $x^6$-direction in the brane configuration in this paper,
and therefore hypermultiplets have real masses with this choice. 
The whole discussion in this paper 
holds in any dimensions $p=1,2,3,4$.

We consider the vacua of this theory. 
Again the s-rule plays an important role: 
at most one D$(p+1)$-brane can stretch between
a D$(p+3)$-brane and a NS$5$-brane. 
Therefore vacua are discrete and isolated,  
and the number of them is 
${}_{\NF} C_{\NC}$ recovering the field theoretical 
result (\ref{number-vacua}). 
The case of hypermultiplets with degenerate masses 
can be considered 
by letting positions of some D$(p+3)$-branes to coincide. 
There appear more massless degrees of freedom 
from strings stretched between D$(p+1)$-branes and 
D$(p+3)$-branes resulting 
in continuous vacua as in the massless case.

\vs{1}

\begin{figure}[thb]
\begin{center}
\includegraphics[width=9cm,clip]{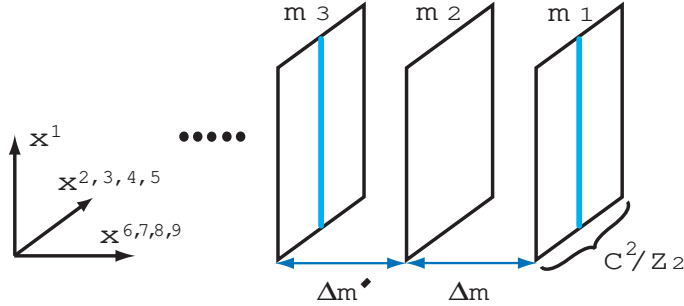}
\end{center}
\caption{\small
The brane configuration for 
massive hypermultiplets before the T-dualization. 
Hypermultiplets coming from 
strings stretched between D$1$- and D$5$-branes 
become massive by placing D$5$-branes with distances 
in the $x^{6,7,8,9}$-coordinates.
}
\label{fig7}
\end{figure}

Next let us consider the original brane configuration 
before the T-dualization.
Masses of hypermultiplets correspond to the positions of 
the D$(p+4)$-branes.  
In the case of $p=1$, the quartets of masses 
(\ref{mass-brane}) are given by 
the D($p+4$)-brane positions in the $x^{6,7,8,9}$-directions 
as seen in Fig.~\ref{fig7}.
We assume real masses for hypermultiplets 
by placing D($p+4$)-branes parallel along the $x^6$-direction. 
The D$(p+4)$-branes are ordered by 
the single coordinate $x^6$ for real masses.
We label D$(p+4)$-branes by $A=1,\cdots,\NF$ such that
a larger $x^6$ position corresponds to a smaller $A$. 
To be consistent with the s-rule in the T-dual picture, 
we find that 
at most one D$p$-brane can be absorbed into 
the ${\bf Z}_2$ fixed point of one D$(p+4)$-brane. 
We thus have ``the exclusion principle of D$p$-branes". 
It will play 
an important role to consider domain wall dynamics in 
the next section. 
From this exclusion principle 
the discrete vacua are labeled by a set of the flavor indices 
$A_r$ ($r=1,\cdots, \NC$) 
corresponding to the $A_r$-th D$(p+4)$-brane into which 
the $r$-th D$p$-brane is absorbed. 
Such a configuration can be read from 
the vacuum expectation values for 
hypermultiplets
\beq
 && H^{1\dagger }H^1 
   = c \, {\rm diag}. 
    (0,\cdots,1,0,\cdots,1,0,*,\cdots,*), \label{vac-H} \\
 && \hs{42} A_1 \hs{11} A_2   \nonumber 
\eeq
where positions of 
the D$p$-branes can be read from the elements 1. 
We label such a vacuum by 
$\langle A_1,\cdots, A_{\NC}\rangle$ 
as in Eq.~(\ref{vacua-labels}).
Again, the number of vacua can be calculated 
as ${}_{\NF} C_{\NC}$ 
recovering the field theoretical result (\ref{number-vacua}). 

\subsection{Duality} \label{Du}
\begin{figure}[thb]
\begin{center}
\includegraphics[width=6cm,clip]{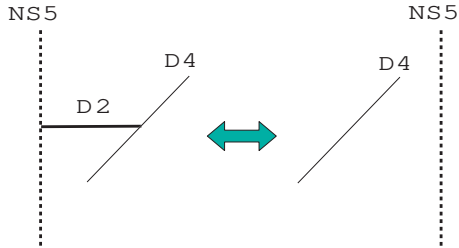}
\end{center}
\caption{\small
The Hanany-Witten effect. 
A D$2$-brane is created (annihilated)  
when NS$5$-brane moves across a D$4$-brane.
}
\label{fig32}
\end{figure}

In this subsection we discuss a duality between 
theories with the gauge groups $U(\Nc)$ and $U(\tilde \Nc)$  
and the same number of hypermultiplets,
where the dual gauge group is defined by 
$\tilde \Nc \equiv \Nf - \Nc$. 
When a NS$5$-brane moves across a D($p+3$)-brane, 
a D($p+1$)-brane is created (annihilated)  
between these branes (see Fig.~\ref{fig32}). 
This phenomenon is known as 
the Hanany-Witten effect~\cite{HW,NOYY2}. 

\begin{figure}[thb]
\begin{center}
\includegraphics[width=9cm,clip]{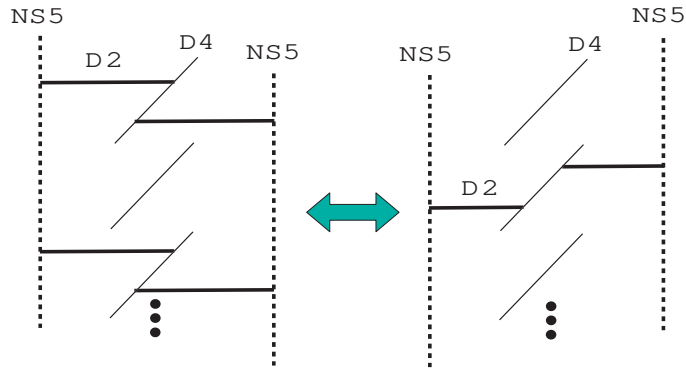}
\end{center}
\caption{\small
The duality between $\Nc$ and 
$\tilde \Nc = \Nf - \Nc$ 
in the brane configuration.
}
\label{fig33}
\end{figure}

Let us exchange 
the positions of two NS$5$-branes both across the D($p+3$)-branes 
in the brane configuration in Fig.~\ref{fig6} 
with noting the Hanany-Witten effect.
Then we obtain the vacuum configuration for  
the dual gauge theory on the D($p+1$)-branes 
with $U(\tilde \Nc)$ gauge group 
with $\Nf$ fundamental hypermultiplets (see Fig.~\ref{fig33}).
From this figure we can see that the sign of the FI-parameter 
is flipped under this duality. 
As a result roles of $H^1$ and $H^2$ 
in hypermultiplets are exchanged 
\beq
  (H^1, H^2, c) \leftrightarrow 
  (\tilde H^2, \tilde H^1, \tilde c = -c)
  \label{relations}
\eeq
with $\tilde H^1$ and $\tilde H^2$ denoting 
a doublet of $\tilde \Nc \times \Nf$ matrices of 
hypermultiplets in the dual theory.\footnote{
The convention to label the index $i=1,2$ on $\tilde H^i$ 
is opposite to that in Appendix D in 
our previous work~\cite{INOS2}. 
Since the sign of the FI-parameter is flipped, 
the present one is more natural.
}   
For instance 
$\tilde H^2$ ($H^1$) and $\tilde H^1$ ($H^2$) 
parametrize the base and cotangent space 
of the vacuum manifold $T^* G_{\Nf,\tilde \Nc}$ 
($T^* G_{\Nf,\Nc}$), respectively, 
after (before) taking the duality.

In the spirit of ${\cal N}=1$ Seiberg duality 
the above duality merely means the possibility of 
continuous deformation of parameters of the theory. 
Namely there is no phase boundary between two theories 
resulting in the identical low-energy effective theory. 
On the other hand Eq.~(\ref{gauge-coupling}) implies that 
the distance $\Delta x^2$ 
of the two NS$5$-branes in the $x^2$-coordinate 
is proportional to the inverse of 
the gauge coupling constant squared 
in the gauge theories 
on the D($p+1$)-branes 
for both brane configurations. 
Therefore this duality becomes exact in the limit of 
$\Delta x^2 \to \pm 0$. 
In fact these gauge theories reduce to 
nonlinear sigma models on 
$T^* G_{\Nf,\Nc}$ in this limit, 
and the duality $\Nc \leftrightarrow \tilde \Nc$ 
becomes exact (\ref{duality-sigma}).

\begin{figure}[thb]
\begin{center}
\includegraphics[width=14cm,clip]{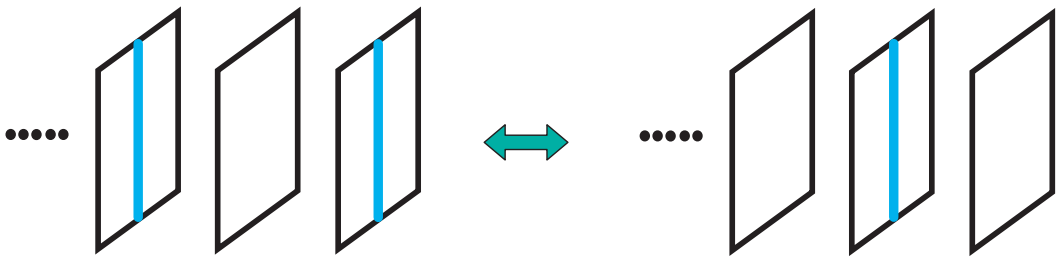}
\end{center}
\caption{\small
The duality between 
$\Nc$ and $\Nf - \Nc$ in the brane configuration 
before the T-dualization.
}
\label{fig30}
\end{figure}

In the original brane picture for the vacua 
in Fig.~\ref{fig7} before taking the T-duality 
the Hanany-Witten effect can be 
understood as a rule to exchange 
existence and non-existence of the D$p$-branes 
at every D($p+4$)-brane 
although creation/annihilation is not so clear
(see Fig.~\ref{fig30}). 
In the strong coupling limit, the duality between 
theories with 
$\Nc$ and $\tilde \Nc = \Nf - \Nc$ 
holds as can be seen in
Eq.~(\ref{massless-Higgs}). 
Since we know that only fields parameterizing 
the base space of the target space 
participate in constructing walls, 
we set $H^2 = 0$ and $\tilde H^1=0$. 
Then the duality is determined by 
[see Eq.~(D.5) in \cite{INOS2}]
\beq
 H^{1\dagger }H^1 + \tilde H^{2\dagger } \tilde H^2  
  =  c {\bf 1}_{\Nf},
   \label{duality-sigma}
\eeq
where $H^1$ is for $U(\Nc)$ gauge theory and 
$\tilde H^2$ is that in $U(\tilde \Nc)$ gauge theory. 
This equation implies the relation between $H^1$ and $\tilde H^2$ 
[see Eq.~(D.4) in \cite{INOS2}]
\beq
 H^1 \tilde H^2{}^{\dagger} = 0.  \label{H1-H2-tilde}
\eeq
The vacuum expectation values of hypermultiplets in 
the vacuum configuration after 
taking a duality are found from 
Eqs.~(\ref{vac-H}) and (\ref{duality-sigma}) 
as 
\beq
 && 
 \tilde H^{2\dagger } \tilde H^2 
   = c \, {\rm diag}. 
    (1,\cdots,0,1,\cdots,0,1,*,\cdots,*) . \\
 && \hs{42} A_1 \hs{11} A_2 \nonumber
\eeq
The elements 1 and 0 in 
Eqs.~(\ref{vac-H}) and this equation 
are exchanged under the duality, 
indicating Fig.~\ref{fig30}.
In the next section we will see that 
the Hanany-Witten effect is generalized to 
kinky configurations.

\section{Non-Abelian Walls as Multiple Kinky D-branes}\label{NWAMKD}


\subsection{Multiple Kinky D-branes} 

\begin{figure}[thb]
\begin{center}
\includegraphics[width=11cm,clip]{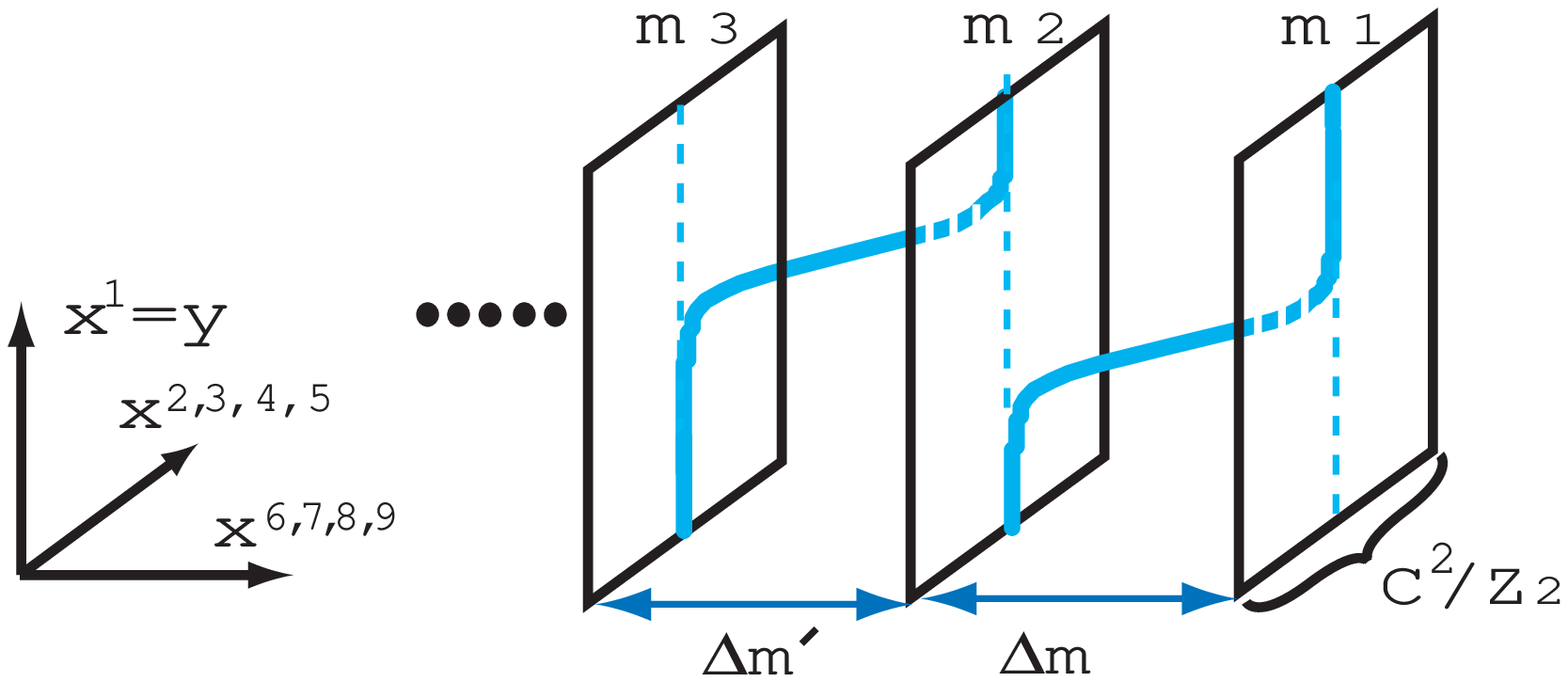}
\end{center}
\caption{\small
Multiple kinky fractional D-branes.
}
\label{fig8}
\end{figure}

As shown above the effective theory on 
the fractional D$p$-branes 
is the SUSY $U(\NC)$ gauge theory 
with massive $\Nf$ hypermultiplets 
and the FI-term.
Since the adjoint scalar field $\Sigma$ represents 
transverse fluctuations of D$p$-branes, 
the diagonal components of (the vacuum expectation value of) 
$\Sigma$, in the gauge of diagonal $\Sigma$,
can be identified as the positions of $\Nc$ D$p$-branes 
along the $x^6$-coordinate.  
In that gauge $\Nc$ diagonal components of $\Sigma$ 
for a wall solution 
are plotted as functions $x^6 = \Sigma_{rr} (x^1)$ 
in Fig.~\ref{fig8}.
They represent multiple kinky D$p$-branes curved  
in the ($x^1$,$x^6$)-plane and  
these curves are determined 
by the BPS equations (\ref{BPSeqs}).

\begin{figure}[thb]
\begin{center}
\includegraphics[width=12cm,clip]{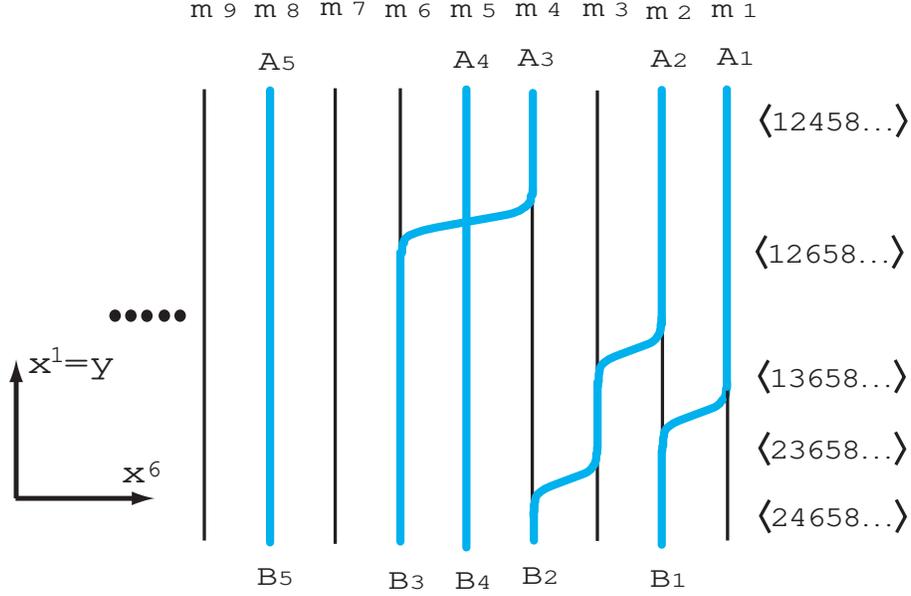}
\end{center}
\caption{\small
Multiple non-Abelian walls as kinky D-branes. 
An example of 
the $\Nc$ kinky D$p$-branes interpolating 
between the vacua 
$\langle A_1 , \cdots, A_{\Nc}\rangle$ 
and $\langle B_1 , \cdots, B_{\Nc}\rangle$ 
is displayed. 
Intermediate vacua between two adjacent walls 
plotted on the right of the figure can be easily 
read from this kinky brane configuration.
}
\label{fig9}
\end{figure}

To be more precise, 
all D$p$-branes are absorbed into 
some $\NC$ D$(p+4)$-branes 
in the limit $y = x^1 \to + \infty$.
The configuration tends to a vacuum state 
$\langle A_1,\cdots, A_{\NC}\rangle$ in this limit 
where we have labeled 
the $r$-th D$p$-brane counted from 
the right of Fig.~\ref{fig8} by $A_r$.   
On the other hand, in the opposite limit $y \to - \infty$, 
they approach another vacuum configuration labeled by 
$\langle B_1,\cdots, B_{\NC}\rangle$. 
The $\NC$ D$p$-branes exhibit kinks somewhere 
in the $y$-coordinate  
as illustrated in Fig.~\ref{fig9}. 
Here we labeled $B_r$ such that 
the $A_r$-th brane at $y \to + \infty$ goes to the $B_r$-th brane 
at $y \to - \infty$. 
The labels $r$ for the vacua are now understood 
to be given for the $r$-th D$p$-brane. 
Hence a set of 
$B_r$ does not have to be ordered. 
If we separate adjacent walls far enough 
the configuration between these walls 
approach a vacuum. 
The brane configurations are very useful to figure out 
such a vacuum as illustrated in Fig.~\ref{fig9}.

It is sometimes troublesome 
to take a gauge of diagonal $\Sigma$. 
Without taking that gauge 
the positions of the D$p$-branes in the $x^6$-coordinate 
can be described in a gauge invariant way 
by solutions $\lambda$ of  
the eigenvalue equation 
\beq
 0 = \det (\lambda {\bf 1}_{\Nc} - \Sigma )
 = \det \left(\lambda 
  - {1\over 2} \Omega^{-1} \del_y \Omega \right) \;. 
\eeq

\begin{figure}
\begin{center}
\begin{tabular}{ccc}
\includegraphics[width=6cm,clip]{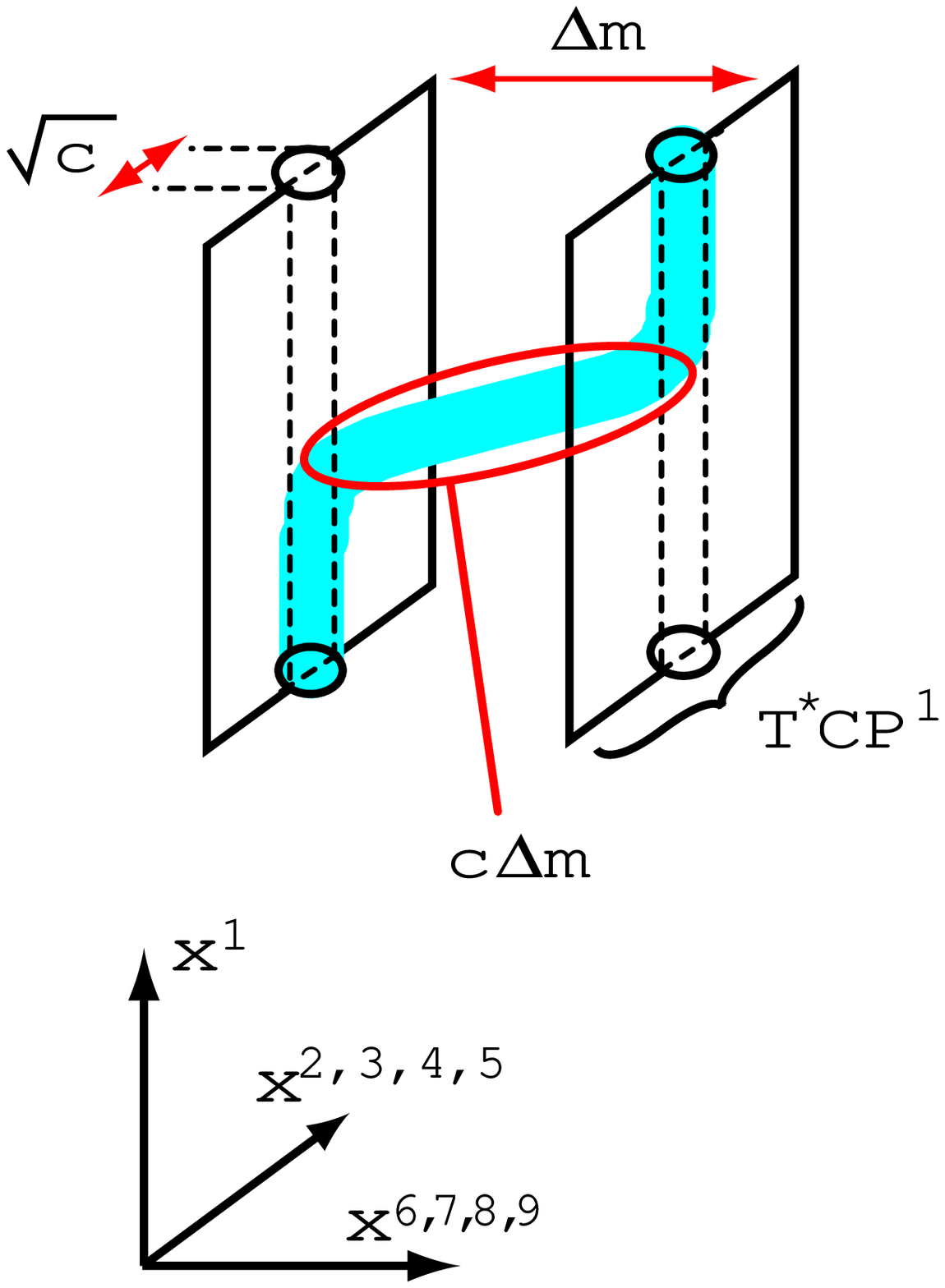} & &
\includegraphics[width=7cm,clip]{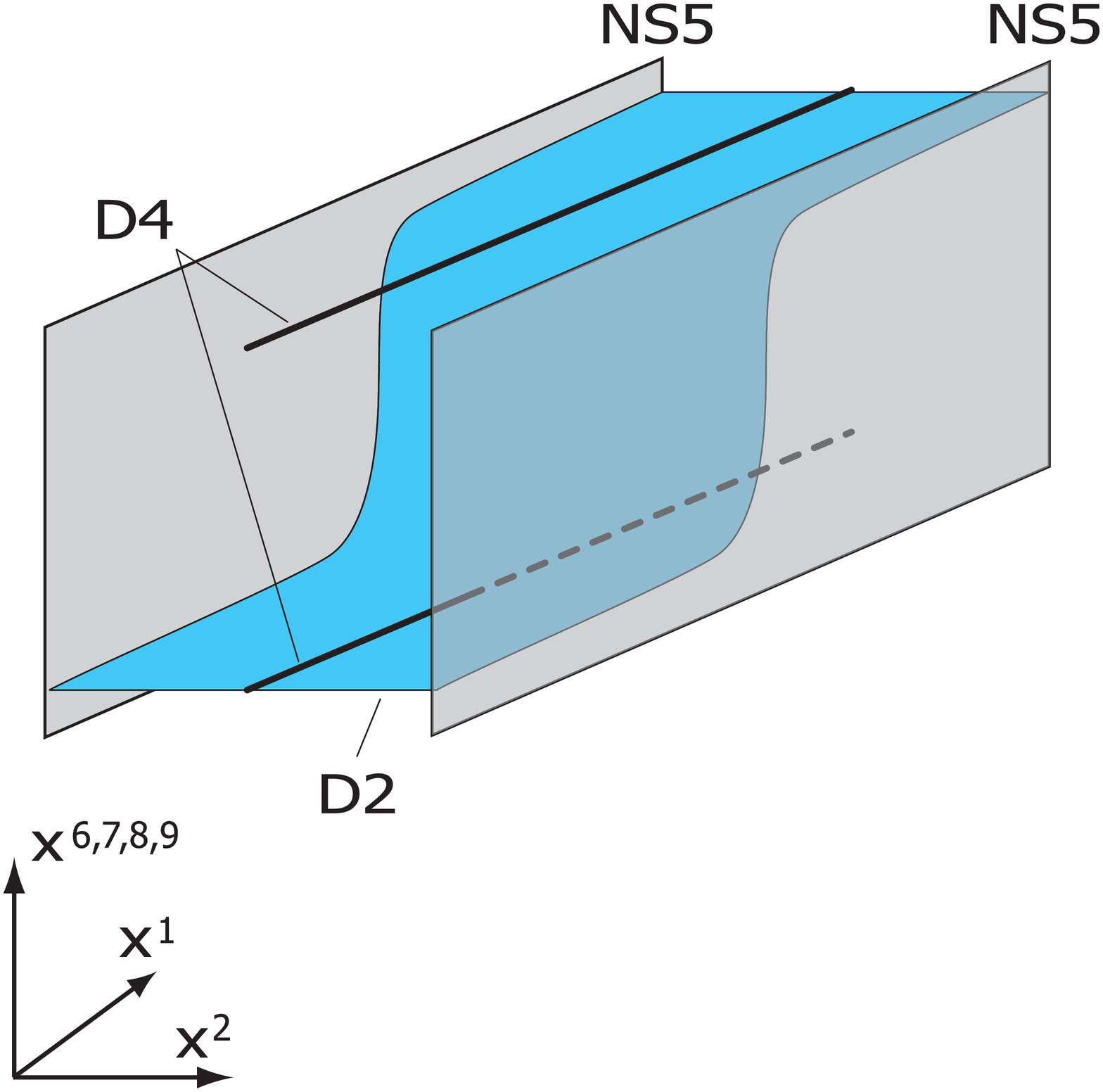} \\
 a) &  \hs{10} & b)  \\
\includegraphics[width=6.5cm,clip]{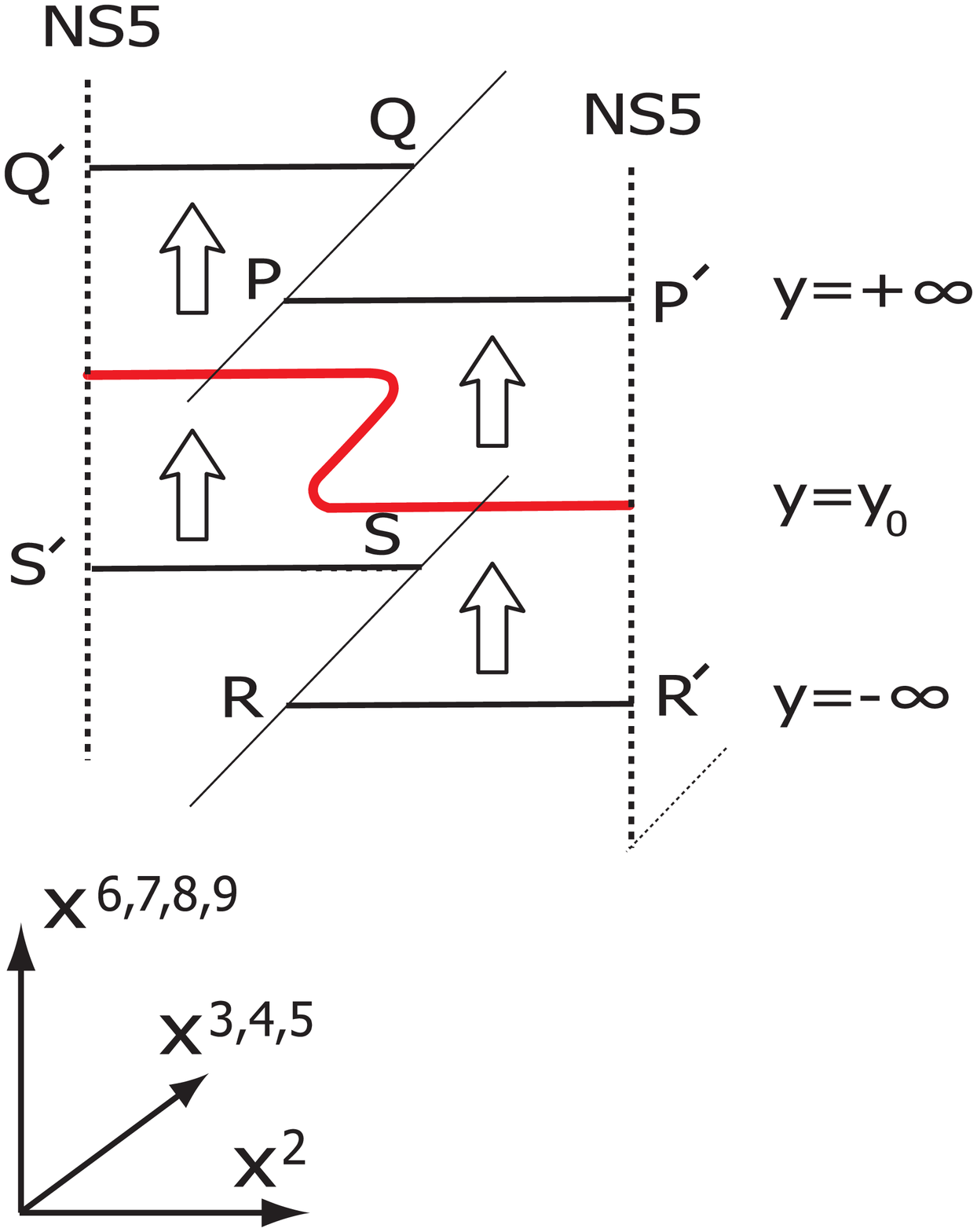} & &
\includegraphics[width=5cm,clip]{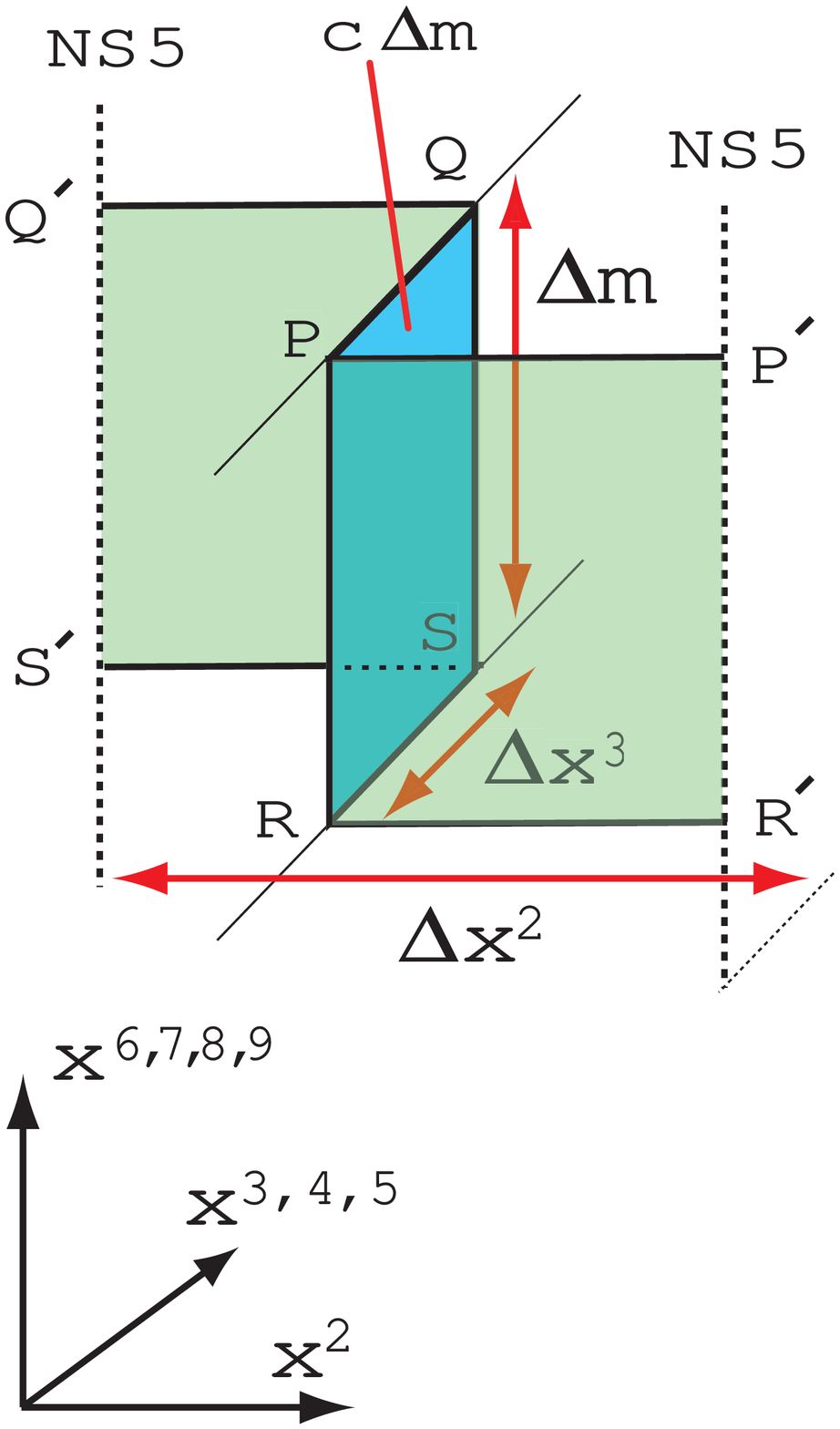} \\
 c) &  \hs{10} & d)
\end{tabular}
\caption{\small 
Calculating the wall tension in the brane configuration 
before a) and after b), c), d) taking T-duality.
a) The wall tension is given by 
the tension of the fractional D$p$-branes 
subtracted by the one for 
the D$p$-branes in the vacuum state. 
b) The wall profile (kinky D($p+1$)-branes) 
in the T-dualized brane configuration 
with suppressing the directions $x^{3,4,5}$ for the FI term. 
c) The wall profile with suppressing the direction $x^1 =y$ 
of the co-dimension of the wall.  
The D($p+1$)-branes stay at the segments RR$'$ and S$'$S for 
$y \to - \infty$. They move up around $y=y_0$ apart from 
the lower D($p+3$)-brane. Here they have to bend to 
the $x^3$ direction to be connected with each other. 
Finally they stay at the segments PR$'$ and Q$'$Q for 
$y \to + \infty$.
d) The movement of the D($p+1$)-brane depending on $y$ is drawn in the 
same figure. 
The wall tension is given by at the square 
PQSR. Tensions proportional to the squares PP$'$R$'$R and 
QQ$'$S$'$S are contributions from vacua.
}
\label{fig25}
\end{center}
\end{figure}

\medskip
We now calculate the wall tension (\ref{eq:tension}) 
using the brane configurations a) before and b), c), d) after 
taking T-duality using figures a) and b), c), d)  
of Fig.~\ref{fig25}, respectively.

a) The orbifold singularity 
in ${\bf C}^2/{\bf Z}_2$ 
is blown up by $S^2 \simeq {\bf C}P^1$ and 
${\bf C}^2/{\bf Z}_2$ becomes 
the $A_1$ gravitational instanton, 
the Eguchi-Hanson space $T^* {\bf C}P^1$.  
Fractional D$p$-branes are D($p+2$)-branes 
two of whose spatial dimensions are wrapped 
around the 2-cycle $S^2$  
with the area proportional to 
the FI-parameter $c$ 
as in Eq.~(\ref{radius}). 
Since fractional D$p$-branes  
for the vacuum state in Fig.~\ref{fig7} 
are actually D($p+2$)-branes 
with size $\sqrt c$ of this $S^2$,  
the tension of a fractional D$p$-brane 
is proportional to the area of the fractional D$p$-brane 
($\sim c \Delta x^1$) which diverges because of 
the world-volume extending to 
the $x^1$-coordinate. 
Kinky fractional D$p$-branes in Fig.~\ref{fig8}
also wrap around the $S^2$ with 
the area $c$ as illustrated in Fig.~\ref{fig25}-a).
The tension of a kinky D$p$-brane 
interpolating between 
adjacent D($p+4$)-branes with 
the distance $\Delta x^6 \sim \Delta m$ 
increases from the one for a vacuum state 
by the finite amount coming from the area of 
the kinky region of the D$p$-brane.
That amount of the tension contributed from a kink 
can be calculated by subtracting 
the tension of the fractional D$p$-brane for vacuum states 
from that of the kinky configuration, to give
\beq
 && \tau_{p+2} [{\rm Area}\, (\mbox{kinky frac. D$p$}) 
 - {\rm Area}\, (\mbox{vacuum frac. D$p$}) ] \non  
 && = \1{g_s l_s^{p+3}} 
     [A (\Delta x^1 + \Delta x^6) - A \Delta x^1]  
 = c \, \Delta m 
\eeq 
with $\tau_{p+2} = 1/ g_s l_s^{p+3}$ 
and $\Delta x^6 = l_s^2 \Delta m$ in Eq.~(\ref{mass-brane}). 
The last equality holds provided that 
the $g_s$ dependence of the area $A$ of $S^2$ in Eq.~(\ref{radius}) 
is given by
\beq
 A = c g_s l_s^{p+1} = {c \over \tau_p}. \label{S2area}
\eeq
For multiple fractional D$p$-branes 
exhibiting multiple kinks, 
the sum of the kink contributions to the tension  
of the kinky D$p$-branes can be obtained as
\beq
  \tau_{p+2} \sum_{r=1}^{\Nc} 
 [{\rm Area}\,(\mbox{$r$-th kinky frac. D$p$})
  - {\rm Area}\,(\mbox{$r$-th vacuum frac. D$p$})]
 = c \sum_{r=1}^{\Nc} (m_{A_r} - m_{B_r}). 
\eeq 
Therefore 
the tension of the fractional D$p$-branes 
in our brane configuration correctly reproduces 
the tension formula (\ref{eq:tension}) of walls. 

b), c), d) The wall tension (\ref{eq:tension}) 
can also be calculated in the brane configuration 
after taking T-duality. 
The directions corresponding to the FI-term 
are not displayed in Fig.~\ref{fig25}-b). 
This figure is not useful to calculate the wall tension.  
Instead, we consider Fig.~\ref{fig25}-c) and d) 
where the co-dimension $y=x^1$ of walls is not displayed. 
The two D($p+1$)-branes ending on 
the lower D($p+3$)-brane along the segment RS
at $y \to -\infty$ are 
expressed by the segments RR$'$ and SS$'$. 
They go to the D($p+1$)-branes, 
ending on the upper D($p+3$)-brane along PQ, 
expressed 
by the segments PP$'$ and QQ$'$ at $y \to +\infty$, 
respectively. 
At intermediate $y \sim y_0$ with $y_0$ the wall position, 
the two D($p+1$)-branes cannot 
end on any D($p+3$)-branes and therefore 
they have to bend to the $x^3$-direction to 
be connected with each other as drawn in Fig.~\ref{fig25}-c). 
By drawing the configurations at all $y$ 
in the same figure, 
we have the configuration illustrated 
in Fig.~\ref{fig25}-d).  
We now calculate the wall tension in Fig.~\ref{fig25}-d).
The tension of the D($p+1$)-brane 
is given by the sum of three pieces,  
Area (PQSR), Area (PP$'$R$'$R) and
Area (QQ$'$S$'$S). 
The contribution from the last two terms
\beq
 \tau_{p+1} [\mbox{Area (PP$'$R$'$R)} 
  + \mbox{Area (QQ$'$S$'$S)}] 
 = \tau_{p+1}\Delta x^2 \Delta x^6
 = {\Delta m \over g^2 l_s^2} 
\eeq  
diverges in the limit $l_s \to 0$,  
where we have used Eqs.~(\ref{gauge-coupling}) 
and (\ref{mass-brane}).
However this is contribution from 
the vacuum energy and we have to subtract it 
to obtain the wall tension. 
This fact can be understood by taking the limit 
$\Delta x^3 \sim c \to 0$. 
In this limit the two D($p+1$)-branes are connected with each other 
to become one D($p+1$)-brane directly stretched between 
the two NS$5$-branes at all $y$. 
The D($p+1$)-brane at each $y$ corresponds to 
a vacuum in the baryonic branch.  
Therefore the D($p+1$)-brane tension contributed from 
the kink is calculated from Eqs.~(\ref{FI-brane}) and (\ref{mass-brane}), 
to yield 
\beq
 \tau_{p+1} \times {\rm Area (PQSR)}
 = \tau_{p+1} \Delta x^3 \Delta x^6 = c \Delta m
\eeq
with $\tau_{p+1} = 1/g_s l_s^{p+2}$. 
We thus have reproduced the wall tension (\ref{eq:tension}) 
in this brane configuration.

A brane configuration for a recently found 
monopole in the Higgs phase 
(a confined monopole) \cite{vm,INOS3} 
has been given by Hanany and Tong \cite{HT2} 
(see also \cite{Auzzi:2004yg}).
It looks very similar with our brane configuration 
for walls in Fig.~\ref{fig25}-d). 
We discuss this issue in Sec.~\ref{CD}.

\subsection{Duality for Walls and 
Extension of the Hanany-Witten Effect} \label{DWEHWT}
In the strong gauge coupling limit 
the duality 
between the theories with the gauge groups 
$U(\NC)$ and $U(\tilde \NC)$ 
with the same number $\Nf$ of hypermultiplets 
holds exactly also for wall solutions 
as shown in \cite{INOS2} by a field theoretical analysis. 
Let us now consider the implication of this field theoretical result 
of duality of the non-Abelian walls to the Hanany-Witten effect in 
the brane configuration. 
Since the gauge coupling constant on the D($p+1$)-branes 
in the T-dualized configuration 
is given in Eq.~(\ref{gauge-coupling}), 
the strong gauge coupling limit corresponds to 
the positions of the two NS$5$-branes with $\Delta x^2 \to \pm 0$ 
as stated above. 
The original Hanany-Witten effect 
explains the duality of the vacua 
at both infinities $y \to \pm \infty$ 
when the positions of the two NS$5$-branes in 
the $x^2$-coordinate are exchanged.
It is, however, difficult to visualize 
what happens for kinky configurations 
in the finite region of $y$. 

\begin{figure}[thb]
\begin{center}
\includegraphics[width=14cm,clip]{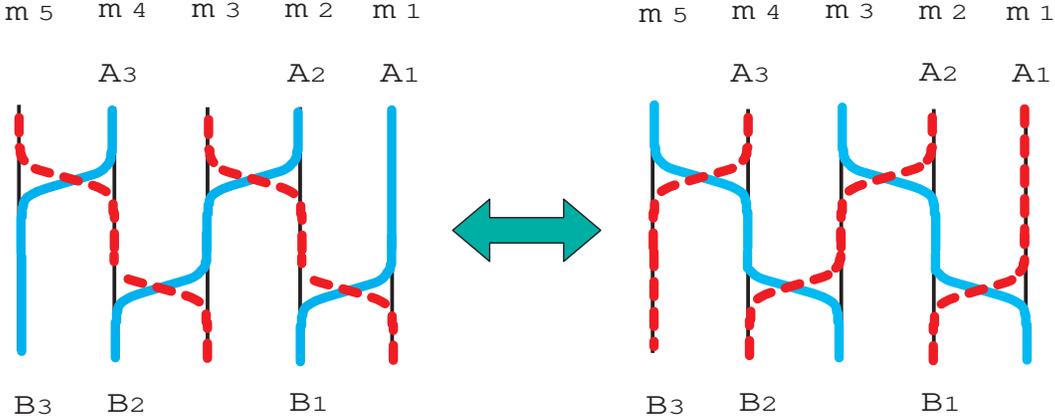}
\end{center}
\caption{\small
Duality between wall configurations in 
theories with gauge groups $U(\Nc)$ and $U(\Nf - \Nc)$. 
Examples of wall configurations in the theory with 
$\Nf=5$ and $\Nc=3$ and its 
dual theory with $\Nf=5$ and $\tilde \Nc=2$ are illustrated. 
The broken curves are wall configurations in 
the dual theory in both figures. 
The dual wall configurations curve oppositely 
due to the flip of the sign of the FI-parameter 
under the duality.
}
\label{fig31}
\end{figure}

Therefore we consider 
the original brane configuration before the T-dualization.
Then the result in \cite{INOS2} implies that 
this duality $\NC \leftrightarrow \tilde \NC$ 
in the strong gauge coupling limit 
can be drawn in Fig.~\ref{fig31}. 
Here wall configurations are given by 
$H^1$ with $H^2=0$ ($\tilde H^2$ with $\tilde H^1=0$) 
for the theory with $U(\Nc)$ [$U(\tilde \Nc)$], 
since flip of the sign of the FI-parameter occurs 
and the roles of $H^1$ and $H^2$ are exchanged 
in both configurations as explained in Eq.~(\ref{relations}). 
We define the moduli matrix for 
the dual theory generating $\tilde H^2$ 
by $\tilde H_0$. 
Here Eqs.~(\ref{def-S}) and (\ref{sol-H}) with 
tilde on all quantities hold for the dual theory
and we have 
$\tilde H_0^1 = 0$, $\tilde H_0^2 = \tilde H_0$ 
for its moduli matrices instead of Eq.~(\ref{H2=0}). 
Then the relation (\ref{H1-H2-tilde}) 
between the hypermultiplets $H^1$ and $\tilde H^2$ 
in both theories 
implies the relation between the moduli matrices 
$H_0$ and $\tilde H_0$, given by
[see Eq.~(D.14) in \cite{INOS2}]
\beq
 H_0 \tilde H_0{}\dagg = 0 . \label{dual-moduli}
\eeq
For a given moduli matrix $H_0$ in the original theory, 
the form of the moduli matrix $\tilde H_0$ for the dual theory 
is determined up to the world-volume symmetry (\ref{art-sym}) 
from this equation.

Wall solutions in dual theories 
can be mapped to each other in the strong gauge coupling limit
as stated above, 
but they are not identical for finite gauge coupling in general. 
Namely, wall solutions under 
exact duality for the two NS$5$-branes with $\Delta x^2 \to \pm 0$ 
are deformed for finite gauge coupling with 
$\Delta x^2 \neq 0$ in different ways 
in both configurations.
Although we did not obtain exact wall solutions 
for finite gauge coupling,  
dual configurations can be determined without loss of exactness 
by the same relation (\ref{dual-moduli}) between 
the moduli matrices for both configurations. 
Although the relation (\ref{dual-moduli}) 
was originally found in the strong gauge coupling limit 
we {\it define} 
the dual configuration for finite gauge coupling by 
that equation. 
We thus conclude that 
{\it the Hanany-Witten effect should be extended 
to kinky D-brane configuration}.

\subsection{Impenetrable Walls and Extension of the S-rule/
Reconnection of D-branes} \label{IWAROD}
\begin{figure}[thb]
\begin{center}
\begin{tabular}{cc}
\includegraphics[width=4cm,clip]{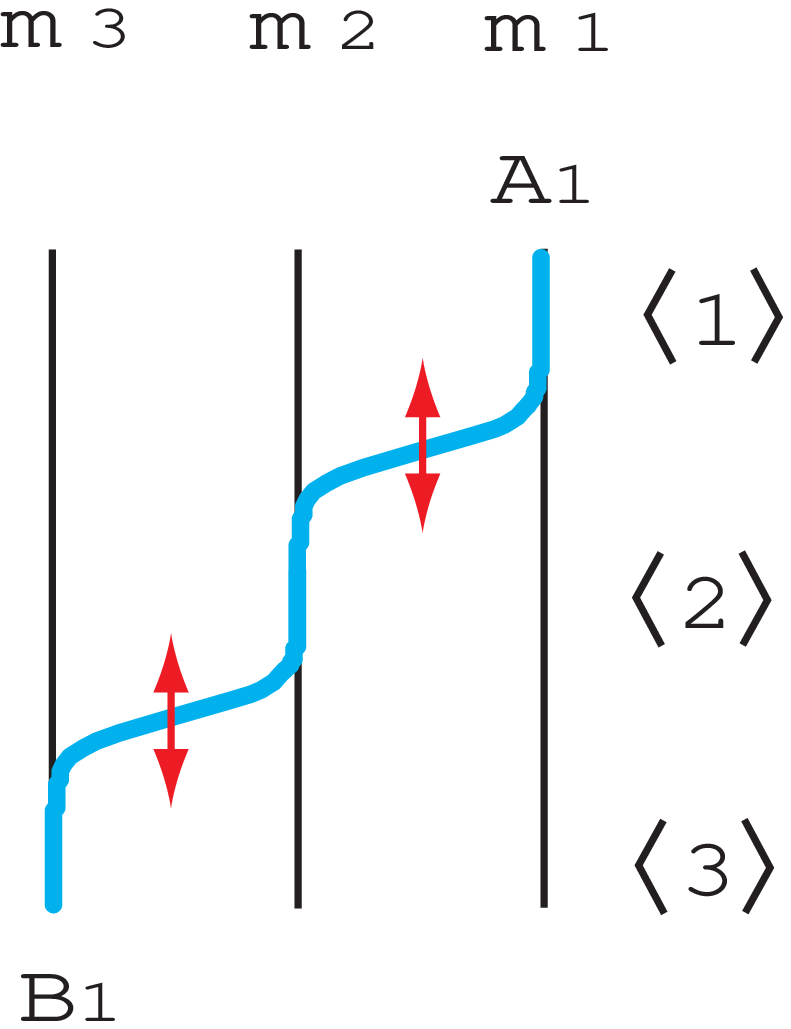}
\hs{20} &
\includegraphics[width=4cm,clip]{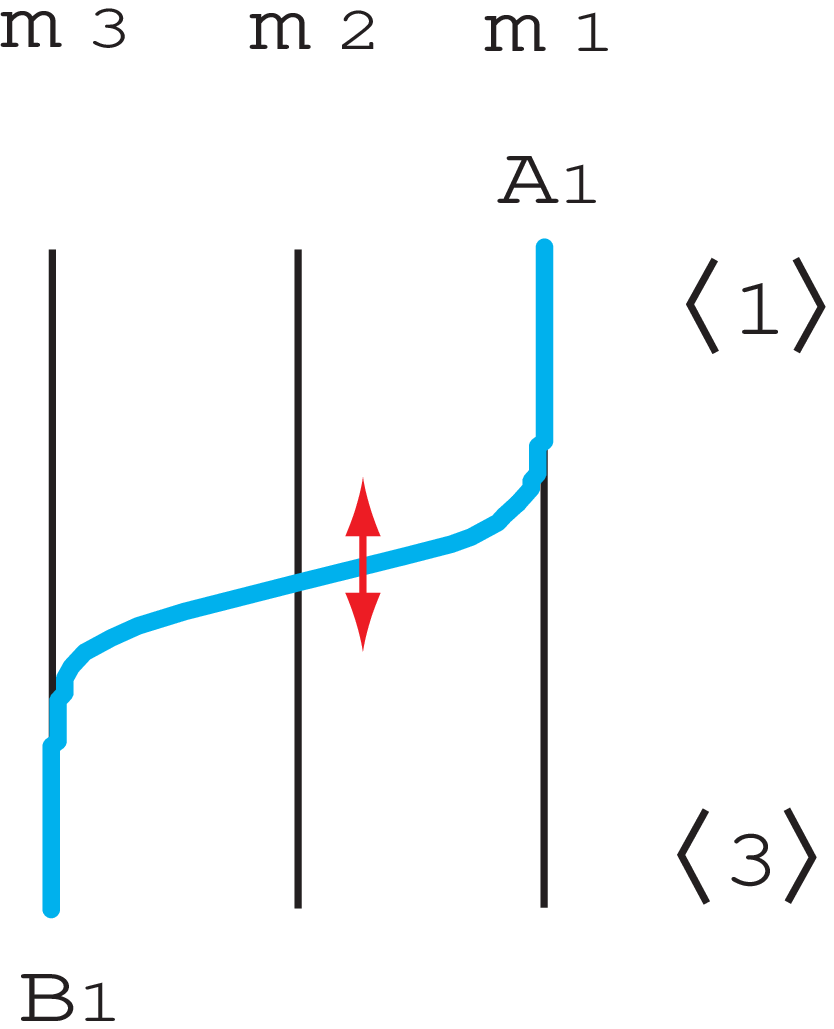}\\
a) \hs{20} & b)
\end{tabular}
\end{center}
\caption{\small
Compressing two walls (a self-compressed brane).
}
\label{fig11}
\end{figure}

We show that 
a lot of interesting phenomena of 
non-Abelian walls found in field theory~\cite{INOS2} 
can be explained very easily 
by means of the brane configurations. 
A double wall configuration for $\NC=1$ and $\NF=3$ 
is shown in Fig.~\ref{fig11}-a). 
These two walls are labeled by 
$\langle 1 \leftarrow  2 \rangle$ 
and $\langle 2 \leftarrow  3 \rangle$ using the vacuum 
labels at $y \to + \infty$ and $y \to - \infty$.  
Positions of these walls cannot be exchanged 
(they are called {\it impenetrable}) 
because the configuration approaches a single wall 
$\langle 1 \leftarrow 3 \rangle$ 
when they are compressed [Fig.~\ref{fig11}-b)]. 
The moduli matrices for these configurations before and after 
compression are
\beq
 && H_0 = \sqrt c \, (1, e^r, e^s ) , \label{simplest-double}\\ 
 && H_0 = \sqrt c \, (1, 0, e^s ) \;, \label{simlest-compressed}
\eeq
respectively, 
with $e^r \in {\bf C}$ and $e^s \in {\bf C}^*$. 
In the strong gauge coupling limit, 
the wall solution for Fig.~\ref{fig11}-a)  
can be obtained by 
substituting the moduli matrix (\ref{simplest-double}) 
to Eq.~(\ref{infinite}) and by using 
Eqs.~(\ref{def-S}) and (\ref{def-Omega}), 
to yield
\beq
 \Sigma = {m_1 e^{2 m_1 y} + m_2 e^{2m_2 y + 2 {\rm Re}(r) } 
          + m_3 e^{2m_3 y + 2 {\rm Re}(s)} \over 
          e^{2 m_1 y}  + e^{2m_2 y + 2 {\rm Re}(r)}
         + e^{2m_3 y + 2 {\rm Re}(s) } } , \hs{5} 
 W_y =0 .
  \label{kinky-U(1)}
\eeq
The positions of kinks can be roughly estimated as 
(see Appendix A in \cite{INOS2})
\beq
 y_1 = {{\rm Re}(r)\over m_1 - m_2}, \hs{5}
 y_2 = {{\rm Re}(s-r)\over m_2 - m_3} \;, \label{positions}
\eeq
which are valid when these two walls are well separated.
The wall configuration in Fig.~\ref{fig11}-b) 
is obtained in the limit of ${\rm Re}(r)  \to - \infty$ 
with the center of position $s$ of two walls fixed. 
[It is also obtained from the moduli matrix 
(\ref{simlest-compressed}).]
There the interpretation of parameters in Eq.~(\ref{positions}) 
as the positions becomes meaningless. 
We call this compressed wall a {\it level-1} compressed wall, 
where the ``level" of compression have been defined by 
the number of zero components between $1$ and 
the non-zero element.

In the same way positions of 
walls made by the same D$p$-brane 
cannot commute and they are impenetrable. 
So the Abelian walls in the $U(1)$ gauge theory are 
all impenetrable.
A level-$l$ compressed wall in $U(\Nc)$ gauge theory 
is generated by a 
moduli matrix in the form of 
\beq
 H_0 = 
 \sqrt{c}\, 
 \left(\begin{array}{ccc}
         & \cdots                            &   \\
  \cdots & 1 \underbrace{0 \cdots  0}_{l} & e^r \cdots \\ 
         & \cdots                            &   
\end{array} \right) \;.
\eeq
We can obtain this configuration by compressing elementary 
walls $l$ times.

\begin{figure}[thb]
\begin{center}
\begin{tabular}{cc}
\includegraphics[width=4cm,clip]{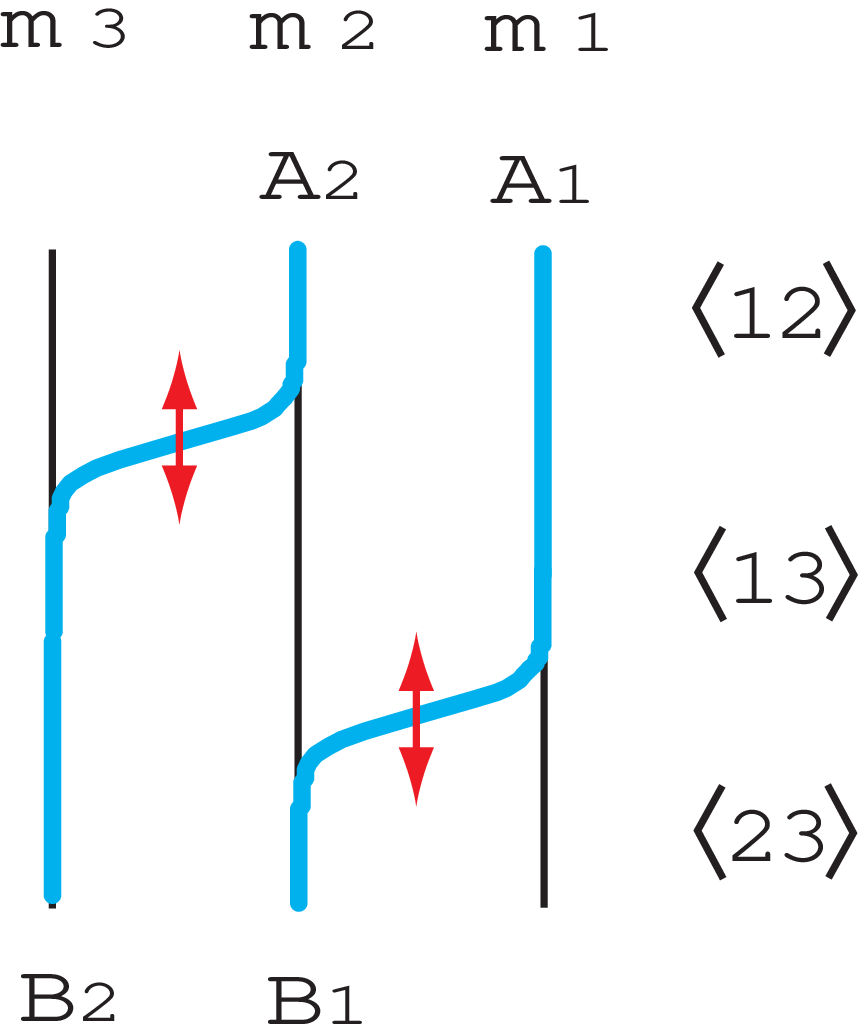}
\hs{20} &
\includegraphics[width=4cm,clip]{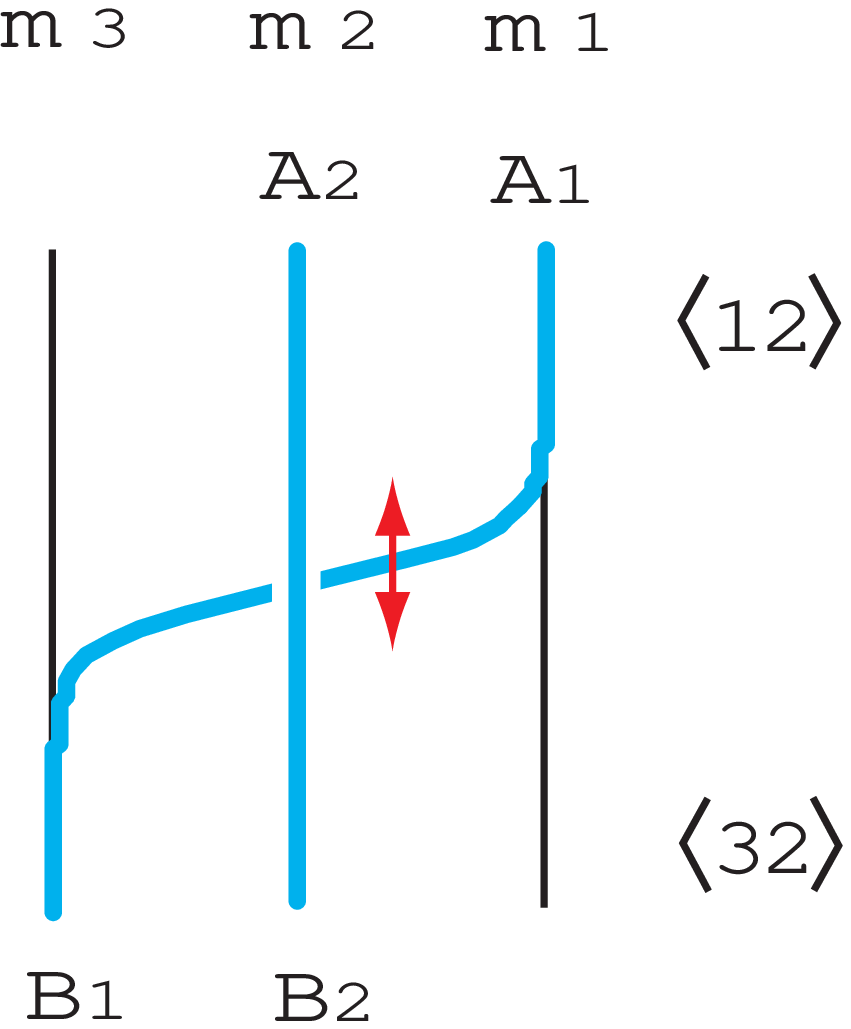} \\
a) \hs{20} & b)
\end{tabular}
\end{center}
\caption{\small
Compressing walls understood by a reconnection 
of two D$p$-branes (crossing brane). 
These configurations are dual to Fig.~\ref{fig11}-a) and -b), 
respectively, with $y \to -y$. 
However physics occurring here is more interesting. 
}
\label{fig13}
\end{figure}

The dual theory of the above example with 
$\Nc \leftrightarrow \tilde \Nc$ 
is the theory with $\NC=2$ and $\NF=3$. 
The double wall configuration in this model was 
discussed in detail in Sec.~4 in Ref.~\cite{INOS2}.
The dual brane configuration of Fig.~\ref{fig11} 
is shown in Fig.~\ref{fig13} (with $y \to -y$).
In this case the process of compressing two walls 
is very interesting. 
A double wall configuration in Fig.~\ref{fig13}-a)  
can be labeled as 
$\langle 12 \leftarrow 23 \rangle$. 
The moduli matrix for that configuration is given by 
\beq
 H_0 = \sqrt{c}\, 
 \left(\begin{array}{ccc}
      1 & e^r & 0 \\
      0 & 1   & e^s
  \end{array} \right) \;, \label{moduli-recom}
\eeq
with $e^r, e^s \in {\bf C}^*$. 
[Also this can be obtained from 
the moduli matrix (\ref{simplest-double}) 
using the relation (\ref{dual-moduli}) with relabeling 
$\tilde H_0$ by $H_0$.]
The wall solution is obtained in Eq.~(4.17) of 
\cite{INOS2} as
\begin{eqnarray}
 && \Sigma ^0 = m{e^{2my}-e^{-2my}\over e^{2my}+e^{-2my}+e^{m R}}
 \sim \left\{\begin{array}{cc}
   \pm m, & 2|y|\gg R, \quad y=\pm |y|\\
   0,& 2|y|\ll R
	  \end{array}\right.\nonumber\\
 && |\Sigma^3 | = 2m{\sqrt{(\cosh({2my})+e^{m R})^2-1}\over 
    e^{2my}+e^{-2my}+e^{m R}}
   \sim \left\{\begin{array}{cc}
     m, & 2|y|\gg R\\
     2m,& 2|y|\ll R
   \end{array}\right., \non 
 && \sqrt{(W^1_y)^2 + (W^2_y)^2} 
   = {\sqrt{4 m^2e^{mR}(e^{2my}+e^{-2my}+e^{m R})} 
        \over (\cosh({2my})+e^{m R})^2-1 } , \; \non
 && \Sigma^1 = \Sigma^2 = W^3_y = 0 , 
  \label{sol-recom}
\end{eqnarray}
where we have set $M={\rm diag.}(m,\,0,\,-m)$ and 
$(s,r)=(mR/2+i\theta /2,-mR/2-i\theta /2)$ for simplicity.
The configuration approaches the vacuum $\langle 13 \rangle$ 
between these two walls. 
Therefore this double wall configuration is approximately 
made of two walls $\langle 12 \leftarrow 13 \rangle$ 
and $\langle 13 \leftarrow 23 \rangle$. 

We now discuss the limit of compressing these two walls 
in Fig.~\ref{fig13}-b) 
in the level of the moduli matrix (\ref{moduli-recom}). 
(This limit for the solution (\ref{sol-recom}) 
in strong coupling will be discussed in 
(\ref{compressed-sol}), below.)
To take this limit 
we first need to transform $H_0$ by a world-volume 
symmetry (\ref{art-sym}) as
\beq
 H_0 \to H_0' 
  = V H_0 
  = \sqrt c \, 
 \left(\begin{array}{ccc}
  1 & 0 & -e^{r+s} \\
  0 & 1 & e^s
  \end{array} \right) \,, \hs{5}
 V =
 \left(\begin{array}{cc}
  1 & -e^r \\
  0 & 1 
  \end{array} \right) \, . \label{wv-tr.for-com.}
\eeq
Taking the limit 
\beq
 e^s \to 0,\hs{5} e^r \to \infty \label{limit}
\eeq 
with keeping 
$-e^{r+s} (\equiv e^t \in {\bf C}^*)$ fixed, 
we obtain the moduli matrix
\beq
 H_0 = \sqrt{c} \,
 \left( \begin{array}{ccc}
  1 & 0 & e^t \\
  0 & 1 & 0 
  \end{array} \right) \,. \label{recomb}
\eeq
We express this moduli matrix 
by $\langle 12 \leftarrow 32 \rangle$. 
In the strong gauge coupling limit 
the wall solutions 
for this moduli matrix can be calculated, to yield 
\begin{eqnarray}
\Sigma =
\left( 
\begin{array}{cc}
\frac{m_1 e^{2 m_1 (y-y_1)} 
       + m_3e^{2m_3(y-y_1)}}{e^{2 m_1(y-y_1)}+e^{2m_3(y-y_1)}}
   & 0 \\
 0 & m_2 
\end{array}
\right), \hs{5} 
W_y = 0, \hs{5} 
 y_1 = {{\rm Re}(t)\over m_1 - m_3}
   \label{compressed-sol}
\end{eqnarray}
where $y_1$ denotes the position of the compressed wall.
This gives a compressed kinky configuration in Fig.~\ref{fig13}-b). 
The configuration is made of the D$p$-brane  
interpolating between the first and the third D($p+4$)-branes 
and the one staying at the second D($p+4$)-brane 
representing a vacuum state [see Fig.~\ref{fig13}-b)].  
They correspond to the first and the second rows in 
the matrix (\ref{recomb}), respectively.\footnote{
The $\Sigma_{11}$ component in this solution 
coincides with the solution 
(\ref{kinky-U(1)}) 
in the limit ${\rm Re}(r) \to - \infty$ with $s$ fixed
in the dual theory. 
Therefore the duality holds exact in this strong gauge coupling limit 
as expected.
}

This phenomenon can be understood 
by recalling the s-rule 
as the exclusion principle for D$p$-branes 
which states that two D$p$-branes cannot be placed 
at the same D($p+4$)-brane. 
Although this s-rule was originally suggested 
for vacuum states,  
we find from this example that 
{\it the s-rule for vacuum states 
is extended to the case of 
kinky D-brane configurations}. 
We may call it the {\it extended s-rule}.

Moreover this phenomenon can be understood 
in terms of the {\it reconnection (recombination)} 
of D-branes.
A reconnection process was analyzed in 
Ref.~\cite{recombination} in the gauge theory on D-branes.\footnote{
We would like to thank Koji Hashimoto for valuable discussions.
}
There a reconnection of 
straightly intersecting D-branes was examined. 
In the case of an intersection of 
a $(p,q)$-string and a D-string with a particular angle 
determined by $p$ and $q$, the configuration is BPS. 
The reconnection occurs by off-diagonal elements of 
gauge fields as a marginal deformation of a moduli parameter 
as shown by K.~Hashimoto and W.~Taylor 
in \cite{recombination}.\footnote{
In the case of an intersection of D$1$-branes, 
the configuration is unstable and contains a tachyon. 
The reconnection occurs with the tachyon condensation 
and the configuration does not contain 
a moduli parameter~\cite{recombination}.
}
Our model gives a similar but complicated example of 
the reconnection of two D$p$-branes in 
a kinky configuration, which still can be analyzed in 
the gauge theory on D-branes. 
In the case of \cite{recombination} 
a singular gauge transformation is needed 
to connect the left and the right sides of 
the configuration in the limit when  the reconnection occurs.  
In fact, completely the same thing occurs in our case also.  
The world-volume transformation $V$ in 
Eq.~(\ref{wv-tr.for-com.}) diverges 
in the limit (\ref{limit}) of the reconnection.  
Correspondingly a singular gauge transformation 
is needed in that limit
as seen as follows.   
We take $\theta=0$ without loss of generality. 
With a gauge choice $\Sigma^1 = \Sigma^2 
= W^3_y = 0$, 
we find that an off-diagonal component of the gauge field
\beq
 W^1_y = {\sqrt{4 m^2e^{mR}(e^{2my}+e^{-2my}+e^{m R})} 
        \over (\cosh({2my})+e^{m R})^2-1 } , \; \hs{5}
 W^2_y = 0 
\eeq
exists before the reconnection. 
This off-diagonal component $W^1_y$ 
can be regarded as a moduli parameter of a marginal 
deformation instead of $R$ because it is a function of $R$. 
This tends to the delta function, $W^1_y \to  \pi \delta (y)$,  
in the limit of the reconnection ($R \to - \infty$). 
To eliminate it we need a singular gauge transformation 
$W_y^1 \rightarrow W_y^{1'}= W_y^1 + \partial_y \Lambda (y) =0$ 
with a gauge parameter of the step function,
$\Lambda(y) = \pi \Theta (-y)$. 
The wall solution becomes the one in Eq.~(\ref{compressed-sol})
[with $M={\rm diag.}(m,\,0,\,-m)$ and 
$(s,r)=(mR/2+i\theta /2,-mR/2-i\theta /2)$].
Therefore we conclude that 
the reconnection process in this model is 
the same as the one of the BPS case in \cite{recombination}. 
This observation suggests that the extended s-rule 
can be understood in terms of the reconnection process.

We make a brief comment here.
The moduli matrix (\ref{recomb}) is one called 
$U(1)$-factorisable as defined in Sec.~3.7 in \cite{INOS2}. 
In the $U(1)$-factorisable cases the BPS equations reduce to 
a set of those of the $U(1)$ gauge theory, 
and therefore they can be solved if those 
in the $U(1)$ gauge theory are solved.
The compressed wall solution, 
generated by the moduli matrix (\ref{recomb}),
in the theory with $\Nc=2$ and $\Nf=3$  for the finite gauge coupling 
coincides with the single wall solution, 
generated by the moduli matrix (\ref{simlest-compressed}), 
in the dual theory with $\Nc=1$ and $\Nf=3$ 
for the {\it same} gauge coupling. 
Namely the kinky D$p$-branes in Fig.~\ref{fig11}-b) and 
Fig.~\ref{fig13}-b) coincide with each other even for 
finite gauge coupling, 
and the vacuum D$p$-brane in the latter 
does not contribute to wall configurations.
Therefore both configurations before and after taking the duality 
coincide with each other 
for the {\it same} distance between the two NS$5$-branes 
[because of the relation (\ref{gauge-coupling})], 
except for the vacuum D$p$-brane in the middle D($p+4$)-brane. 
Thus the extended Hanany-Witten effect happens to give 
an {\it exact} duality in this case.\footnote{ 
We cannot expect that such an exact duality holds 
in finite gauge coupling in general. 
Even for this model, 
when walls are not compressed, 
configurations for both sides 
[Figs.~\ref{fig11}-a) and \ref{fig13}-a)] 
will be deformed in different ways.
}

This example shows that  
existence of a compressed wall or a reconnected wall 
reduces the number of the moduli parameters 
in the same topological sector. 
In other words, 
the full dimensionality of the topological sector 
can be counted only when 
the configuration is generic, 
namely with no compressed walls 
nor reconnected branes, 
as will be discussed in the subsection \ref{MSINAW}.

\subsection{Penetrable Walls}
All walls are impenetrable in the $U(1)$ gauge theory 
as stated in the last subsection. 
A characteristic feature of non-Abelian walls is 
that there also exist pairs of walls 
whose positions can be exchanged under 
a marginal deformation 
by changing moduli parameters slowly. 
They are called {\it penetrable walls}. 
This can occur because 
the size of the matrix $\Sigma$ is greater than two 
for a non-Abelian gauge group. 
However the case of $\Nc=2$ and $\Nf=3$ does not allow 
such walls because the theory is dual to $\Nc=1$ 
as shown in the last subsection. 
So we consider the case of $\NC=2$ and $\NF=4$ 
to exhibit this phenomenon. 
We show that the brane picture explains 
this phenomenon very easily although 
it is complicated in the field theory.

\begin{figure}[thb]
\begin{center}
\begin{tabular}{cc}
\includegraphics[width=5cm,clip]{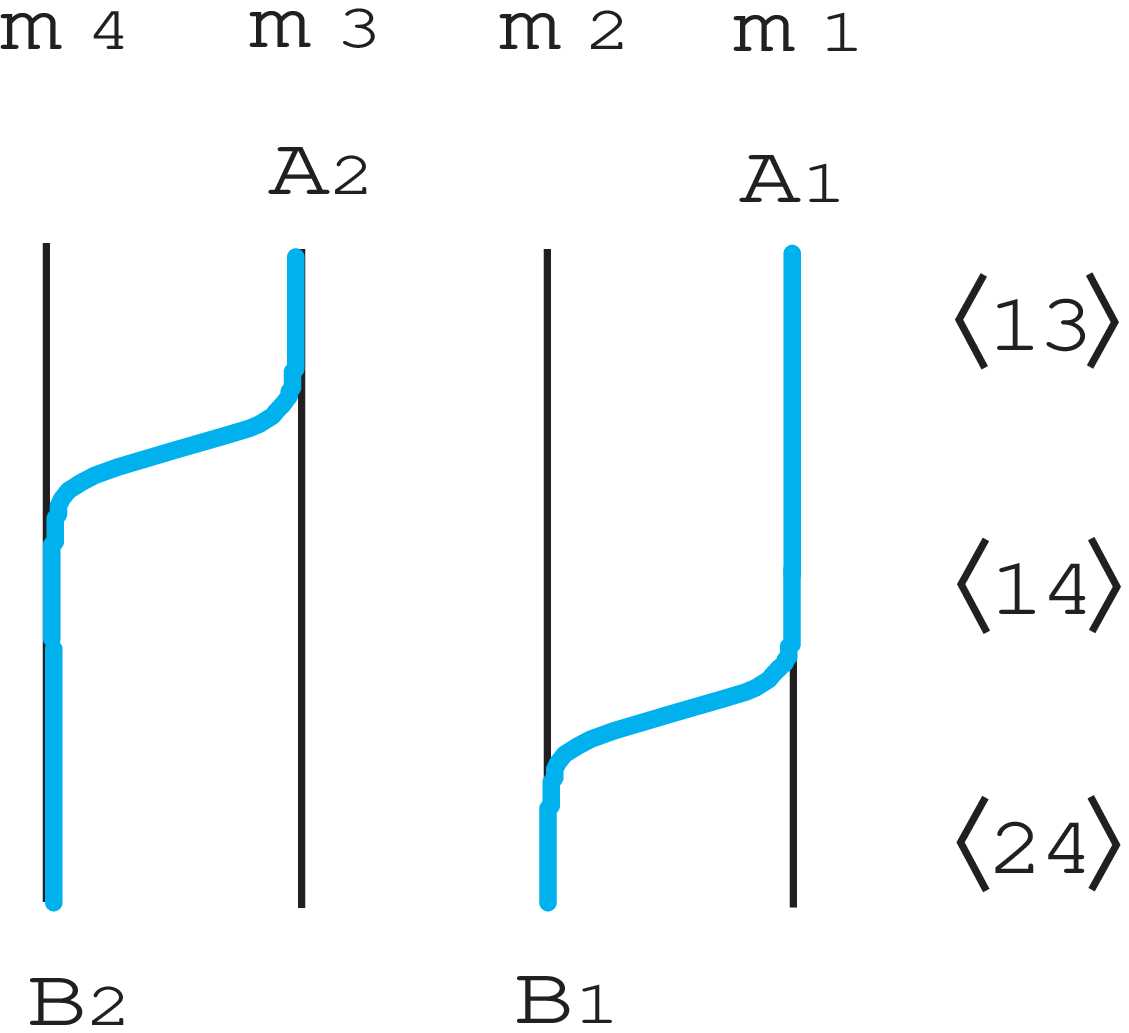}
\hs{20} & 
\includegraphics[width=5cm,clip]{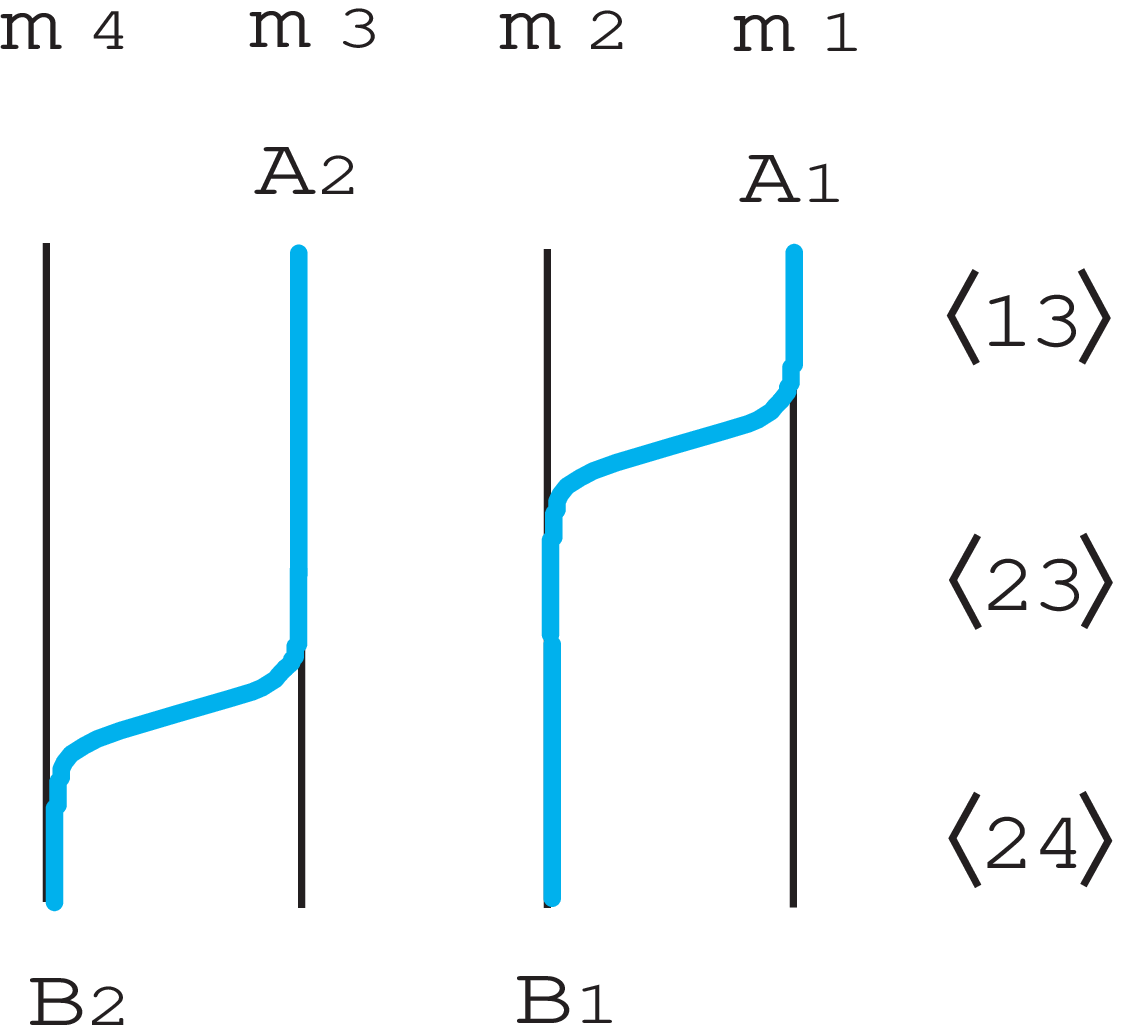} \\
 a) \hs{20} & b)
\end{tabular}
\end{center}
\caption{\small
A pair of penetrable walls $\langle 13 \leftarrow 24 \rangle$ 
in $\NC=2$, $\NF=4$ before [a)] and after [b)] exchanging 
wall positions.
}
\label{fig16}
\end{figure}

There exist pairs of impenetrable walls  
as in the $\NF=3$ case in the last subsection.
A wall configuration with penetrable walls is 
generated by the moduli matrix
\beq
 H_0 = 
 \left(\begin{array}{cccc}
  1 & e^r & 0 & 0   \\
  0 & 0   & 1 & e^s
  \end{array} \right) \label{pene-1} 
\eeq
with $e^r, \, e^s \in {\bf C}^*$. 
The wall solution in the strong gauge coupling limit 
is obtained~\cite{INOS2}, to give
\begin{eqnarray}
 \Sigma =
  \left( 
   \begin{array}{cc}
    {m_1e^{2m_1y}+m_2e^{2m_2y+2{\rm Re}(r)}
    \over e^{2m_1y}+e^{2m_2y+2{\rm Re}(r)}} & 0\\
    0 & {m_3e^{2m_3y}+m_4e^{2m_4y+2{\rm Re}(s)}
        \over e^{2m_3y}+e^{2m_4y+2{\rm Re}(s)}}
  \end{array}
  \right) , \hs{5} 
 W_y = 0 .  \label{pene-2}
\end{eqnarray}
This brane configuration is illustrated in Fig.~\ref{fig16}.
The positions of the kink interpolating 
between the first and the second D($p+4$)-branes and 
the one between the third and the fourth D($p+4$)-branes 
are (exactly) given by
\begin{eqnarray}
 y_1 = {{\rm Re}(r)\over m_1-m_2},\quad 
 y_2={{\rm Re}(s)\over m_3-m_4},  \label{posi-pene}
\end{eqnarray} 
respectively.
In this brane picture, the existence of 
penetrable walls becomes very manifest  
although this phenomenon is complicated in the field theory. 
The first and the second rows in the matrix (\ref{pene-1}) 
correspond to the kinky D$p$-brane connecting 
the first and the second D($p+4$)-branes 
and the one connecting the third and the fourth ones, 
respectively.
When the positions of two walls are exchanged, 
the configuration in Fig.~\ref{fig16}-a) becomes
that in  Fig.~\ref{fig16}-b), and vice versa. 
Both configurations in Fig.~\ref{fig16}-a) and b)
are labeled by $\langle 13 \leftarrow 24 \rangle$,  
but {\it different vacua 
$\langle 13 \rangle$ and $\langle 23 \rangle$ 
appear} in the intermediate $x^1$ 
in Fig.~\ref{fig16}-a) and b), respectively.
This is why it is complicated in field theory;  
By exchanging positions of two walls, 
the wall at bigger (smaller) $y$ connecting 
$\langle 13 \rangle$ and $\langle 14 \rangle$
($\langle 14 \rangle$ and $\langle 24 \rangle$)
in Fig.~\ref{fig16}-a) is transformed to 
the one at smaller (bigger) $y$ connecting 
$\langle 23 \rangle$ and $\langle 24 \rangle$
($\langle 13 \rangle$ and $\langle 23 \rangle$) 
in Fig.~\ref{fig16}-b). 
Therefore walls before and after exchange 
are not identical in ordinary definition 
of walls in field theory, 
since they connect different vacua. 
However from the brane picture we find 
that the identities of walls are maintained by 
this exchange.  

\begin{figure}[thb]
\begin{center}
\begin{tabular}{cc}
\includegraphics[width=5cm,clip]{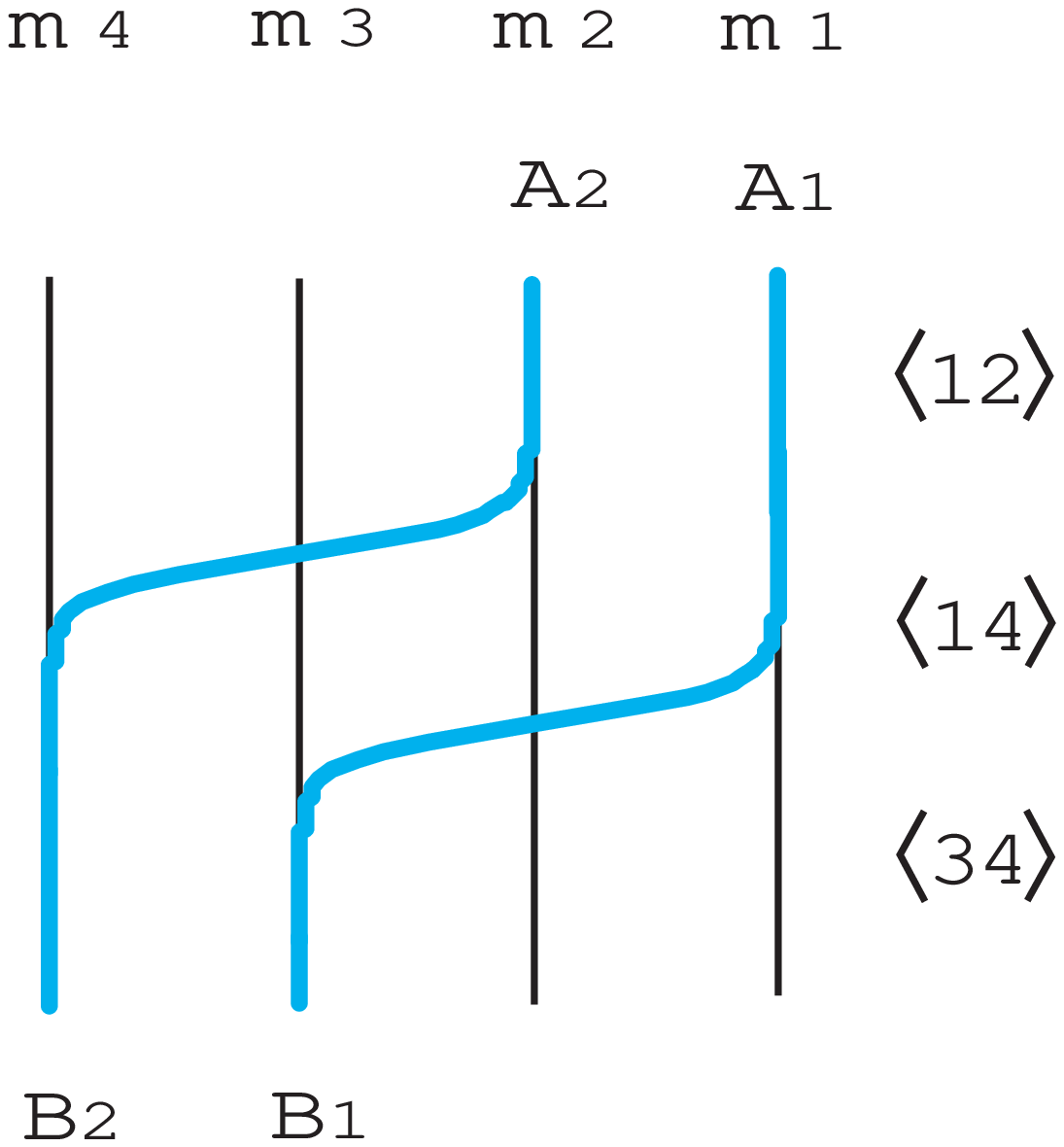}
\hs{20} & 
\includegraphics[width=5cm,clip]{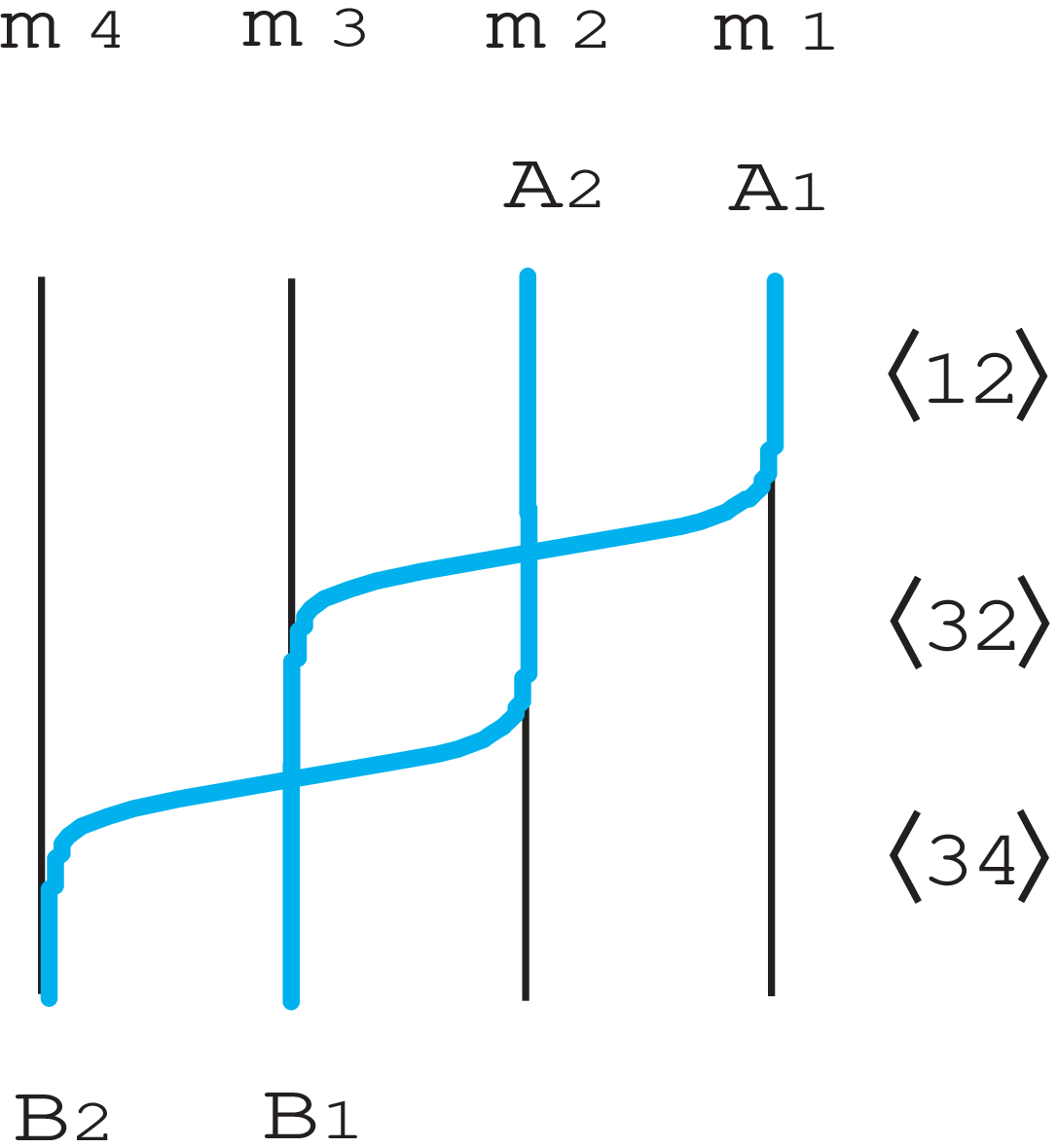} \\
 a) \hs{20} & b) 
\end{tabular}
\end{center}
\caption{\small
Another pair of penetrable walls in $\NC=2$, $\NF=4$ 
$\langle 12 \leftarrow 34 \rangle$ 
before [a)] and after [b)] exchanging wall positions.
}
\label{fig1819}
\end{figure}

\begin{figure}[thb]
\begin{center}
\begin{tabular}{cc}
\includegraphics[width=5cm,clip]{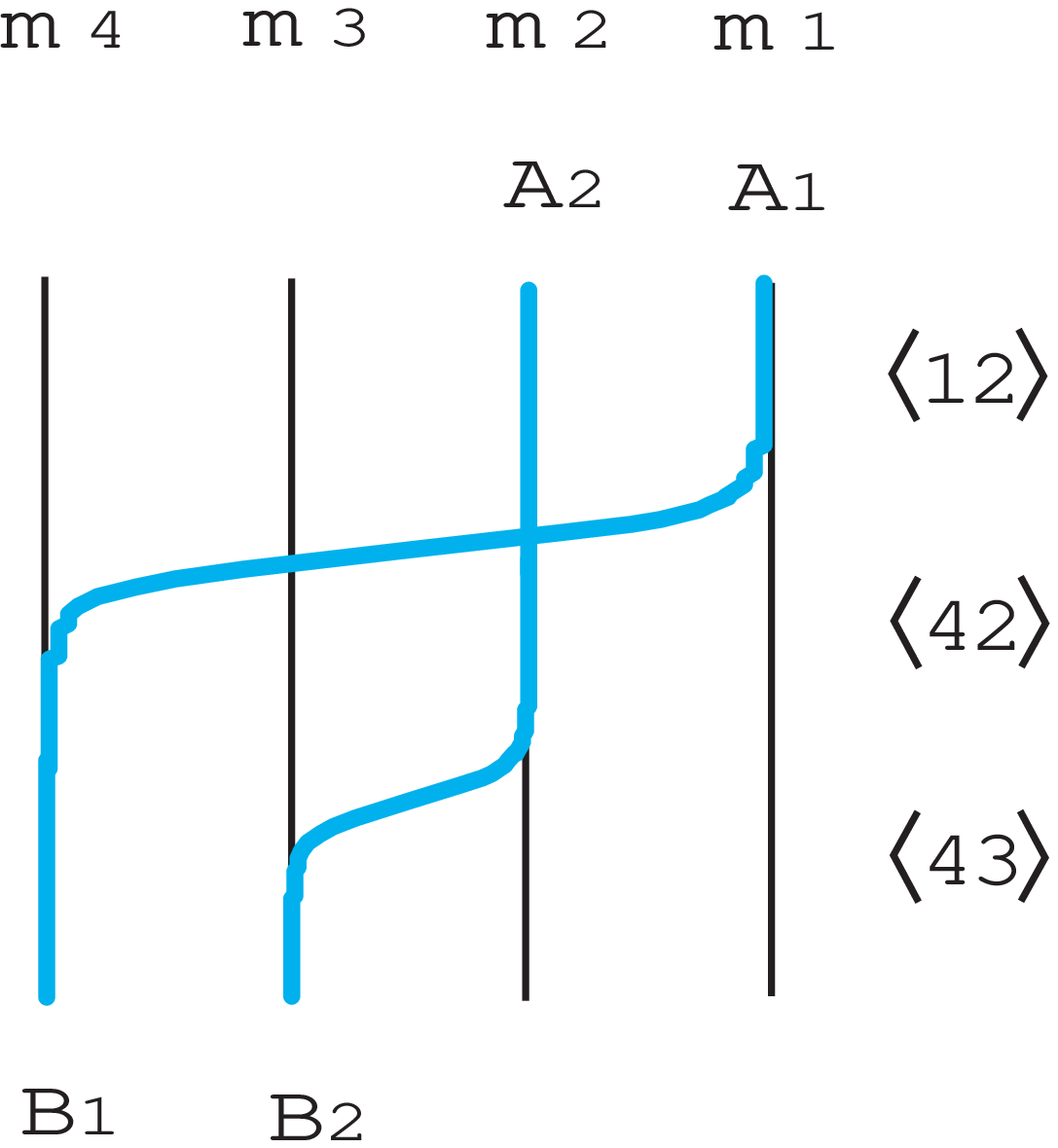}
\hs{20} &
\includegraphics[width=5cm,clip]{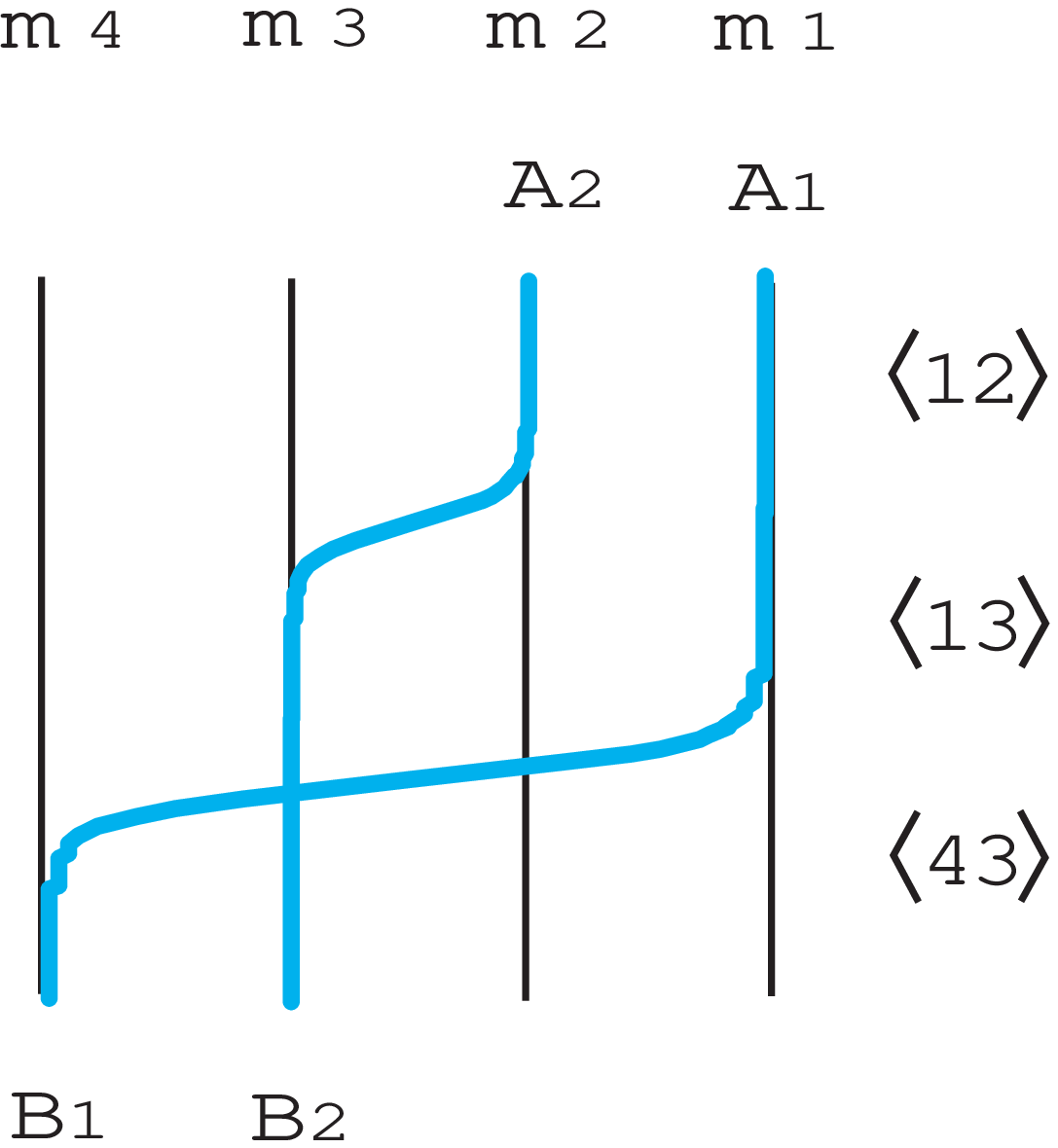} \\
 a) \hs{20} & b) 
\end{tabular}
\end{center}
\caption{\small 
The third pair of penetrable walls 
$\langle 12 \leftarrow 43\rangle$ in $\NC=2$, $\NF=4$ 
before [a)] and after [b)] exchanging wall positions.
}
\label{fig2021}
\end{figure}

There exist two more configurations admitting penetrable walls 
in the topological sector 
${\cal M}^{\langle 12 \rangle \leftarrow \langle 34 \rangle}$ 
in this model with $\NC=2$ and $\NF=4$.
They are generated by the moduli matrices 
\beq
 H_0 = 
 \left(\begin{array}{cccc}
  1 & 0 & e^r & 0   \\
  0 & 1 & 0   & e^s
 \end{array} \right) , \hs{20}
 H_0 = 
 \left(\begin{array}{cccc}
  1 & 0 & 0   & e^u   \\
  0 & 1 & e^v & 0
  \end{array} \right)  \label{pene-3}
\eeq
with $e^r, e^s, e^u, e^v \in {\bf C}^*$.
The brane configurations corresponding to these matrices 
are illustrated in Figs.~\ref{fig1819} and \ref{fig2021}, 
respectively. 
In Fig.~\ref{fig1819} two level-1 
compressed walls [the first and the second rows 
in the first matrix in (\ref{pene-3})] 
are penetrable. 
Note that they are penetrable only when both of them 
are compressed.  
In Fig.~\ref{fig2021} an elementary wall 
[the second row in the second matrix in (\ref{pene-3})] 
and a level-2 compressed single wall 
[the first row in the same matrix] 
are penetrable. 
For both cases there appear different vacua before and 
after exchanging positions of two walls.

\subsection{Moduli Space and Indices for Non-Abelian Walls} \label{MSINAW}
In this subsection we calculate dimensions of the moduli space and 
its topological sectors in our brane picture.
To this end we should classify compressed walls 
into two classes. 
The first type is a compressed wall 
made of single D$p$-brane as in Fig.~\ref{fig11}-b).  
This reduces the number of moduli parameters 
in the topological sector 
so we should not consider it to count dimensions. 
We call this type a ``self-compressed brane''. 
The second type is a compressed wall configuration 
made of two D$p$-branes as in Fig.~\ref{fig13}-b). 
This occurs if and only if 
some of $B_r$ are not ordered 
as $B_1 \leq B_2 \leq \cdots \leq B_{\NC}$~\cite{INOS2}.
We call this type a ``crossing brane''. 
The latter can be distinguished from the former 
by the presence of the reconnection.

\begin{figure}[thb]
\begin{center}
\includegraphics[width=5cm,clip]{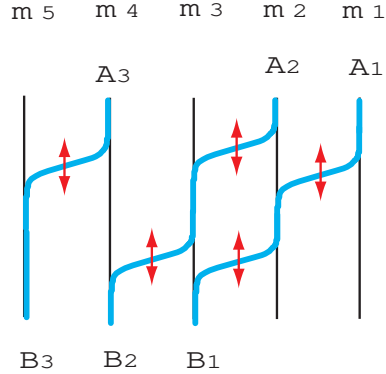}
\end{center}
\caption{\small Maximally kinky configuration/Counting 
the dimension of 
the moduli space of walls by the brane configuration.
The dimension of this example of the topological sector 
${\cal M}^{\langle 1,2,4 \rangle \leftarrow \langle 3,4,5 \rangle}$ 
can be counted to be $2 \times 5 = 10$ 
where the factor two comes from 
the translational and $U(1)$ zero modes.
}
\label{fig22}
\end{figure}

The dimension of any topological sector can be counted 
if there are no compressed walls namely 
if there are no self-compressed branes
and no crossing branes. 
We call such a configuration 
a ``maximally kinky configuration" 
in that sector (see Fig.~\ref{fig22}). 
The dimension of a given topological sector 
can be counted by considering the maximally kinky 
configuration in that sector, and it is given by 
two times the number of kinks 
where the factor two exists 
because each kink carries  the translational and $U(1)$ zero modes.
See Fig.~\ref{fig22} as an example. 

\begin{figure}[thb]
\begin{center}
\includegraphics[width=9cm,clip]{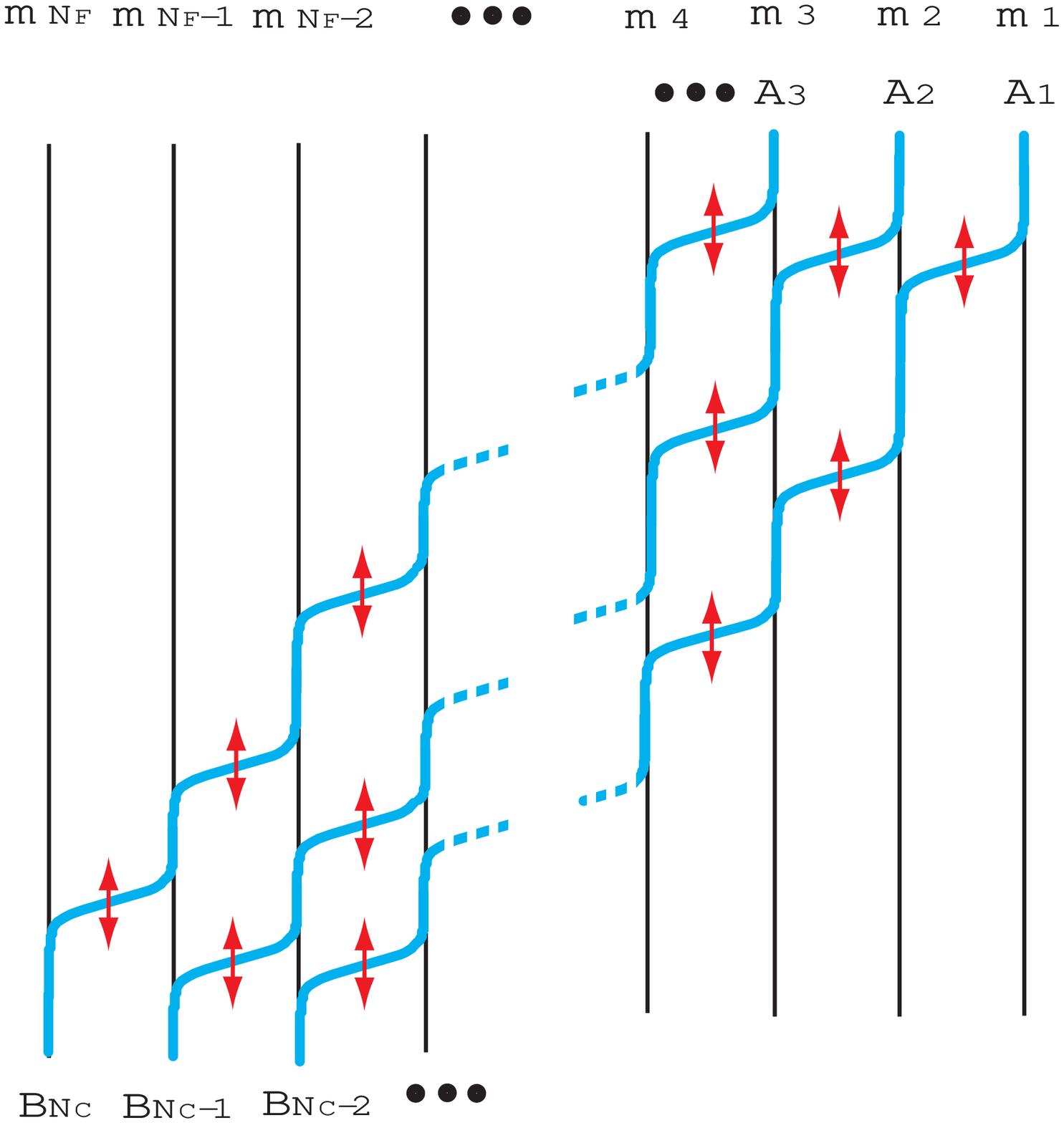}
\end{center}
\caption{\small
Counting the dimension of 
the maximal topological sector of the moduli space.
It can be counted by the configuration interpolating between 
the right-most vacuum 
$\langle A_1,\cdots,A_{\Nc}\rangle 
= \langle 1,2,\cdots,\Nc \rangle$ and 
the left-most vacuum 
$\langle B_1,\cdots,B_{\Nc}\rangle 
= \langle \Nf-\Nc+1,\cdots,\Nf \rangle$. 
}
\label{fig17}
\end{figure}

\medskip
The maximal topological sector is 
the topological sector with the maximal dimension. 
It connects the right-most vacuum $\langle 1, 2, \cdots, \NC \rangle$
and the left-most vacuum 
$\langle  \NF-\NC , \NF-\NC +1 , \cdots, \NF-1,\NF \rangle$.
%
%
The brane configuration 
for the maximally kinky configuration 
in the maximal topological sector
is illustrated in Fig.~\ref{fig17}.
Then the dimension of the maximal topological sector, 
or the dimension of the total moduli space, 
can be calculated 
from Fig.~\ref{fig17}, to yield 
\beq
 \dim {\cal M}_{\rm wall} 
&=& 2 \left[ 2 \sum_{k=1}^{N_{\rm C}} k 
       +  N_{\rm C} (N_{\rm F} - 1 - 2 N_{\rm C}) \right] \non 
  &=& 2 \NC (\NF - \NC) = \dim G_{\NF,\NC}.  \label{dim-wall-mod}
\eeq
Therefore we recover the field theoretical result 
(\ref{dim-wall-mod0}) obtained in \cite{INOS2}.
Comparing Fig.~\ref{fig17} for the wall moduli 
and Fig.~\ref{fig5} for the massless Higgs branch, 
we can understand that the dimension of the former is 
the half of the latter.\footnote{
Although there figures are very similar 
we do not know a deep connection between them yet.
}

\bigskip

\begin{figure}[thb]
\begin{center}
\includegraphics[width=9cm,clip]{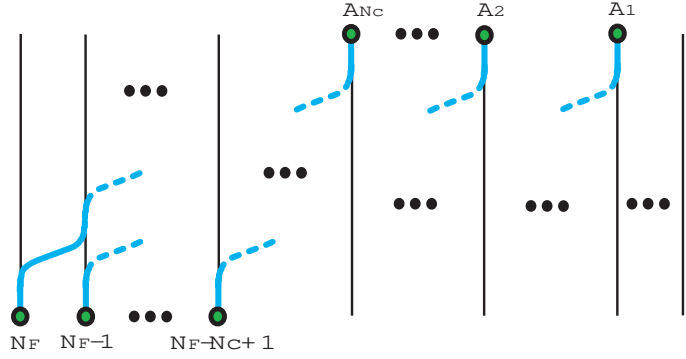}
\end{center}
\caption{\small 
The definition of the indices of vacua.
The index for the vacuum $\langle A_1,\cdots,A_{\Nc}\rangle$ 
is defined by the dimension of 
the configuration interpolating that vacuum and 
the left-most vacuum $\langle \Nf-\Nc+1,\cdots,\Nf \rangle$.
}
\label{fig29}
\end{figure}

\begin{figure}[thb]
\begin{center}
\includegraphics[width=9cm,clip]{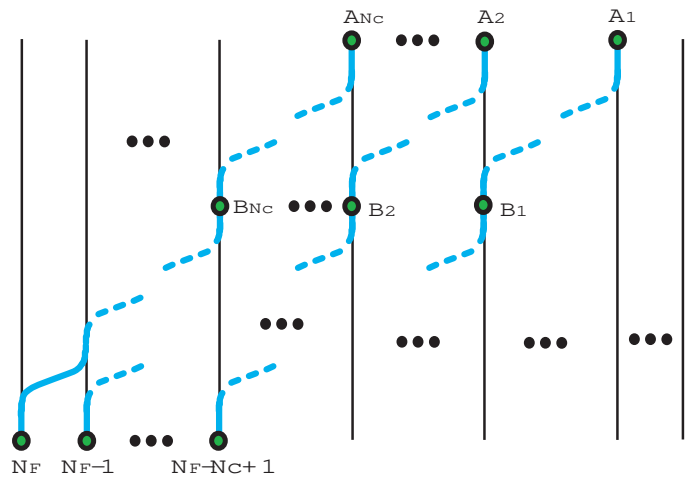}
\end{center}
\caption{\small 
Counting the dimension of a given topological sector 
of the moduli space of walls 
using the indices. 
It can be calculated as the difference of the indices of 
two vacua by adding 
the configuration interpolating between 
the vacuum $\langle  B_1,\cdots,B_{\Nc} \rangle$ 
and the the left-most vacuum 
$\langle \NF-\NC+1 , \cdots,\NF \rangle$ 
into the bottom of that sector.
}
\label{fig26}
\end{figure}

We now introduce more elegant method to 
count dimensions of the topological sectors. 
The index of a given vacuum $\langle A_1,\cdots,A_{\Nc} \rangle$ 
is defined by the real dimension of 
the maximally kinky configuration in the topological sector 
which connects that vacuum and the left-most vacuum 
$\langle  \NF-\NC+1, \NF-\NC +2 , \cdots ,\NF \rangle$. 
The moduli matrix for such a configuration is given by 
\begin{eqnarray}
&&\hspace{5.2em}A_1\hspace{3.1em}A_r \hspace{2.7em} \nn\\
 H_0&=&\sqrt{c}\left(
\begin{array}{ccccccccccc}
\cdots 0&1&*\cdots &*&\cdots &&\cdots e^{v_1} \;\;\;\;\; 0& 
    \cdots & \hs{3} \cdots \hs{3} 0 \\
 & &\vdots  & &        & &\vdots  \hs{10} \ddots &\;\; \ddots  &   \\
 & &\cdots 0&1&*\cdots & &\cdots  &\hs{-5} e^{v_r} &   
      \underbrace{0 \;  \cdots \; 0}_{\Nc - r }  \\
 & &\vdots  & &        & &\vdots  &  \hs{10} \ddots  &   & \\
 & &  & &\cdots 0&1&*\cdots &\cdots & e^{v_{\Nc}}   \\
\end{array}\right) {}^{< r} \;, 
 \label{index-moduli} \\
&&\hspace{13.8em}A_{N_{\rm C}}\hspace{5.2em} \nn
\end{eqnarray}
and corresponding brane configuration is illustrated 
in Fig.~\ref{fig29}. 
Then the index can be calculated by either the matrix 
(\ref{index-moduli}) or Fig.~\ref{fig29}, to give 
\beq
 \nu_{\langle A_1,\cdots,A_{\Nc} \rangle} 
  = 2 \sum_{r=1}^{N_{\rm C}} (N_{\rm F} - N_{\rm C} + r - A_r)
  = 2 N_{\rm C} (\Nf - N_{\rm C}) 
  + N_{\rm C} (N_{\rm C} + 1) 
  - 2 \sum_{r=1}^{N_{\rm C}} A_r  \;.
\eeq
Actually, this index can be 
shown to coincide with the Morse index of the wall moduli space, 
or the base manifold $G_{N_{\rm F},N_{\rm C}}$ 
of the massless Higgs branch 
$T^* G_{N_{\rm F},N_{\rm C}}$~\cite{EINOST}. 
Using this index we can calculate 
the dimension of arbitrary topological sector 
${\cal M}^{\langle A_1,\cdots,A_{\Nc} \rangle 
 \leftarrow 
\langle B_1,\cdots,B_{\Nc} \rangle}$.  
If we join a sector connecting 
$\langle B_1,\cdots,B_{\Nc} \rangle$
and the left-most vacuum,  
${\cal M}^{\langle  B_1,\cdots,B_{\Nc} \rangle \leftarrow 
\langle \NF-\NC+1 , \cdots,\NF \rangle}$, 
to the end of that sector (see Fig. \ref{fig26}), 
we get the index  
$\nu_{\langle A_1,\cdots,A_{\Nc} \rangle}$.
Therefore the real dimension 
corresponding to translation and $U(1)$ zero modes is given by 
the difference between 
the indices for these vacua: 
\beq
 \dim
 {\cal M}^{
   \langle A_1,\cdots,A_{\Nc} \rangle 
     \leftarrow 
   \langle B_1,\cdots,B_{\Nc} \rangle} 
 = \nu_{\langle A_1,\cdots,A_{\Nc} \rangle}
  - \nu_{\langle B_1,\cdots,B_{\Nc} \rangle} 
 = 2 \sum_{r=1}^{N_{\rm C}} (B_r - A_r) \;.
  \label{dim-topo-sec}
\eeq
This result recovers that of the field theory. 
For instance 
the dimension (\ref{dim-wall-mod}) of the wall moduli space 
is given by 
$\nu_{\langle 1,\cdots,N_{\rm C} \rangle}$.

\bigskip

As discussed in Sec.~\ref{IWAROD} 
the number of the moduli parameters reduces 
from the dimension of the topological sector 
(the dimension of the generic moduli matrix for that sector)
whenever there exist crossing branes. 
These kinds of configurations are not suitable 
to count the dimension of that topological sector 
as stated above.  
We can recognize them by looking at
the standard form (\ref{standard-form}) 
of the moduli matrix; 
They occur when $B_r$ are not ordered because 
we have to eliminate some components to fix the 
world-volume symmetry (\ref{art-sym}) as stated below 
(\ref{standard-form}).
In \cite{INOS2} such a standard form 
of the moduli matrix was shown to 
parameterize a patch with less dimensions 
in the same topological sector 
with the generic moduli matrix given by ordered $B_r$. 
The reduced number of dimension of such 
a patch is given by the number of inversion: 
the number of sets of ($i,j$) with 
$\sig(i) < \sig(j)$ and $i>j$,
where $\sig$ is an element of 
the permutation group in Eq.~(\ref{decomposed}).
To count this number 
by using brane configurations 
we define the {\it intersection number} 
by the number of intersections of D$p$-branes 
with the sign determined as follows.
When the $r$-th D$p$-brane 
(connecting $A_r$ at $y\to \infty$ and $B_r$ at $y \to -\infty$)
intersects anti-clockwise with the $s$-th D$p$-brane, 
we give $+1$ ($-1$) as an 
intersection number at that intersection point 
if $r < s$ ($r > s$). 
In either case the $r$-th D$p$-brane must be compressed 
by the s-rule. 
Then the total intersection number $I$ of a given configuration 
is defined as the sum of the intersection numbers 
of all intersection points in that configuration, 
$I = \sum_{p} {\rm sign} (s-r)$ with $p$ intersection points. 
By its definition $I$ is non-negative and $I=0$ 
corresponds to the generic moduli matrix, 
generating the maximally kinky configuration in general. 
The important thing is that 
$I$ is invariant 
under the change of the positions of kinks 
provided that reconnections do not occur.
For instance the configuration in 
Fig.~\ref{fig1819}-b) intersect twice with different sign 
and therefore it can be changed to 
Fig.~\ref{fig1819}-a) without reconnection 
and is generated by the generic moduli matrix. 
On the other hand if a reconnection occurs the ordering of 
$B_r$ changes.
We find that {\it the number of reduced moduli 
(the number of inversion) 
coincides with the total intersection number $I$ of the configuration.}
For instance both figures in Fig.~\ref{fig16} (Fig.~\ref{fig1819}) 
have the intersection number zero, 
and therefore they are described by 
the generic moduli matrix  
with ordered $B_r$ (with some zero elements) 
in ${\cal M}^{\langle 13 \rangle \leftarrow \langle 24 \rangle}$ 
(${\cal M}^{\langle 12 \rangle \leftarrow \langle 34 \rangle}$). 
On the other hand both the configurations 
in Fig.~\ref{fig2021} have the intersection number one 
and therefore they are not described by the generic moduli matrix 
in that sector.

Moduli matrices in the standard form 
with various ordering of $B_r$ 
contain moduli parameterizing coordinate patches 
in the topological sector 
${\cal M}^{\langle A_1, \cdots, A_{\Nc} \rangle 
\leftarrow  \langle B_1, \cdots, B_{\Nc} \rangle}$ 
as in Eq.~(\ref{decomposed}).
The generic moduli matrix generating 
the maximally kinky configuration corresponds 
to the patch 
$U^{\langle A_1, \cdots, A_{\Nc} \leftarrow 
B_1, \cdots, B_{\Nc} \rangle}$ 
with ordered $B_r$ and with $I=0$.   
Other patches with non-ordered $B_r$ are 
obtained by causing some reconnections. 
Such the minimum number of reconnections 
is given by the intersection number $I$.
However all configurations in a given topological sector 
${\cal M}^{\langle A_1, \cdots, A_{\Nc} \rangle 
 \leftarrow  \langle B_1, \cdots, B_{\Nc} \rangle}$
cannot be classified by only the intersection number 
or the number of the reconnection 
because there exist configurations with the same 
intersection number $I$, which cannot 
be transformed to each other.  
The real dimension of arbitrary patch in Eq.~(\ref{decomposed}) 
can be calculated, to yield
\beq
 \dim U^{\langle A_1, \cdots, A_{\Nc} 
 \leftarrow  B_{\sig(1)}, \cdots, B_{\sig(\Nc)} \rangle} 
 &=& \dim {\cal M}^{\langle A_1, \cdots, A_{\Nc} \rangle 
 \leftarrow  \langle B_1, \cdots, B_{\Nc} \rangle} 
  - 2 I(\sigma)  \non
 &=&  2 \sum_{r=1}^{\Nc} (B_r - A_r ) - 2 I(\sigma) .
\eeq 

\medskip
Before closing this section 
we discuss 
Nambu-Goldstone (NG) modes in this system. 
Wall configurations break the translational symmetry 
as well as the $U(1)^{\NF-1}$ flavor symmetries. 
Correspondingly there appear 
massless NG modes on the world-volume theory on the walls.
The complex number of moduli parameters 
coincides with the (real) number of walls and 
is greater than $\Nf -1$ in general. 
The real parts of the moduli parameters 
correspond to wall positions.  
The $\Nf -1$ imaginary parts of the moduli parameters 
are NG bosons for broken $U(1)^{\NF-1}$ symmetry.  
The rest of imaginary parts are zero modes 
which are not dictated by the spontaneously broken symmetry, 
and are called quasi-NG (QNG) bosons~\cite{QNG}.
Together with ordinary NG bosons 
they constitute complex scalar fields 
needed for unbroken four SUSY. 
The number of QNG bosons was calculated as 
\beq 
 N_{\rm QNG} = (\NC - 1) (\tilde\NC -1)  \label{QNG}
\eeq
in the maximal topological sector~\cite{INOS2}. 
We can now easily count the number of QNG bosons in 
arbitrary topological sector in the brane picture 
because only one kink carries a $U(1)$ NG zero mode 
(the overall phase) but 
the others are QNG zero modes
in each intermediate space between adjacent D$(p+4)$-branes.
For instance the number of the QNG bosons in Fig.~\ref{fig22} 
is counted as one because there exist two kinks between 
the second and the third D($p+4$)-branes.
In the case of the maximal topological sector 
the field theoretical result (\ref{QNG}) is recovered from 
Fig.~\ref{fig17}.

\section{Conclusion and Discussion} \label{CD}

We have realized BPS non-Abelian multi-walls 
in SUSY $U(\Nc)$ gauge theory with $\Nf$ fundamental 
hypermultiplets by a brane configuration made of 
$\Nc$ kinky fractional D$p$-branes interpolating between 
$\Nf$ D($p+4$)-branes separated by 
amounts corresponding to hypermultiplet mass differences.
The tension formula (\ref{eq:tension}) for non-Abelian 
walls has been correctly reproduced 
as the kinky D-brane tension.
The duality between wall configurations 
in the theories with gauge groups $U(\Nc)$ and $U(\tilde \Nc)$ 
found in field theory in the strong gauge coupling limit 
has been explained by exchange of the two NS$5$-branes 
and the Hanany-Witten effect 
extended to the kinky configurations. 
Compressing two walls made of two D$p$-branes 
has been explained by the extended s-rule, 
and the reconnection (recombination) 
of these two D$p$-branes has been found to occur.
We have shown that the dimensions of the moduli space 
for non-Abelian walls and its topological sectors 
can be counted by the brane configuration.  
In particular we have defined the index which is  
useful to count their dimensions. 
The total moduli space 
${\cal M}_{\rm wall} \simeq G_{\Nf,\Nc}$ for non-Abelian walls 
has been understood to have the dimension of 
the base space of the Higgs branch 
in the massless limit, 
${\cal M}^{M=0}_{\rm vacua} \simeq T^* G_{\Nf,\Nc}$, of the theory, 
by comparing these two brane configurations 
(Figs.~\ref{fig5} and \ref{fig17}).
Some more correspondences between field theoretical results 
and the brane configuration have been clarified.

Let us discuss the several issues in the following.


\underline{A brane configuration for a monopole in the Higgs phase}. 
Recently a monopole in the Higgs phase 
(a confined monopole) has been found \cite{vm,INOS3}, 
and it has been shown to be a 1/4 BPS state realized as a kink in 
the vortex effective theory. 
The brane configuration for a monopole in the Higgs phase 
has been given by Hanany and Tong in Fig.~3 in 
Appendix A of \cite{HT2}. 
(The same configuration has been discussed 
in \cite{Auzzi:2004yg}.) 
It looks very similar with our brane configuration 
for walls in Fig.~\ref{fig25}-d), 
but is not identical to ours as explained 
as follows. 
The configuration is possible for $p \geq 3$. 
In the case of $p=3$ 
the $U(\Nc)$ gauge theory 
with $\Nf$ hypermultiplets in four dimensions 
is realized on D$4$-branes 
obtained by taking T-duality in the $x^8$- and 
$x^9$-directions in Fig.~\ref{fig6}. 
$k$ vortices in that theory are realized as $k$ D$2$-branes 
stretched between D$4$-branes~\cite{HT},  
and the total configuration becomes 
\beq
 \mbox{$k$   D$2$:} && 0 \hs{4}3 \hs{8} 8       \non
 \mbox{$\NC$ D$4$:} && 012  \hs{10}          89 \non
 \mbox{$\NF$ D$6$:} && 01 \hs{2} 345 \hs{4} 89 \non
 \mbox{2 NS$5$:}    && 01 \hs{8} 6789  \;.
\eeq
In Fig.~\ref{fig25}-d) vortices as D$2$-branes are 
at the segment PQ or RS. 
In \cite{HT2} a monopole in the Higgs phase has been 
realized as a D($p-1$)-brane whose position at the $x^6$-coordinate 
depends on $y=x^1$ such that it is at RS when $y \to - \infty$ 
and at PQ when $y \to + \infty$.
At intermediate $y$ the D($p-1$)-brane along $\Delta x^3$ 
cannot end and therefore it bends and 
extends to the $x^2$-coordinate 
to attach itself to the two NS$5$-branes.
Then the configuration 
becomes very similar with Fig.~\ref{fig25}-d). 
However they are different because 
branes ending on the NS$5$-branes are 
the D($p-1$)-branes 
for their case of kinks inside vortices but 
the D($p+1$)-branes for our case of kinks themselves. 
The mass of the monopole can be calculated 
by the tension of the D($p-1$)-brane. 
In this case, the contribution from Area (PQSR) 
is the energy of the vortex 
and we have to subtract it, contrary to our case 
for walls. 
Therefore the monopole mass is calculated 
from $\Delta x^2 \Delta x^6$ 
which is the sum of 
Area (PP$'$R$'$R) and
Area (QQ$'$S$'$S), to give
\beq
 \tau_{p-1} \Delta x^2 \Delta x^6 
 = {\Delta m \over g^2}
\eeq
where we have used Eqs.~(\ref{gauge-coupling})
and (\ref{FI-brane}).
This correctly reproduces the mass of 
a monopole in the Higgs phase~\cite{vm,INOS3}. 
This also coincides with the mass of an ordinary monopole 
in the Coulomb phase 
because we get the brane configuration~\cite{HW} 
(see also \cite{HT2,Auzzi:2004yg})
for a monopole in the Coulomb phase in the limit 
$\Delta x^3 \sim c \to 0$. 

In Ref.~\cite{INOS3} 
we have given the exact solutions for 
1/4 BPS states including all 
of walls, monopoles and vortices. 
A brane configuration for these mixed states 
should be realized as a mixture of ours and 
the one in \cite{HT2}.

\begin{figure}[thb]
\begin{center}
\includegraphics[width=5cm,clip]{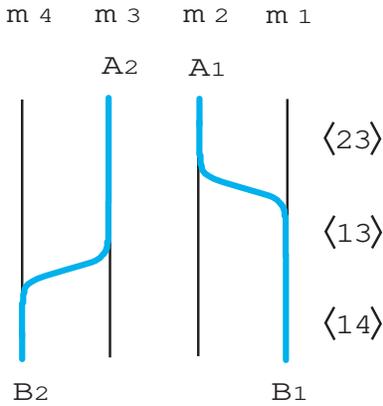}
\end{center}
\caption{\small 
A brane configuration for a non-BPS wall configuration.
}
\label{fig24}
\end{figure}

\medskip
\underline{Non-BPS walls}. 
Brane-anti-brane systems are usually unstable. 
Kink-anti-kink configurations made of 
a single D$p$-brane are also unstable 
and decay to a vacuum state.
However we can construct 
{\it stable} non-BPS wall configurations 
by multiple kinky D$p$-branes  
at least one of which is anti-BPS.  
This is possible if the flavor $\Nf$ is greater than four 
and the color $\Nc$ and $\tilde \Nc$ are greater than two. 
The simplest solution can be obtained from 
the BPS double wall solution (\ref{pene-2}) 
in the model with $\Nf =4$ and $\Nc=2$, 
by flipping the sign of $y$ in one BPS wall solution,  
say the (1,1)-element of $\Sigma$ in 
Eq.~(\ref{pene-2}) and the upper-left block 
in the hypermultiplets $H$, to make it anti-BPS: 
\begin{eqnarray}
 \Sigma =
  \left( 
   \begin{array}{cc}
    {m_1e^{- 2m_1y} + m_2e^{-2m_2y - 2{\rm Re}(r)}
    \over e^{-2m_1y}+e^{-2m_2y - 2{\rm Re}(r)}} & 0\\
    0 & {m_3e^{2m_3y}+m_4e^{2m_4y+2{\rm Re}(s)}
        \over e^{2m_3y}+e^{2m_4y+2{\rm Re}(s)}}
  \end{array}
  \right) , \hs{5} 
 W_y = 0   \label{non-BPS}
\end{eqnarray}
where we have rewritten $r$ as $-r$.
The corresponding brane configuration is illustrated 
in Fig.~\ref{fig24}. 
We did not examine if the two walls are 
non-interactive or interactive with each other  
where interaction may arise from the fluctuation 
in the off-diagonal elements in $\Sigma$ 
and/or the off-diagonal blocks in $H$.
However we can show that this configuration 
is likely stable as follows. 
The solution (\ref{non-BPS}) 
has two complex moduli parameters $r$ and $s$, 
and positions of the two walls 
are still given by Eq.~(\ref{posi-pene}).  
If the relative position (distance) is large enough, 
$R \to \infty$, the system reduces to two separated 
(anti-)BPS walls and gives the stable minimum with 
respect to the small fluctuations, 
because the BPS energy bound is saturated 
at one region and the anti-BPS energy bound 
at another region, separately.  
Therefore we find that there does not exist  
repulsive force between walls for 
the whole region of $R$. 
This situation is completely different from 
the one in \cite{Binosi:2000jb} where 
there exists repulsive force between 
BPS and anti-BPS walls. 
Then we have two possibilities.
If that stable situation continues to hold even for small $R$ 
the above solution is an absolute minimum, 
and they are penetrable. 
If we have a negative potential for small fluctuation 
at small $R$, it is likely that we have stable bound states.
We thus conclude that the non-BPS configuration is most 
likely to be stable. 
A concrete discussion is desired. 

More subtle configuration 
which is stable for small 
fluctuations and metastable for large fluctuations 
has been considered in the strong gauge coupling limit 
of the model with $\Nf=2$ and $\Nc=1$~\cite{EMS}. 
Effective action for these non-BPS walls 
was constructed by the nonlinear realizations 
and the Green-Schwarz method \cite{Clark:2002bh}. 
Discussing implications of these states by means of 
brane configurations in string theory 
is very interesting and remains as a future problem.

\medskip
\underline{Degenerate masses}. 
Only $U(1)$ Nambu-Goldstone (NG) bosons are localized 
on walls in our model with 
non-degenerate masses for hypermultiplets, 
because $SU(\Nf)$ flavor group is explicitly 
broken down to $U(1)^{\Nf-1}$ by non-degenerate masses. 
If all masses for hypermultiplets 
are degenerate, the moduli space of vacua becomes 
the continuously connected manifold 
${\cal M}^{M=0}_{\rm vacua} \simeq T^* G_{\Nf,\Nc}$ 
parametrized by (quasi-) NG bosons for 
spontaneously broken $SU(\Nf)$.
However walls cannot exist in this case. 
If there are non-degenerate masses even partially, 
the vacuum manifold consists of disconnected manifolds, 
and walls can exist.
In the case of partially degenerate masses, global 
symmetry contains non-Abelian groups, instead of 
$U(1)^{\Nf-1}$. 
If vacuum corresponding a degenerate mass appears 
at a boundary ($y \to \pm \infty$), we need to specify 
the boundary condition with respect to such a non-Abelian 
global symmetry, 
since different points in the space of the global symmetry group 
are physically distinct points. 
In such a situation, the NG modes for the broken 
non-Abelian global symmetry is not a physical 
normalizable modes localized on walls 
(but are the bulk modes), 
since their wave functions extend 
to the infinity. 
If vacua with degenerate mass appear at both infinities 
($y \to \pm \infty$), the relative rotation modes of two 
global symmetry transformations are localized on the wall 
and are physical NG bosons.
In this way, one can obtain non-Abelian NG bosons 
localized on the wall.
A field theoretical analysis for 
the example of
$\Nc =2$ and $\Nf =4$ with the mass matrix 
$M= {\rm diag}.\, (m,m,-m,-m)$
was discussed by Shifman and Yung \cite{Shifman:2003uh}. 
It admits two walls 
and $U(2)$ NG bosons appear when they are coincident.
More detailed analysis for general models with degenerate masses 
will be reported elsewhere~\cite{EINOS}.

\begin{figure}
\begin{center}
\includegraphics[width=2cm,clip]{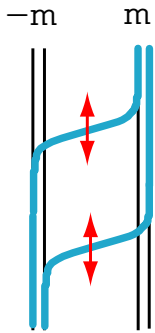}
\caption{\small 
Two walls in the model with $\Nc=2$ and $\Nf=4$ and 
the hypermultiplet masses
$M= {\rm diag}.\, (m,m,-m,m)$. 
They are penetrable because the s-rule allows 
two D$p$-branes can exist inside 
two coincident D($p+4$)-branes.
}
\label{fig-degenerate}
\end{center}
\end{figure}

In the brane picture the vacua are realized by 
some coincident D($p+4$)-branes. 
Flavor symmetry is enhanced to non-Abelian 
by the freedom of 
open strings on the coincident D($p+4$)-branes. 
The s-rule allows at most $n$ D$p$-branes to be placed at 
$n$ coincident D($p+4$)-branes. 
This drastically changes the dynamics of walls 
realized as kinky D$p$-branes.  
For instance, as a result of the s-rule,  
two walls in the theory \cite{Shifman:2003uh} 
with the degenerate masses 
$M= {\rm diag}.\, (m_1,m_2,m_3,m_4) 
= {\rm diag}.\, (m,m,-m,-m)$ 
are penetrable (see Fig.~\ref{fig-degenerate}),  
although they become impenetrable in generic configuration 
with mass perturbation 
$M= {\rm diag}.\, (m + \Delta m,m,-m,- m- \Delta m)$ 
by small mass $\Delta m$.
Moreover kinky D$p$-branes traveling 
through the coincident D($p+4$)-branes 
can break the non-Abelian flavor symmetry 
arisen from coincidence of D($p+4$)-branes. 
Thus corresponding non-Abelian NG bosons can be localized on walls.
We expect that almost all discussions in this paper 
such as counting dimensions 
hold for these cases with degenerate hypermultiplet masses.

\medskip
\underline{Complex, triplet and quartet masses}.
In this paper we have assumed real masses for hypermultiplets. 
Actually D$p$-branes with  $p=4,3, 2, 1$ allow 
real, complex, triplet and quartet masses, respectively. 
Considering these general masses does {\it not} change 
our discussion drastically because 1/2 BPS states can 
become kinky along only one spatial coordinate.
However these more general mass parameters 
become important if we consider BPS states 
preserving less than 1/2 SUSY. 
For instance a domain wall junction was constructed in \cite{KS} 
as a 1/4 BPS state in the $D=4$, ${\cal N}=2$ SUSY QED 
with three hypermultiplets with complex masses, 
embedding the known exact solutions~\cite{junction}. 
The theory is realized on a D$3$-brane in the 
brane configuration 
\beq
 \mbox{$1$ D$3$:} && 0123     \non
 \mbox{$3$ D$7$:} && 01234567 \non
 \mbox{${\bf C}^2/{\bf Z}_2$ ALE:}  && \hs{8.2} 4567 \;.
\eeq
Positions of the three D$7$-branes in the $(x^8,x^9)$-plane 
correspond to three complex masses for hypermultiplets.
A junction depends on two coordinates, say $x^2$ and $x^3$ 
of the D$3$-brane world volume $x^0,x^1,x^2,x^3$. 
The D$3$-brane position in $x^8$ and $x^9$ depends 
on $x^2$ and $x^3$. 
The value of the complex adjoint scalar $\Sigma$ of 
the vector multiplet corresponds to 
the D$3$-brane position 
in the ($x^2$,$x^3$)-plane. 
At three directions in $x^2,x^3 \to \infty$ 
with keeping $x^2/x^3$ fixed, 
there appear three D$7$-branes corresponding 
to the three possible vacua. 
Between each three pairs of D$7$-branes, we obtain three 
different walls represented by kinky D$3$-branes. 
A junction interpolates among three vacua (D$7$-branes) 
as a function of the entire $(x^2,x^3)$-plane. 
Therefore the junction is represented by a curved D$3$-brane, 
interpolating among three D$7$-branes. 
The world volume is ${\bf R}^{1,1} \times \Sigma$ with 
a two dimensional surface $\Sigma$. 
The triplet or quartet masses may be required 
to construct 1/8 BPS states.

\medskip 
\underline{Other representations/gauge groups}.
Generalization to theories with other representations 
is an interesting problem. 
An adjoint hypermultiplet can be introduced 
if we do not divide the D$(p+4)$-brane world-volume by ${\bf Z}_2$.
The adjoint hypermultiplet represents positions of 
the D$p$-branes inside the D$(p+4)$-branes, 
and the vacua realize 
the ADHM moduli space for instantons. 
\begin{figure}
\begin{center}
\includegraphics[width=6cm,clip]{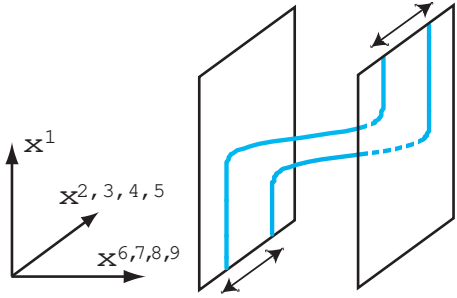}
\caption{\small 
Kinky D-branes for theory with 
an adjoint hypermultiplet. 
Walls can move to the $x^{2,3,4,5}$-directions parametrized 
by adjoint hypermultiplets.  
Therefore they are mostly penetrable. 
}
\label{fig2}
\end{center}
\end{figure}
It gives (likely) non-normalizable moduli parameters 
in wall solutions which may be fixed by the boundary conditions 
and may not be localized on walls 
(see Fig.~\ref{fig2}).\footnote{
Walls in this theory are being analyzed by David Tong.
Similar non-normalizable moduli parameter for 
wall solutions 
was discussed in a simpler model~\cite{Naganuma:2000gu}.
} 
Since the positions of the D$p$-branes 
inside the D$(p+4)$-branes do not coincide in general,  
non-Abelian walls made of different kinky D$p$-branes
are mostly penetrable with each other 
although all walls made of one kinky D$p$-brane 
are still impenetrable. 
If we consider masses for the adjoint hypermultiplet also, 
the vacuum structure is drastically changed. 
This is realized by the Scherk-Schwarz dimensional 
reduction along a common direction of 
D$p$- and D($p+4$)-brane world-volume using 
the rotational symmetry of four co-dimensions 
of the D$p$-branes in the D($p+4$)-branes.\footnote{
We would like to thank Yuji Tachikawa for pointing this out.
}
The massive adjoint hypermultiplet may participate 
in constituting walls.

On the other hand if we divide 
the D$(p+4)$-brane world-volume 
by ${\bf Z}_{n+1}$ ($A_{n}$-type) or 
other discrete subgroups of $SU(2)$ ($D_n$- or $E$-type), 
we obtain the quiver gauge theories of $A$ ($D$ or $E$)-types 
with a product gauge group $U(N_1) \times U(N_2) \times \cdots $ 
and bifundamental hypermultiplets. 
The behaviors of the fundamental hypermultiplets 
can be considered to be similar with our model, 
but those of the bifundamental hypermultiplets are not clear. 
These may allow richer structures for non-Abelian walls.

Other gauge groups such as $SO(N)$ or $Sp(N)$ 
can be discussed by introducing orientifold planes. 
However one immediate question is how to introduce 
naturally a $U(1)$ factor in the gauge group and 
a FI term which are needed to obtain 
discrete vacua.

\medskip 
\underline{Breaking to ${\cal N}=1$ SUSY}. 
Here let us consider $D=4$, ${\cal N}=2$ SUSY gauge 
theory for $p=3$ in the T-dualized brane configuration.
We have the corresponding brane configurations, 
if we replace D$2$-branes by D$4$-branes occupying 01267 
and D$4$-branes by D$6$-branes occupying 0134567
in Fig~\ref{fig4}. 
First pull out all the D6-branes to the spatial infinity at $x^2=+\infty$. 
Then there appear $\Nf - \Nc$ semi-infinite D$4$-branes 
attached on the right of the right NS$5$-brane 
by the Hanany-Witten effect~\cite{HW}. 
By rotating one of NS$5$-branes holomorphically in the $v$-$w$ plane,
where $v=x^8+ix^9$ and $w=x^3+ix^4$,
the ${\cal N}=2$ SUSY gauge theory 
reduces to ${\cal N}=1$ SUSY gauge 
theory~\cite{Elitzur:1997fh,Barbon:1997zu}. 
In the field theory side this deformation corresponds to 
adding a large mass for the adjoint scalar field $\Sigma$. 
Relation with the known 
results for walls~\cite{Acharya:2001dz,Ritz:2002fm} 
in ${\cal N}=1$ SUSY gauge theory 
would be an interesting subject.
In particular 
the Seiberg duality is realized by exchanging rotated NS$5$-branes. 
Investigating implications of duality 
on walls is interesting.


\medskip
\underline{Direct derivation of the wall effective action}. 
It is desirable to replace kinky D-branes by other 
orthogonal straight D-branes connecting D($p+4$)-branes. 
This is because it immediately leads to 
the effective field theory on non-Abelian walls 
in the D-brane picture;  
If we could do this it would be possible to construct 
the moduli space for non-Abelian walls 
by the vacuum conditions on the effective theory 
as other solitons like instantons~\cite{Wi}, 
monopoles~\cite{brane-monopole} and vortices~\cite{HT}. 
It might be possible to understand not only their 
dimensionality but also why the wall 
moduli space ${\cal M}_{\rm wall} \simeq G_{\Nf,\Nc}$ 
is a special Lagrangian submanifold of 
the Higgs branch 
${\cal M}_{\rm vacua}^{M=0} \simeq T^* G_{\Nf,\Nc}$ 
in the massless limit. 
Also this may be related to the problem of 
gauge field localization, 
because the world-volume theory on D-branes 
contains localized gauge fields by its definition.

\medskip 
\underline{Higher derivative corrections}. 
Let us discuss two points on higher derivative corrections. 
The first point is the higher derivative corrections 
coming from the Dirac-Born-Infeld (DBI) action 
on D$p$-branes. 
In the case of monopoles, 
BPS solutions in the field theory are not modified 
(up to some order) 
when the theory is promoted to the DBI action 
which includes higher derivative terms.\footnote{
We would like to thank K.~Hashimoto.} 
Therefore they can be understood as 
solitons in the DBI theory called the BIons~\cite{BIon}. 
It is very interesting to investigate if wall solutions 
are also unchanged when we promote the theory to the DBI action 
and if we can interpret them as BIons with co-dimension one.

The second point is higher derivative corrections 
in the effective action on walls.
The Manton's effective theory on the non-Abelian walls 
was constructed as (\ref{moduli_metric}) in \cite{INOS1,INOS2} 
in the strong gauge coupling limit: 
\begin{eqnarray}
 {\cal L}_{\rm walls}^{g \to \infty} 
  = c \int d^4 \theta \int dy \ \log \det \Omega_{g \to \infty} 
  = c \int d^4 \theta \int dy \ \log \det (H_0 e^{2 M y} H_0{}^\dagger) \; 
 \label{moduli_metric}
\end{eqnarray}
where we have promoted the moduli matrix 
to $D=4$, ${\cal N}=1$ chiral superfield 
$H_0 (x,\theta,\thb)$ satisfying 
$\bar D_{\dot\alpha} H_0 =0$, 
and $d^4 \theta$ denotes integration over superspace.\footnote{ 
For finite gauge coupling there exists $1/g^2$ correction to 
this K\"ahler potential~\cite{EINOS}.}
It is valid at low energies up to two derivative terms 
because it is constructed under the assumption of 
slowly moving collective coordinates. 
The higher derivative corrections to this action 
is expected to exist
but they are not known in general. 
If we could realize walls by straight D-branes 
as stated above but not by kinky D-branes,  
higher derivative corrections to this action
would be realized by the DBI action.

\medskip 
\underline{Embedding to SUGRA}. 
Wall solutions for $U(1)$ gauge theory 
in the strong gauge coupling limit 
were embedded into $D=5$ SUGRA~\cite{SUGRA}. 
There it was shown that matter parts of solutions are not changed. 
These works may be extended to 
full kinky D-brane configuration in non-Abelian gauge theory 
and/or at finite gauge coupling constant, 
coupled with SUGRA. 
It would be interesting to investigate the AdS/CFT correspondence 
in these configurations.

\section*{Acknowledgments}

We would like to thank Yuji Tachikawa and 
David Tong for valuable comments.
MN is grateful to Thomas E. Clark, 
Arkady Vainshtein and Tonnis ter Veldhuis 
for useful comments, and 
especially to Koji Hashimoto for 
valuable discussions on the reconnection of D-branes 
and comments on the DBI action.
This work is supported in part by Grant-in-Aid for Scientific 
Research from the Ministry of Education, Culture, Sports, 
Science and Technology, Japan No.13640269 (NS) 
and 16028203 for the priority area ``origin of mass'' 
(NS). 
The works of K.Ohashi and M.N. are 
supported by Japan Society for the Promotion 
of Science under the Post-doctoral Research Program. 
M.E. and Y.I. gratefully acknowledge 
support from a 21st Century COE Program at 
Tokyo Tech ``Nanometer-Scale Quantum Physics" by the 
Ministry of Education, Culture, Sports, Science 
and Technology.  
M.E. gratefully acknowledges 
support from the Iwanami Fujukai Foundation.
K.Ohta is supported in part by Special Postdoctoral Researchers
Program at RIKEN.


\newcommand{\J}[4]{{\sl #1} {\bf #2} (#3) #4}
\newcommand{\andJ}[3]{{\bf #1} (#2) #3}
\newcommand{\AP}{Ann.\ Phys.\ (N.Y.)}
\newcommand{\MPL}{Mod.\ Phys.\ Lett.}
\newcommand{\NP}{Nucl.\ Phys.}
\newcommand{\PL}{Phys.\ Lett.}
\newcommand{\PR}{ Phys.\ Rev.}
\newcommand{\PRL}{Phys.\ Rev.\ Lett.}
\newcommand{\PTP}{Prog.\ Theor.\ Phys.}
\newcommand{\hep}[1]{{\tt hep-th/{#1}}}


\begin{thebibliography}{100}
\bibitem{ADHM} 
M.~F.~Atiyah, N.~J.~Hitchin, V.~G.~Drinfeld and Yu.~I.~Manin,
Phys.\ Lett.\ A {\bf 65}, 185 (1978).

\bibitem{Nahm}  
W.~Nahm,
Phys.\ Lett.\ B {\bf 90}, 413 (1980).


\bibitem{SW}
N.~Seiberg and E.~Witten,
Nucl.\ Phys.\ B {\bf 426}, 19 (1994)
[Erratum-ibid.\ B {\bf 430}, 485 (1994)]
[arXiv:hep-th/9407087]; 
Nucl.\ Phys.\ B {\bf 431}, 484 (1994)
[arXiv:hep-th/9408099];
N.~Seiberg,
Nucl.\ Phys.\ B {\bf 435}, 129 (1995)
[arXiv:hep-th/9411149].

\bibitem{HT}
A.~Hanany and D.~Tong,
JHEP {\bf 0307}, 037 (2003)
[arXiv:hep-th/0306150].

\bibitem{HT2}
A.~Hanany and D.~Tong,
JHEP {\bf 0404}, 066 (2004)
[arXiv:hep-th/0403158].

\bibitem{ENS}
M.~Eto, M.~Nitta and N.~Sakai,
Nucl.\ Phys.\ B {\bf 701}, 247 (2004) 
[arXiv:hep-th/0405161]. 

\bibitem{Auzzi:2003fs}
  R.~Auzzi, S.~Bolognesi, J.~Evslin, K.~Konishi and A.~Yung,
  Nucl.\ Phys.\ B {\bf 673}, 187 (2003)
  [arXiv:hep-th/0307287].


\bibitem{Shifman:1997hg}
  M.~A.~Shifman,
  Phys.\ Rev.\ D {\bf 57}, 1258 (1998)
  [arXiv:hep-th/9708060]; 
  M.~A.~Shifman and M.~B.~Voloshin,
  Phys.\ Rev.\ D {\bf 57}, 2590 (1998)
  [arXiv:hep-th/9709137].

\bibitem{Acharya:2001dz}
B.~S.~Acharya and C.~Vafa,
arXiv:hep-th/0103011.

\bibitem{Ritz:2002fm}
A.~Ritz, M.~Shifman and A.~Vainshtein,
Phys.\ Rev.\ D {\bf 66}, 065015 (2002)
[arXiv:hep-th/0205083];
%
Phys.\ Rev.\ D {\bf 70}, 095003 (2004)
[arXiv:hep-th/0405175];
%
A.~Ritz,
JHEP {\bf 0310}, 021 (2003)
[arXiv:hep-th/0308144].

\bibitem{GTT2} J.~P.~Gauntlett, D.~Tong and P.~K.~Townsend,  
               Phys.\ Rev.\ D {\bf 64}, 025010 (2001)  
               [arXiv:hep-th/0012178]. 
%
\bibitem{To}  D.~Tong, 
               Phys.\ Rev.\ D {\bf 66}, 025013 (2002)  
               [arXiv:hep-th/0202012]; 
               JHEP {\bf 0304}, 031 (2003) 
               [arXiv:hep-th/0303151];
               K.~S.~M.~Lee, 
               Phys.\ Rev.\ D {\bf 67}, 045009 (2003) 
               [arXiv:hep-th/0211058]. 

\bibitem{INOS1}
Y.~Isozumi, M.~Nitta, K.~Ohashi and N.~Sakai, 
Phys. Rev. Lett. {\bf 93}, 161601 (2004)
[arXiv:hep-th/0404198].

\bibitem{INOS2} 
  Y.~Isozumi, M.~Nitta, K.~Ohashi and N.~Sakai,
  Phys.\ Rev.\ D {\bf 70}, 125014 (2004)
  [arXiv:hep-th/0405194].

%
\bibitem{INOS4}
Y.~Isozumi, M.~Nitta, K.~Ohashi and N.~Sakai, 
in the proceedings of 12th International Conference on 
Supersymmetry and Unification of Fundamental Interactions (SUSY 04), 
Tsukuba, Japan, 17-23 Jun 2004,  
edited by K. Hagiwara {\it et al.} (KEK, 2004) p.1 - p.16 [arXiv:hep-th/0409110]; 
to appear in the proceedings of ``NathFest'' at PASCOS conference, 
Northeastern University, Boston, Ma, August 2004  [arXiv:hep-th/0410150]. 

%
\bibitem{INOS3}
  Y.~Isozumi, M.~Nitta, K.~Ohashi and N.~Sakai,
  Phys.\ Rev.\ D {\bf 71}, 065018 (2005)
  [arXiv:hep-th/0405129].

\bibitem{wv}
              J.~P.~Gauntlett, R.~Portugues, D.~Tong, and P.K.~Townsend, 
               Phys.\ Rev.\ D {\bf 63}, 085002 (2001)  
               [arXiv:hep-th/0008221];  
M.~Shifman and A.~Yung,
Phys.\ Rev.\ D {\bf 67}, 125007 (2003)
[arXiv:hep-th/0212293].

\bibitem{vm}
D.~Tong,
Phys.\ Rev.\ D {\bf 69}, 065003 (2004)
[arXiv:hep-th/0307302];
R.~Auzzi, S.~Bolognesi, J.~Evslin and K.~Konishi,
Nucl.\ Phys.\ B {\bf 686}, 119 (2004)
[arXiv:hep-th/0312233]; 
M.~Shifman and A.~Yung,
Phys.\ Rev.\ D {\bf 70}, 045004 (2004)
[arXiv:hep-th/0403149]. 

\bibitem{Kneipp}
M.~A.~C.~Kneipp and P.~Brockill,
Phys.\ Rev.\ D {\bf 64}, 125012 (2001) [arXiv:hep-th/0104171];
M.~A.~C.~Kneipp,
Phys.\ Rev.\ D {\bf 68}, 045009 (2003) [arXiv:hep-th/0211049]; 
Phys.\ Rev.\ D {\bf 69}, 045007 (2004) [arXiv:hep-th/0308086];
arXiv:hep-th/0401234. 

\bibitem{Auzzi:2004yg}
  R.~Auzzi, S.~Bolognesi and J.~Evslin,
  JHEP {\bf 0502}, 046 (2005)
  [arXiv:hep-th/0411074].

\bibitem{EINOS1}
M.~Eto, Y.~Isozumi, M.~Nitta, K.~Ohashi and N.~Sakai, 
arXiv:hep-th/0412048

\bibitem{HW}
A.~Hanany and E.~Witten,
Nucl.\ Phys.\ B {\bf 492}, 152 (1997)
[arXiv:hep-th/9611230].

\bibitem{Witten:1997sc}
E.~Witten,
Nucl.\ Phys.\ B {\bf 500}, 3 (1997)
[arXiv:hep-th/9703166].
%
\bibitem{HOO}
K.~Hori, H.~Ooguri and Y.~Oz,
Adv.\ Theor.\ Math.\ Phys.\  {\bf 1}, 1 (1998)
[arXiv:hep-th/9706082].

\bibitem{NOYY}
T.~Nakatsu, K.~Ohta, T.~Yokono and Y.~Yoshida,
Nucl.\ Phys.\ B {\bf 519}, 159 (1998)
[arXiv:hep-th/9707258].

\bibitem{NOYY2}
T.~Nakatsu, K.~Ohta, T.~Yokono and Y.~Yoshida,
Mod.\ Phys.\ Lett.\ A {\bf 13}, 293 (1998)
[arXiv:hep-th/9711117].

\bibitem{Giveon:1998sr}
A.~Giveon and D.~Kutasov,
Rev.\ Mod.\ Phys.\  {\bf 71}, 983 (1999)
[arXiv:hep-th/9802067].

\bibitem{Wi}
E.~Witten,
Nucl.\ Phys.\  {\bf B460}, 541 (1996)
[arXiv:hep-th/9511030]; 
M.~R.~Douglas,
arXiv:hep-th/9512077.

\bibitem{brane-monopole}
M.~B.~Green and M.~Gutperle,
Phys.\ Lett.\  B {\bf 377}, 28 (1996)
[arXiv:hep-th/9602077]; 
D.~E.~Diaconescu,
Nucl.\ Phys.\ B {\bf 503}, 220 (1997)
[arXiv:hep-th/9608163].

\bibitem{NS}
N.~Nekrasov and A.~Schwarz,
Commun.\ Math.\ Phys.\  {\bf 198}, 689 (1998)
[arXiv:hep-th/9802068]; 
K.~Furuuchi,
Prog.\ Theor.\ Phys.\  {\bf 103}, 1043 (2000)
[arXiv:hep-th/9912047].

\bibitem{Hashimoto:1999zw}
A.~Hashimoto and K.~Hashimoto,
JHEP {\bf 9911}, 005 (1999)
[arXiv:hep-th/9909202];
K.~Hashimoto, H.~Hata and S.~Moriyama,
JHEP {\bf 9912}, 021 (1999)
[arXiv:hep-th/9910196]; 
D.~J.~Gross and N.~A.~Nekrasov,
JHEP {\bf 0007}, 034 (2000)
[arXiv:hep-th/0005204].

\bibitem{Gross:2000ss}
D.~J.~Gross and N.~A.~Nekrasov,
JHEP {\bf 0103}, 044 (2001)
[arXiv:hep-th/0010090];
M.~Hamanaka,
Phys.\ Rev.\ D {\bf 65}, 085022 (2002)
[arXiv:hep-th/0109070];
arXiv:hep-th/0303256.

\bibitem{LT}
N.~D.~Lambert and D.~Tong,
Nucl.\ Phys.\ B {\bf 569}, 606 (2000)
[arXiv:hep-th/9907098].

\bibitem{Eguchi:1978xp}
T.~Eguchi and A.~J.~Hanson,
Phys.\ Lett.\ B {\bf 74}, 249 (1978);
T.~Eguchi, P.~B.~Gilkey and A.~J.~Hanson,
Phys.\ Rept.\  {\bf 66}, 213 (1980).



\bibitem{KN}
P.~B.~Kronheimer and H.~Nakajima, 
Math. Ann.  {\bf 288}, 263 (1990); 
H.~Nakajima,  
Duke Math. J. {\bf 76}, 365 (1994).

\bibitem{DM}
M.~R.~Douglas and G.~W.~Moore,
arXiv:hep-th/9603167; 
C.~V.~Johnson and R.~C.~Myers,
Phys.\ Rev.\ D {\bf 55}, 6382 (1997)
[arXiv:hep-th/9610140].

\bibitem{recombination}
K.~Hashimoto and S.~Nagaoka,
JHEP {\bf 0306}, 034 (2003)
[arXiv:hep-th/0303204]; 
K.~Hashimoto and W.~Taylor,
JHEP {\bf 0310}, 040 (2003)
[arXiv:hep-th/0307297];
S.~Nagaoka,
JHEP {\bf 0402}, 063 (2004)
[arXiv:hep-th/0312010].

%


\bibitem{HKLR}
N.~J.~Hitchin, A.~Karlhede, U.~Lindstr\"om and M.~Ro\v{c}ek,
Commun.\ Math.\ Phys.\  {\bf 108}, 535 (1987).

\bibitem{LR}
U.~Lindstr\"om and M.~Ro\v{c}ek,
Nucl.\ Phys.\ B {\bf 222} (1983) 285.

\bibitem{ANS}
  M.~Arai, M.~Nitta and N.~Sakai,
  Prog.\ Theor.\ Phys.\  {\bf 113}, 657 (2005)
  [arXiv:hep-th/0307274]; 
to appear in the Proceedings of the 3rd International Symposium on Quantum Theory and Symmetries (QTS3), September 10-14, 2003, 
[arXiv:hep-th/0401084];
to appear in the Proceedings of the International Conference on ``Symmetry Methods in Physics (SYM-PHYS10)'' held at Yerevan, Armenia, 13-19 Aug. 2003
[arXiv:hep-th/0401102]; 
to appear in the Proceedings of  
SUSY 2003 held at the University of Arizona, Tucson, AZ, June 5-10, 2003
[arXiv:hep-th/0402065].


\bibitem{Carlino:2000ff}
G.~Carlino, K.~Konishi and H.~Murayama,
JHEP {\bf 0002}, 004 (2000)
[arXiv:hep-th/0001036]; 
Nucl.\ Phys.\ B {\bf 590}, 37 (2000)
[arXiv:hep-th/0005076].



\bibitem{KS} K.~Kakimoto and N.~Sakai, 
          Phys.\ Rev.\ D {\bf 68}, 065005 (2003)  
          [arXiv:hep-th/0306077]. 


\bibitem{IOS1} Y.~Isozumi, K.~Ohashi, and N.~Sakai, 
  JHEP\ {\bf 0311}, 060 (2003) 
  [arXiv:hep-th/0310189]. 

\bibitem{IOS2} Y.~Isozumi, K.~Ohashi, and N.~Sakai, 
 JHEP\ {\bf 0311}, 061 (2003)
  [arXiv:hep-th/0310130]. 



\bibitem{AF}
L.~Alvarez-Gaume and D.~Z.~Freedman,
Commun.\ Math.\ Phys.\  {\bf 91}, 87 (1983).


\bibitem{AT} 
 E.~Abraham and P.~K.~Townsend, 
             Phys.\ Lett.\ B {\bf 291}, 85 (1992).
%
%
%
\bibitem{ANNS} 
M.~Arai, M.~Naganuma, M.~Nitta, and N.~Sakai, 
   Nucl.\ Phys.\ B {\bf 652},  35 (2003) [arXiv:hep-th/0211103]; 
``BPS Wall in N=2 SUSY Nonlinear Sigma Model with Eguchi-Hanson Manifold''
in Garden of Quanta - In honor of Hiroshi Ezawa, 
Eds. by J.~Arafune et al. 
(World Scientific Publishing Co. Pte. Ltd. Singapore, 2003) 
pp 299-325, [arXiv:hep-th/0302028]; 
M. Arai, E.~Ivanov and J.~Niederle, 
              Nucl.\ Phys.\ B {\bf 680},  23 (2004) 
              [arXiv:hep-th/0312037]. 
%
\bibitem{GTT1}
 J.~P.~Gauntlett, D.~Tong, and P.K.~Townsend, 
               Phys.\ Rev.\ D {\bf 63} 085001 (2001)  
               [arXiv:hep-th/0007124].    
%
\bibitem{Naganuma:2001pu}
M.~Naganuma, M.~Nitta and N.~Sakai,
Grav.\ Cosmol.\  {\bf 8}, 129 (2002)
[arXiv:hep-th/0108133]; 
 R.~Portugues and P.~K.~Townsend, 
               JHEP\ {\bf 0204}, 039 (2002) [arXiv:hep-th/0203181]. 

\bibitem{EINOST}
M.~Eto, Y.~Isozumi, M.~Nitta, K.~Ohashi, K.~Ohta, 
N.~Sakai and Y.~Tachikawa, 
unpublished.




\bibitem{QNG}
W.~Buchmuller, S.~T.~Love, R.~D.~Peccei and T.~Yanagida, 
Phys.\ Lett.\ B {\bf 115}, 233 (1982);
W.~Buchmuller, R.~D.~Peccei and T.~Yanagida,
Phys.\ Lett.\ B {\bf 124}, 67 (1983);
Nucl.\ Phys.\ B {\bf 227}, 503 (1983);  
K.~Higashijima, M.~Nitta, K.~Ohta and N.~Ohta,
Prog.\ Theor.\ Phys.\ {\bf 98}, 1165 (1997) 
[arXiv:hep-th/9706219];
M.~Nitta,
Int.\ J.\ Mod.\ Phys.\ A {\bf 14}, 2397 (1999)
[arXiv: hep-th/9805038]; 
K.~Higashijima and M.~Nitta,
Prog.\ Theor.\ Phys.\  {\bf 103}, 635 (2000)
[arXiv:hep-th/9911139];
K.~Furuta, T.~Inami, H.~Nakajima and M.~Nitta,
Prog.\ Theor.\ Phys.\  {\bf 106}, 851 (2001)
[arXiv:hep-th/0106183].


\bibitem{Binosi:2000jb}
D.~Binosi and T.~ter Veldhuis,
Phys.\ Rev.\ D {\bf 63}, 085016 (2001)
[arXiv:hep-th/0011113].


\bibitem{EMS}
M.~Eto, N.~Maru and N.~Sakai,
Nucl.\ Phys.\ B {\bf 696}, 3 (2004)
[arXiv:hep-th/0404114]; 
in the proceedings of 12th International Conference on 
Supersymmetry and Unification of Fundamental Interactions (SUSY 04), 
Tsukuba, Japan, 17-23 Jun 2004,  
edited by K. Hagiwara {\it et al.} (KEK, 2004) p.849 - p.852 
[arXiv:hep-th/0409152].

\bibitem{Clark:2002bh}
T.~E.~Clark, M.~Nitta and T.~ter Veldhuis,
Phys.\ Rev.\ D {\bf 67}, 085026 (2003)
[arXiv:hep-th/0208184];
Phys.\ Rev.\ D {\bf 69}, 047701 (2004)
[arXiv:hep-th/0209142];
Phys.\ Rev.\ D {\bf 70}, 105005 (2004) 
[arXiv:hep-th/0401163];
  Phys.\ Rev.\ D {\bf 71}, 025017 (2005)
  [arXiv:hep-th/0409030]; 
  Phys.\ Rev.\ D {\bf 70}, 125011 (2004)
  [arXiv:hep-th/0409151].
  

\bibitem{Shifman:2003uh}
M.~Shifman and A.~Yung,
Phys.\ Rev.\ D {\bf 70}, 025013 (2004)
[arXiv:hep-th/0312257].


\bibitem{EINOS}
M.~Eto, Y.~Isozumi, M.~Nitta, K.~Ohashi and N.~Sakai, 
in preparation.


\bibitem{junction}
H.~Oda, K.~Ito, M.~Naganuma and N.~Sakai,
Phys.\ Lett.\ B {\bf 471}, 140 (1999)
[arXiv:hep-th/9910095]; 
%
K.~Ito, M.~Naganuma, H.~Oda and N.~Sakai,
Nucl.\ Phys.\ B {\bf 586}, 231 (2000)
[arXiv:hep-th/0004188];
%
Nucl.\ Phys.\ Proc.\ Suppl.\  {\bf 101}, 304 (2001)
[arXiv:hep-th/0012182];
%
M.~Naganuma, M.~Nitta and N.~Sakai,
Phys.\ Rev.\ D {\bf 65}, 045016 (2002)
[arXiv:hep-th/0108179]; 
in the proceedings of 
3rd International Sakharov Conference On Physics, 
24-29 Jun 2002, Moscow, Russia, 
edited by A. Semikhatov {\it et al.} (Scientific World Pub., 2003) 
p.537 - p.549 
[arXiv:hep-th/0210205].


\bibitem{Naganuma:2000gu}
M.~Naganuma and M.~Nitta,
Prog.\ Theor.\ Phys.\  {\bf 105}, 501 (2001)
[arXiv:hep-th/0007184].


\bibitem{Elitzur:1997fh}
S.~Elitzur, A.~Giveon and D.~Kutasov,
Phys.\ Lett.\ B {\bf 400}, 269 (1997)
[arXiv:hep-th/9702014].

\bibitem{Barbon:1997zu}
J.~L.~F.~Barbon,
Phys.\ Lett.\ B {\bf 402}, 59 (1997)
[arXiv:hep-th/9703051].

\bibitem{BIon}
C.~G.~Callan,Jr. and J.~M.~Maldacena, 
Nucl.\ Phys.\ B {\bf 513}, 198 (1998);  
G.~W.~Gibbons,
Nucl.\ Phys.\ B {\bf 514}, 603 (1998); 
A.~Hashimoto, 
Phys.\ Rev.\ D {\bf 57}, 6441 (1998). 

\bibitem{SUGRA}
M.~Arai, S.~Fujita, M.~Naganuma and N.~Sakai,
Phys.\ Lett.\ B {\bf 556}, 192 (2003)
[arXiv:hep-th/0212175];
to appear in the proceedings of International Seminar on Supersymmetries and Quantum Symmetries SQS 03, Dubna, Russia, 24-29 Jul 2003,  
[arXiv:hep-th/0311210]; 
to appear in the proceedings of SUSY 2003: SUSY in the Desert: 11th Annual International Conference on Supersymmetry and the Unification of Fundamental Interactions, Tucson, Arizona, 5-10 Jun 2003.
[arXiv:hep-th/0402040]; 
M.~Eto, S.~Fujita, M.~Naganuma and N.~Sakai,
Phys.\ Rev.\ D {\bf 69}, 025007 (2004)
[arXiv:hep-th/0306198].



\end{thebibliography}
\end{document}